\documentstyle[elsart12]{article}

\def\eqalign#1{\null\,\vcenter{\openup1\jot \mathsurround=0pt
     \ialign{\strut\hfil$\displaystyle{##}$&$\displaystyle{{}##}$\hfil
      \crcr#1\crcr}}\,}

\def\be{\begin{equation}} 
\def\ee{\end{equation}}
\def\d{\partial}
\def\half{{\mathchoice{{\textstyle{1\over 2}}}{1\over 2}{1\over 2}{1
\over 2}}} 
\def\quarter{{\mathchoice{{\textstyle{1\over 4}}}{1\over 4}{1\over 4}
{1 \over 4}}}

\def\phionebar{\bar\phi_1}
\def\phitwobar{\bar\phi_2}

\def\Phidag{\Phi^\dagger}

\def\Dbw{\mathop{D}\limits^{\leftrightarrow}} 
\def\({\left(}
\def\){\right)} 
\def\[{\left[} 
\def\]{\right]}
\def\RR{{\rm I\!\!\, R}}
\def\CC{{\rm I\!\!\! C}}
\def\ZZ{{\rm \angle \!\!\! Z}}

\include{epsf}

\begin{document}

\

\vskip 1 truecm

\centerline{\Large \bf Semilocal and Electroweak Strings}

\

\

\centerline{Ana Ach\'ucarro$^{1,2}$ and Tanmay Vachaspati$^3$}

\

\centerline{
$^1$ Department of Theoretical Physics, UPV-EHU, 48080 Bilbao, Spain.}

\centerline{
$^2$ Institute for Theoretical Physics, University of Groningen, 
The Netherlands.}

\centerline{
$^3$Physics Department, Case Western Reserve University, Cleveland, 
OH 44106, USA.}

\

\

\

\centerline{\it Abstract}

\smallskip

We review a class of non-topological defects in the standard 
electroweak model, and their implications. 
Starting with the semilocal string, which provides a
counterexample to many well known properties of topological vortices,
we discuss electroweak strings and their stability with and without
external influences such as magnetic fields. Other known properties of
electroweak strings and monopoles are described in some detail and
their potential relevance to future particle accelerator experiments
and to baryon number violating processes is considered. We also review
recent progress on the cosmology of electroweak defects and the connection
with superfluid helium, where some of the effects discussed here
could possibly be tested.

\
\
\vfil
%\leftline{EHU-FT/9808, CWRU-P34-1998}
\eject
\tableofcontents \vfill\eject
\vfill\eject

\section{Introduction}
\label{intro}

In a classic paper from 1977 \cite{Nam77}, a decade after the $SU(2)_L \times
U(1)_Y$  model of electroweak interactions had been proposed
\cite{GlaSalWei}, Nambu made the observation that, while the 
Glashow-Salam-Weinberg (GSW) model does not admit isolated, regular
magnetic monopoles, there could be monopole-antimonopole pairs joined
by short segments of a vortex carrying Z-magnetic field 
(a {\it Z-string}). The monopole
and antimonopole would tend to annihilate but, he argued, longitudinal
collapse could be stopped by rotation. He dubbed these configurations
{\it dumbells}\footnote{or {\it monopolia}, after analogous
configurations in superfluid helium \cite{Mak77}} and estimated their
mass at a few TeV. A number of papers advocating other, related,
soliton-type solutions 
\footnote{One example, outside the scope of the present review,
are so-called vorticons, proposed by Huang and Tipton,
which are closed loops of string with one quantum of Z boson trapped
inside.} 
in the same energy range followed \cite{early},
but the lack of topological stability led to the idea finally being
abandoned during the eighties. 

Several years later, and completely independently, it was observed
that the coexistence of global and gauge symmetries can lead to stable
non-topological strings called ``semilocal strings'' \cite{VacAch91} in
the $\sin^2 \theta_w = 1$ limit of the GSW model that Nambu had
considered.  Shortly afterwards it was proved that Z-strings were 
stable near this limit \cite{Vac92}, and the whole subject made a
comeback.  This report is a review of the current status of research
on electroweak strings.

Apart from the possibility that electroweak strings may be the first
solitons to be observed in the standard model, there are two
interesting consequences of the study of electroweak and semilocal
strings.  One is the unexpected connection with baryon number and
sphalerons. The other is a deeper understanding of the connection
between the topology of the vacuum manifold (the set of ground states
of a classical field theory) and the existence of stable
non-dissipative configurations, in particular when global and local
symmetries are involved simultaneously.

In these pages we assume a level of familiarity with the general
theory and basic properties of topological defects, in particular with
the homotopy classification. There are some
excellent reviews on this subject in the literature to which we refer
the reader \cite{reviews,Col85a,SalVol87}. 
On the other hand, electroweak and
semilocal strings are non-topological defects, and  this forces us to take a
slightly different point of view from most of the existing literature. Emphasis
on stability properties is mandatory, since one cannot be sure from the start
whether these defects will actually form. With very few exceptions, this
requires an analysis on a case by case basis.

Following the  
discussion in \cite{Col85}, one should begin with the definition of
dissipative configurations.
 Consider a classical field
theory with energy density $T_{00} \geq 0$ such that $T_{00} = 0$
everywhere for the ground states (or ``vacua'') of the theory. A
solution of a classical field theory is said to be dissipative if
\be \lim_{t \to \infty} {\rm max}_{\bf x} T_{00} ({\bf x},t) = 0
\ee
 
We will consider theories with spontaneous symmetry breaking from a Lie
group $G$ (which we assume to be finite-dimensional and compact) to a subgroup
$H$; the space
$\cal V$ of ground states of the theory
is usually called the vacuum manifold and, in the absence of
accidental degeneracy, is given by ${\cal V}=G/H$.

The classification of topological defects is based on the homotopy
properties of the vacuum manifold. If the vacuum manifold contains
non-contractible $n$-spheres then field configurations in $n+1$ spatial
dimensions whose asymptotic values as $r \to \infty$ ``wrap around'' those
spheres are necessarily non-dissipative, since continuity of the
scalar field guarantees that, at all times,  at least in one point in space the scalar
potential (and thus the energy) will be non-zero.  The
region in space where energy is localized is referred to as a {\it
topological defect}. 
Field configurations whose asymptotic values are in the same 
homotopy class are said to be in the same {\em topological sector} or to 
have the same {\em winding number}.

In three spatial dimensions, it is customary to use the names {\it
monopole}, {\it string}\footnote{The names {\it cosmic string} and
{\it vortex} are also common. Usually, ``vortex'' refers to the
configuration in two spatial dimensions, and ``string'' to the
corresponding configuration in three spatial dimensions; the adjective
``cosmic'' helps to distinguish them from the so-called fundamental
strings or superstrings. }
and {\it domain wall} to refer to defects that are pointlike,
one-dimensional or two-dimensional respectively. Thus, one can have
topological domain walls only if $\pi_0({\cal V}) \neq 1$, topological
strings only if $\pi_1({\cal V})\neq 1$ and topological monopoles only
if $\pi_2({\cal V}) \neq 1$.  
Besides, defects in different topological sectors
cannot be deformed into each other without introducing 
singularities or supplying an infinite amount of energy. 
This is the origin of the homotopy classification of topological
defects. We should point out that the topological classification of
textures based on $\pi_3({\cal V})$ has a very different character,
and will not concern us here; in particular, configurations from
different topological sectors can be continuously deformed into each
other with a finite cost in energy. 
In general, textures unwind until they reach the vacuum sector and
therefore they are dissipative.

It is well known, although not always sufficiently stressed, that the
precise relationship between the topology of the vacuum and
the existence of stable defects is subtle.  First of all, note that a
trivial topology of the vacuum manifold does not imply the {\it
non-existence} of stable defects.

Secondly, we have said that a non-trivial homotopy of the vacuum
manifold can result in non-dissipative solutions  but, in general,
these solutions need not be time independent nor stable to small
perturbations.  One exception is the field theory of a single scalar
field in 1+1 dimensions, where a disconnected vacuum manifold
(i.e. one with $\pi_0({\cal V}) \neq 1$) is sufficient to prove the
existence of time independent, classically stable ``kink'' solutions
\cite{GolJac75, Col85}. But this is not the norm. The $O(3)$ model, for
instance, has topological global monopoles
\cite{BarVil89} which are time independent, 
but they are
unstable to angular collapse even in the
lowest non-trivial winding sector\cite{Gol89}.

It turns out that the situation is particularly subtle in theories
where there are global and gauge symmetries involved simultaneously.
The prototype example is the semilocal string, described in
section \ref{semilocal}.  In the semilocal string model, the
classical dynamics is governed by a single parameter $\beta =
m_s^2/m_v^2$ that measures the square of the ratio of 
the scalar mass, $m_s$, to the
vector mass, $m_v$ (this is the same parameter that distinguishes type
I and type II superconductors). It turns out that:

- when $\beta>1$ the semilocal model provides a counterexample to the
widespread belief that quantization of magnetic flux is tantamount to
its localization, i.e., confinement. The vector boson is massive and
we expect this to result in confinement of magnetic flux to regions of
width given by the inverse vector mass. However, this is not the case!
As pointed out by Hindmarsh \cite{Hin92} and Preskill \cite{Pre92}, this
is a system where magnetic flux is topologically conserved and
quantized, and there is a finite energy gap between the non-zero flux
sectors and the vacuum, and yet there are {\it no stable vortices}.

- when $\beta <1$ strings are stable\footnote{We want to stress that,
contrary to what is often stated in the literature, the semilocal
string with $\beta <1$ is {\it absolutely stable}, and not just {\it
metastable}.} even though the vacuum manifold is simply connected,
{$\pi_1 ({\cal V}) = 1$}.  Semilocal vortices with $\beta < 1$
are a remarkable example of a non-topological defect
which is stable both perturbatively and to semiclassical tunnelling
into the vacuum
\cite{PreVil92}.

As a result, when { the global symmetries of a semilocal
model} are gauged, dynamically stable non-topological solutions can
still exist for certain ranges of parameters very close to stable
semilocal limits. In the case of the standard electroweak model, for
instance, strings are (classically) stable only when $\sin^2 \theta_w
\approx 1$ and the mass of the Higgs is smaller than the mass of the Z
boson.

We begin with a description of the Glashow-Salam-Weinberg model,
in order to set our notation and conventions, and a brief discussion
of topological vortices (cosmic strings). It will be sufficient for
our purposes to review cosmic strings in the Abelian Higgs model, with
a special emphasis on those aspects that will be relevant to
electroweak and semilocal strings. We should point out that these
vortices were first considered in condensed matter by Abrikosov
\cite{Abr57} in the non-relativistic case, in connection with type II
superconductors. Nielsen and Olesen were the first to consider them in
the context of relativistic field theory, so we will follow a standard
convention in high energy physics and refer to them as Nielsen-Olesen
strings
\cite{NieOle73}.

Sections \ref{semilocal} to
\ref{zoo} are dedicated to semilocal and electroweak strings,  and other 
embedded defects in the standard GSW model. Electroweak strings in
extensions of the GSW model are discussed in section
\ref{ewstringinextensions}.

In section \ref{stability} the stability of straight, infinitely long
electroweak strings is analysed in detail (in the absence of
fermions). Sections \ref{superconductivity} to
\ref{ewstringandsphaleron} investigate fermionic superconductivity 
on the string, the effect of fermions on the string stability, and the
scattering of fermions off electroweak strings.  The surprising connection
between strings and baryon number, and their relation to sphalerons,
is described in sections
\ref{ewstringbaryonnumber} and \ref{ewstringandsphaleron}.
Here we also discuss the possibility of
string formation in particle accelerators (in the form of dumbells, as
was suggested by Nambu in the seventies) and in the early universe. 

Finally, section \ref{he3section} describes a condensed matter analog
of electroweak strings in superfluid helium which may be used to test
our ideas on vortex formation, fermion scattering and baryogenesis.

A few comments are in order: 

$\bullet$ Unless otherwise stated we take spacetime to be flat, 3+1
dimensional Minkowski space; the gravitational properties of embedded
strings are expected to be the same as those of Nielsen-Olesen strings
\cite{GibOrtRuiSam92} and will not be considered here.  A limited discussion
of possible cosmological implications can be found in sections 
\ref{networks} and \ref{cosmologicalapplications}.

$\bullet$ We concentrate on regular defects in the standard model of
electroweak interactions. Certain extensions of the
Glashow-Salam-Weinberg model are {briefly} considered in
section 6 but otherwise they are outside the scope of this review; the
same is true of singular solutions.  In particular, we do not discuss
isolated monopoles in the GSW model
\cite{GibOrtRuiSam92,ChoMai96}, which are necessarily singular.

$\bullet$ No family mixing effects are discussed in this review 
and we also ignore $SU(3)_c$ colour interactions, even though their 
physical affects are expected to be very interesting, in particular in
connection with baryon production by strings (see section
\ref{ewstringbaryonnumber}). 

$\bullet$ Our conventions are the following: spacetime has signature
$(+,-,-,-)$. Planck's constant and the speed of light are set to one,
$\hbar = c = 1$.  The notation $(x)$ is shorthand for all spacetime
coordinates $(x^0, x^i),
\ i = 1,2,3$; whenever the $x$-coordinate is meant, it will be
stated explicitly. We also use the notation $(t, {\vec x})$. 

$\bullet$ Complex conjugation and hermitian conjugation are both
indicated with the same symbol, ($^\dagger$), but it should be clear
from the context which one is meant. For fermions, ${\overline \psi} =
\psi^\dagger \gamma^0$, as usual. 
Transposition is indicated with the symbol $(^T)$.

$\bullet$ One final word of caution: a gauge field is a Lie Algebra
valued one-form $A = A_\mu dx^\mu = A_\mu^a T^a dx^\mu$, but it is
also customary to write it as a vector.
In cylindrical coordinates $(t,\rho,\varphi,z)$,
$ A = A_t dt + A_\rho d\rho + A_\varphi d\varphi + A_z dz$ is often written
${\vec {A}} = 
{A}_t \hat{t} + 
{A}_\rho \hat{\rho} 
+ ({A}_\varphi / \rho)  \hat{\varphi} + {A}_z \hat{z} $, 
In spherical coordinates, $(t,r,\theta,\varphi)$, 
$ A = A_t dt + A_r dr
+ A_\theta d\theta + A_\varphi d\varphi $ is also written ${\vec
{A}} = {A}_t \hat{t} + {A}_r \hat{r} + ({A}_\theta /r) 
\hat{\theta} + ({A}_\varphi / r\sin\theta)  \hat{\varphi} $.  
We use both notations throughout.

\subsection{The Glashow-Salam-Weinberg model.}
\label{WS}

In this section we set out our conventions, which mostly follow those of
\cite{CheLi91}.

The standard (GSW) model of electroweak interactions is described by
the Lagrangian
\be L = L_b \ + \sum_{\rm families} L_f  \ + \  L_{fm}
\label{GWSaction}\ee 
The first term describes the bosonic sector, comprising a neutral
scalar
$\phi^0$, 
a charged scalar
$\phi^+$, a massless photon
$A_\mu$, and three massive vector bosons, two of them charged
($W^\pm_\mu$) and the  neutral $Z_\mu$. 

The last two terms describe the dynamics of
the fermionic sector, which consists of the  three  families of quarks and 
leptons
\be 
     \pmatrix{\nu_e \cr e \cr u\cr d\cr} \qquad
     \pmatrix{\nu_\mu \cr \mu \cr c\cr s\cr}\qquad
	    \pmatrix{\nu_\tau \cr \tau \cr t\cr b\cr}
\ee

\subsubsection{The bosonic sector}

The bosonic sector  
describes an $SU(2)_L\times U(1)_Y$ invariant theory with a scalar field
$\Phi$ in the fundamental representation of $SU(2)_L$. It
is described by the Lagrangian:
\begin{equation}
L_b = L_W + L_Y + L_{\Phi} - V(\Phi )
\label{GSWb}
\end{equation}
with
\begin{equation}
\eqalign{
L_W &= - {1 \over 4} W_{\mu \nu}^{\ \  a} W^{\mu \nu a} \qquad, \qquad a = 1,2,3
\cr
L_Y &= - {1 \over 4} Y_{ \mu \nu} Y^{\mu \nu} \cr}
\label{2.3}
\end{equation}
where 
$\ W_{\mu \nu}^{\ \ a} = \partial_\mu W_\nu^a - \partial_\nu W_\nu^a + g
\epsilon^{abc} W_\mu^b W_\nu^c \ 
$ 
and $\ Y_{\mu \nu}  = \partial_\mu Y_\nu - \partial_\nu Y_\mu \ $ are the field
strengths for the $SU(2)_L$ and $U(1)_Y$ gauge fields  respectively. 
Summation over repeated $SU(2)_L$ indices is
understood, and there is no need to distinguish between upper and
lower ones. $\epsilon^{123} = 1$. Also,
\begin{equation}
L_\Phi = |D_\lambda \Phi |^2 \equiv
   \biggl |\biggl (\partial _\lambda - 
             {{ig} \over 2} \tau ^a W_\lambda ^a - 
                  {{ig'} \over 2} Y_\lambda \biggr) \Phi \biggr| ^2
\label{2.4}
\end{equation}
\begin{equation}
V(\Phi ) = \lambda (\Phi ^{\dag} \Phi - \eta ^2 /2 )^2 \ ,
\label{2.5}
\end{equation}
where $\tau^a$ are the Pauli matrices,
\be
\tau^1 = \pmatrix {0 \ \ 1 \cr 1 \ \ 0}\ \ , \qquad
\tau^2 = \pmatrix {0 \ -i \cr i \ \ \ 0}\ \  , \qquad
\tau^3 = \pmatrix {1 \ \ \ 0 \cr 0 \ -1} \ \ ,
\ee
from which one constructs the weak isospin generators $T^a = {1 \over
2}\tau^a $ satisfying $[ T^a, T^b ] = i\epsilon^{abc} T^c$. 

The classical field equations of motion for the bosonic sector of the
standard model of the electroweak interactions are (ignoring fermions):
\be
D^\mu D_\mu \Phi + 2\lambda \left ( \Phi^{\dag} \Phi -
                                 {{\eta^2} \over 2} \right ) \Phi = 0
\label{phieqn}
\ee
\be
D_\nu W^{\mu \nu a} = j^{\mu a}_W = {i \over 2} g \left [
   \Phi^{\dag} \tau^a D^\mu \Phi - (D^\mu \Phi )^{\dag} \tau^a \Phi
                                              \right ]
\label{weqn}
\ee
\be
\partial_\nu Y^{\mu \nu} = j^{\mu}_Y = {i \over 2} g' \left [
      \Phi^{\dag} D^\mu \Phi - (D^\mu \Phi )^{\dag} \Phi
                                                     \right ] \ \ ,
\label{ye}
\ee
where
$D_\nu W^{\mu \nu a} =  \partial_\nu W^{\mu \nu a} +
                           g \epsilon^{abc} W_\nu ^b W^{\mu \nu c} \ $.

When the Higgs field $\Phi$ acquires a vacuum expectation
value (VEV), the symmetry breaks from $SU(2)_L\times U(1)_Y$ to
$U(1)_{em}$.  In particle physics it is standard practice to work in
unitary gauge and take the VEV of the Higgs to be
$\langle\Phi ^T \rangle = \eta(0,1)/\sqrt 2$.  In that case the
unbroken $U(1)$ subgroup, which describes electromagnetism, is
generated by the charge operator
\be
Q \ \equiv \  T^3 + {Y \over 2} \ = \  
  \pmatrix{1 \ \ 0 \cr 0 \ \ 0}
\ee
and the two components of the Higgs doublet are charge
eigenstates
\begin{equation}
\Phi = \pmatrix{\phi^+\cr \phi^0\cr}\ \ .
\end{equation}
$Y$ is the
hypercharge operator, which acts on the Higgs like the $2 \times 2$  identity
matrix. Its eigenvalue on the various matter  fields
can be read off from the covariant derivatives $D_\mu = \partial_\mu -
ig W_\mu^a T^a - i g' Y_\mu (Y/2) $ which are listed explicitly in
equations (\ref{2.4}) and (\ref{DPsi})-(\ref{Ddr}).

In unitary gauge, the $Z$ and $A$ fields
are defined as
\begin{equation}
Z_\mu \equiv \cos\theta_wW_\mu^3 - \sin\theta_w Y_\mu \ ,
\ \ \ \
A_\mu \equiv \sin\theta_wW_\mu^3+ \cos\theta_w Y_\mu \ ,
\label{2.9}
\end{equation}
and $W_\mu^\pm \equiv (W_\mu^1 \mp i W_\mu^2) / \sqrt 2 $ are the W bosons.
The weak mixing angle $\theta_w$ is given by $\tan\theta_w \equiv {{g'} / g} $;
electric charge is 
$e = g_z \sin\theta_w \cos\theta_w$ with $g_z \equiv (g^2 +
{g'} ^2 )^{1/2} $.

However, unitary gauge is not the most convenient choice in the
presence of topological defects, where it is often singular.  Here we
shall need a more general definition in terms of an arbitrary Higgs
configuration $\Phi (x)$:
\be
Z_\mu \equiv \cos\theta_w ~n^a (x) W_\mu ^a - \sin\theta_w ~Y_\mu \ ,
\ \ \ \
A_\mu \equiv \sin\theta_w ~n^a (x) W_\mu ^a + \cos\theta_w ~Y_\mu \ ,
\label{zmuamu}
\ee
where
\be
n^a (x) \equiv  -{{\Phi^{\dag}(x) \tau^a \Phi (x)} \over {\Phi^{\dag}(x)
\Phi}(x)}
\ .
\label{nadef}
\ee
 is a unit vector by virtue of the Fierz identity $\ \sum_a (\Phidag
\tau^a \Phi)^2 = (\Phidag \Phi)^2\ $. In what follows we omit writing
the $x$-dependence of $n^a$ explicitly. Note that $n^a$ is ill defined
when $\Phi = 0$, so in particular at the defect cores.

The generators associated with the photon and the Z-boson are, 
respectively,
\be 
Q = n^aT^a + Y/2 \ \ , 
\qquad\qquad T_Z = \cos^2\theta_w n^aT^a - \sin^2 \theta_w {Y\over 2} = 
n^aT^a - \sin^2 \theta_w Q \ \ ,
\label{QandTzgenerators}\ee
while the generators associated with the (charged) W bosons are
determined, up to a phase, by the conditions
\be [Q, T^\pm] = \pm T^\pm  \qquad\qquad 
[T^+,T^-] = n^aT^a = T_Z + \sin^2 \theta_w Q \ \ , \qquad\qquad (T^+)^\dagger = T^-
\ee  
(note that, if $n^a = (0,0,1)$ as is the case in unitary gauge, one
would take $T^\pm = (T^1 \pm i T^2)/\sqrt 2$.)

 There are several different choices for defining the electromagnetic
field strength but, following Nambu, we choose:
\be
A_{\mu \nu} = sin\theta _w ~n^a W_{\mu \nu} ^a + 
cos\theta _w ~Y_{\mu \nu}
\label{amunu}
\ee
where, $W_{\mu \nu}^a$ and $Y_{\mu \nu}$ are field strengths. The
different choices for the definition of the field strength agree in
the region where $D_\mu \Phi = 0$ where $D_\mu$ is the covariant
derivative operator; in particular this is different from the well
known 't Hooft definition which is standard for monopoles
\cite{tHoPol74}. (For a recent discussion of the 
various choices see, e.g. \cite{HinJam94,Hin94,Tor98}). 
And the combination of $SU(2)$ and
$U(1)$ field strengths orthogonal to $A_{\mu \nu}$ is defined to be
the $Z$ field strength:
\be
Z_{\mu \nu} = cos\theta _w ~n^a W_{\mu \nu} ^a - 
sin\theta_w ~Y_{\mu \nu} \ .
\label{zmunu}
\ee

\subsubsection{The fermionic sector}
\label{fermionicsector}

The fermionic Lagrangian is given by a sum over families plus
family mixing terms ($L_{fm}$). Family mixing effects are outside the scope of
this review, and  we will not  consider them any further. 
Each family includes lepton and quark sectors
\begin{equation}
L_f = L_l + L_q
\label{2.6}
\end{equation}
which for, say, the first family are 
\begin{equation}
L_l = - i {\bar \Psi} \gamma^\mu D_\mu \Psi 
      - i {\bar e}_R \gamma^\mu D_\mu e_R
      + h({\bar e}_R \Phi^{\dag} \Psi + {\bar \Psi} \Phi e_R ) \ \ , 
\qquad\qquad {\rm where} \ \ \Psi = \pmatrix{\nu_e \cr e\cr}_L
\label{2.7}
\end{equation}
\be
\eqalign{
L_q =& -i ({\bar u} , {\bar d})_L \gamma^\mu D_\mu 
         \pmatrix{u\cr d\cr}_L 
      -i {\bar u}_R \gamma^\mu D_\mu u_R
      -i {\bar d}_R \gamma^\mu D_\mu d_R 
\cr
& -G_d \biggl [ ({\bar u}, {\bar d})_L \pmatrix{\phi^+\cr \phi^0\cr} d_R
+{\bar d}_R (\phi^{-} , {\phi^*}) \pmatrix{u\cr d\cr}_L \biggr ]
\cr
& - G_u \biggl [ ({\bar u}, {\bar d})_L \pmatrix{-{\phi^*}\cr
 \phi^{-}\cr} u_R + {\bar u}_R (-\phi^0, \phi^+) \pmatrix{u\cr d\cr}_L 
       \biggr ] 
\cr}
\label{2.8}
\ee
where $\phi^*$ and $\phi^-$ are the complex conjugates of $\phi^0$ and
$\phi^+$ respectively.  $h,
\ G_d$ and
$G_u$ are Yukawa couplings.  The indices $L$ and $R$ refer to left- and
right-handed components and, rather than list their charges under the various
transformations, we give here all covariant derivatives explicitly:
\begin{equation}
D_\mu \Psi = D_\mu \pmatrix{\nu \cr e\cr}_L =
\biggl ( \partial_\mu - {{ig} \over 2} \tau^a W_\mu ^a +
{{ig'} \over 2}  Y_\mu \biggr ) \pmatrix{\nu \cr e\cr}_L 
\label{DPsi}
\end{equation}
\begin{equation}
D_\mu e_R = ( \partial_\mu + ig' Y_\mu ) e_R 
\label{Der}
\end{equation}
\begin{equation}
D_\mu \pmatrix{u\cr d\cr}_L = 
\biggl ( \partial_\mu - {{ig} \over 2} \tau^a W_\mu ^a -
{{ig'} \over 6}  Y_\mu \biggr ) \pmatrix{u\cr d\cr}_L
\label{Dul}
\end{equation}
\begin{equation}
D_\mu u_R = ( \partial_\mu - {{i2g'} \over 3} Y_\mu ) u_R 
\label{Dur}
\end{equation}
\begin{equation}
D_\mu d_R = ( \partial_\mu + {{ig'} \over 3} Y_\mu ) d_R
\label{Ddr}
\end{equation}

\noindent $\bullet$ One final comment: 

Electroweak strings are non-topological and their stability turns out
to depend on the values of the parameters in the model.  In this paper
we will consider the electric charge $e$, Yukawa couplings and the VEV
of the Higgs, $\eta / \sqrt 2$, to be given by their measured values,
but the results of the stability analysis will be given as a function
of the parameters $\sin^2 \theta_w$ and $\beta = (m_H/ m_Z)^2$ (the
ratio of the Higgs mass to the Z mass squared); we remind the reader
that $\sin^2 \theta_w \approx 0.23$, $m_Z \equiv g_z\eta/2 = 91.2$
GeV, $m_W \equiv g\eta/2 = 80.41$ GeV and current bounds on the Higgs
mass $m_H \equiv \sqrt{2\lambda} \eta$ are $m_H > 77.5 $ GeV.

\section{Review of Nielsen-Olesen topological strings}
\label{NO}

We begin by reviewing Nielsen-Olesen (NO) vortices in the Abelian
Higgs model,  with emphasis on those aspects that are
relevant to the study of electroweak strings. More detailed
information can be found in existing reviews \cite{reviews}. 

\subsection{The Abelian Higgs model}

The theory contains a complex scalar field
$\Phi$ and a $U(1)$ gauge field  which becomes massive through the
Higgs mechanism. By analogy with the GSW model, we will call this field 
$Y_\mu$.
The
action is

\be {\cal S} = \int d^4x \left[ |D_\mu\Phi|^2
- \lambda \Biggl(\Phidag\Phi - {\eta^2\over 2}\Biggr)^2 - \quarter
Y_{\mu\nu}Y^{\mu\nu} \right]
\label{aHaction}
\ee where $D_\mu = \d_\mu -iqY_\mu$ is the $U(1)$-covariant derivative,
and $Y_{\mu \nu} = \d_\mu Y_\nu -\d_\nu Y_\mu $ is the $U(1)$ field
strength. The theory is invariant under $U(1)$ gauge transformations:
\be \Phi (x) \to e^{iq\chi (x)} \Phi (x) = \hat\Phi (x) \ , \qquad Y_\mu
(x) \to Y_\mu (x) + \d_\mu \chi (x) = \hat Y_\mu (x) \ee 
which give
$D_\mu \Phi (x) \to \hat D_\mu \hat \Phi(x) = e^{iq\chi (x)} D_\mu \Phi $.

The equations of motion derived from this Lagrangian 
 are:  
\be
\eqalign{
&D_\mu D^\mu\Phi + 2 \lambda(|\Phi|^2 - {\eta^2\over 2})\Phi =  0 \cr
&\d^\mu Y_{\mu\nu}  = - iq\Phidag{\Dbw}_{\nu}\Phi \cr} 
\label{eom}
\ee
 
Before we proceed any further, we should point out that, up to an
overall scale, the classical dynamics of the Abelian Higgs model is
governed by a single parameter, $\beta = 2 \lambda / q^2$,
the (square of the) ratio of the scalar mass to the vector mass
\footnote{ $\beta$ is also 
the parameter that distinguishes superconductors or type I ($\beta
<1$) from type II ($\beta >1$)}. The action
(\ref{aHaction}) contains three parameters, $(\lambda, \eta, q)$,
which combine into the scalar mass $\sqrt {2\lambda} \eta = m_s \equiv
l_s^{-1} $, the vector mass $q \eta = m_v \equiv l_v^{-1} $, and an overall
energy scale given by the vacuum expectation value of the Higgs,
$\eta/ \sqrt 2$. 
The rescaling  
\be \Phi(x) = {{\eta} \over {\sqrt{2}} } \hat{\Phi}(x) \ \ , 
\qquad\qquad x = {\sqrt{2} \over q \eta} \hat{x} 
\ \ , \qquad\qquad   
Y_\mu = {\eta\over \sqrt{2}} \hat {Y}_\mu
\label{natural}
\ee
changes the action to 
\be 
{S} = {1\over q^2 } \int d^4 x\[ |D_\mu\Phi|^2 
 - \half
\beta (\Phidag\Phi -1)^2 
- \quarter Y_{\mu\nu}Y^{\mu\nu}
\] \ \ , 
\ee 
where now ${D}_\mu = {\d}_\mu - i {Y}_\mu$ and we have
omitted hats throughout for simplicity. In physical terms this corresponds to
taking $l_v$ as the unit of length (up to a factor of $\sqrt2$) and absorbing 
the $U(1)$ charge $q$ into the definition of the gauge field, thus
\be
\Phi \to 
{\Phi \over <|\Phi|>}   \ \ , 
\qquad\qquad  x \to  {x \over \sqrt{2} l_v}   \ \ , 
\qquad\qquad eY_\mu \to {Y_\mu \sqrt{2} l_v} \ \ ,
\qquad\qquad E \to {E \over <|\Phi|^2>} \ \ .
\label{natural2}
\ee

The energy associated with (\ref{aHaction})  
is
\be {\cal E} = \int d^3x \[|D_0\Phi|^2 + |
D_i\Phi|^2 + \lambda \Biggl(\Phidag\Phi - {\eta^2\over 2}\Biggr)^2\ +
\half {\vec E}^2 +
\half {\vec B}^2\]
\ee 
where the electric and magnetic fields are given by $F_{0i} = E_i $
and $F_{ij} =
\epsilon_{ijk} B^k$ respectively ($i,j,k = 1,2,3$).
Modulo gauge transformations, the ground states are given by
$Y_\mu = 0$, $\Phi = \eta e^{iC}/{\sqrt 2}$, 
where $C$ is
constant. Thus,  the vacuum manifold is the circle
\be 
{\cal V}  = \{ \Phi \in \CC \ | \ \Phidag \Phi - {\eta^2 \over 2} = 0 \} 
\cong S^1 \ .
\label{vacman}
\ee

A necessary condition for a configuration to have finite energy is
that the asymptotic scalar field configuration as $r \to \infty$ must
lie entirely in the vacuum manifold. Also, $D_\mu \Phi $ must tend to
zero, and this condition means that scalar fields at neighbouring
points must be related by an infinitesimal gauge
transformation. Finally, the gauge field strengths must also vanish
asymptotically. Note that, in the Abelian Higgs model, the last
condition follows from the second, since $ 0 = [D_\mu, D_\nu]\Phi =
-iqY_{\mu\nu} \Phi$ implies $Y_{\mu \nu} = 0$ 
But this need not
be the case when the Abelian Higgs model is embedded in a larger
model.

Vanishing of the covariant derivative term implies that, at large $r$,
the asymptotic configuration $\Phi(x)$ must lie on a gauge orbit; 
\be
\Phi(x) = g(x) \Phi_0 \ , \qquad {\rm where} \ 
g(x) \in  G \qquad {\rm and} \qquad\Phi_0 \in \cal V \ .
\label{gaugeorbit}
\ee
where $\Phi_0$ is a reference point in $\cal V$.
Note that, since all symmetries are gauge symmetries, 
the set of points that can be reached from $\Phi_0$ through a gauge
transformation (the gauge orbit of $\Phi_0$) spans the entire vacuum
manifold. Thus, ${\cal V} = G/H = G_{\rm local} / H_{\rm local}$,
where $G_{\rm local }$ indicates the group of gauge -- {\it i.e.} 
local -- symmetries.  On the other hand the spaces $\cal V$ and 
$G_{\rm local}/H_{\rm local}$ 
need not coincide in models with both local and
global symmetries, and this fact will be particularly relevant in the
discussion of semilocal strings.

\subsection{Nielsen-Olesen vortices}

In what follows we use cylindrical coordinates $(t, \rho, \varphi,
z)$.  We are interested in a static, cylindrically symmetric configuration
corresponding to an infinite, straight string along the $z$-axis.

The ansatz of Nielsen and Olesen \cite{NieOle73} for a string with winding
number $n$ is 
\be \Phi = {\eta \over \sqrt 2} f (\rho) e^{in\varphi}\ \ , 
\qquad\quad   q Y_\varphi =
 n v (\rho) \ \ , 
\qquad\quad Y_\rho = Y_t = Y_z= 0
\label{NOansatz}\ee
(that is, $Y = v(\rho) ~d\varphi \ $ or $ \ \vec Y = {\hat \varphi}
~v(\rho) /\rho \ $),  \ with boundary conditions

\be f(0) = v(0) = 0\ , \qquad f(\rho) \to 1 \ ,  
\qquad v(\rho) \to 1 \qquad {\rm  as}  \ \rho \to \infty .
\label{NObc}
\ee

Note that, since 
$Y_z = Y_t = 0$, and all other fields are independent of $t$ and $z$, the
electric field is zero, and the only surviving component of the
magnetic field $\vec B$ is in the $z$ direction.

Substituting this ansatz into the equations of motion we obtain the equations
that the functions $f$ and $v$ must satisfy:

\be
\eqalign{
f''(\rho) + {{f'(\rho)} \over \rho} - {n^2f (\rho) \over {\rho^2}} [1-v(\rho)]^2 + 
\lambda \eta^2 (1-f(\rho)^2) f(\rho) &= 0 \cr
v''(\rho) - {{v'(\rho)} \over \rho} + {{q^2 \eta^2}} f^2(\rho)[1-v(\rho)] &= 0
\cr}
\label{NOeqs}
\ee

\begin{figure}[tbp]
\caption{\label{fvgraphs} The functions $f_{NO}$, $v_{NO}$ 
for a string with winding number $n=1$ (top panel) and $n=50$ (bottom
panel), for $\beta \equiv 2\lambda/q^2 = 0.5$. 
The radial coordinate has been rescaled as in 
eq. (\ref{natural}),  $\hat{\rho} = q\eta \rho/\sqrt{2}$.
}
\vskip 1truecm
\epsfxsize = \hsize \epsfbox{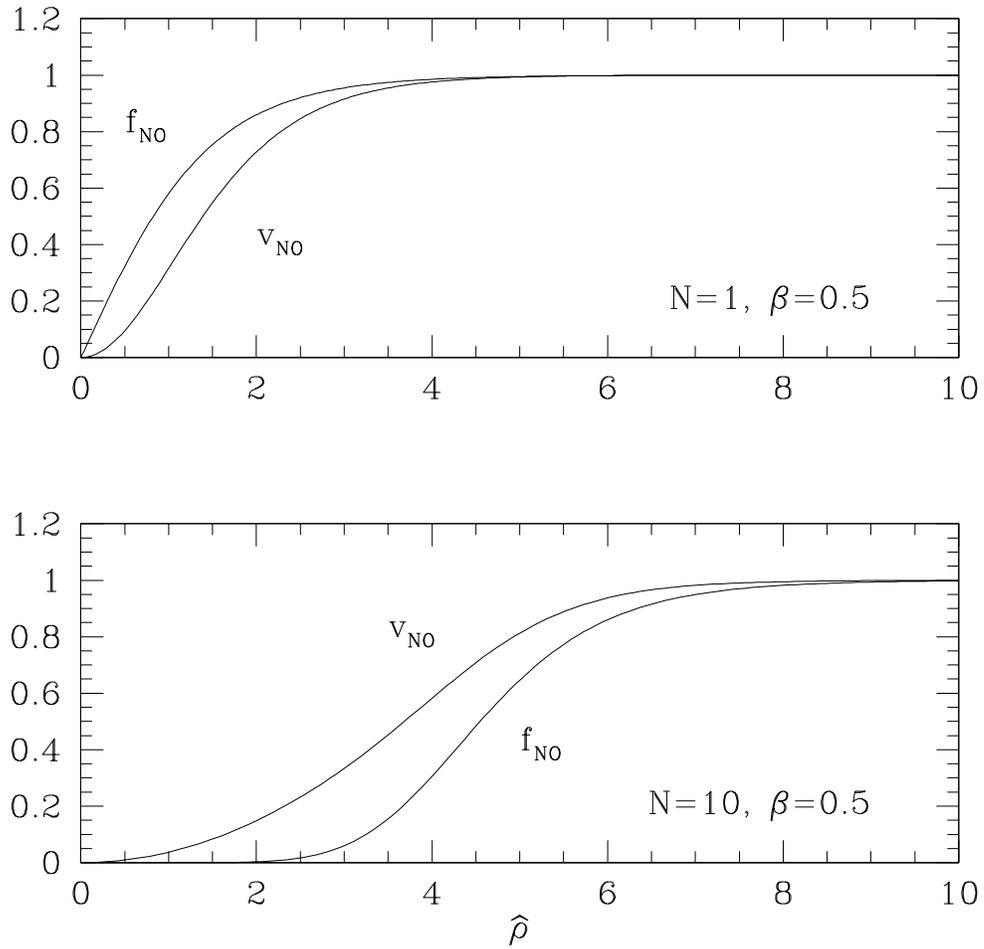}
\end{figure}

In what follows we will denote the solutions to the system
(\ref{NOeqs},\ref{NObc}) by $f_{NO}$ and $v_{NO}$; they are not known
analytically, but have been determined numerically; for $n=1$, 
$\beta = 0.5$, they have the profile in Fig. \ref{fvgraphs}.

At small $\rho$, the functions $f$ and $v$ behave as $\rho^n$ and
$\rho^2$ respectively; as $\rho \to \infty$, they approach their
asymptotic values exponentially with a width given by the inverse
scalar mass, $m_s$, and the inverse vector mass, $m_v$, respectively,
if $\beta < 4$. For $\beta >4$ the fall-off of both the scalar and the
vector is controlled by the vector mass \cite{Peri93}.

One case in which it is possible to find analytic expressions for the
functions $f_{NO}$ and $v_{NO}$ is in the limit $n\to \infty$
\cite{AchGreHarKui94}. Inside the core of a large $n$ vortex, the
functions $f$ and $v$ are \be f(\rho) = \left ( {q \over 4n} m_s m_v
\rho^2 \right )^{n\over 2} e^{-qm_sm_v \rho^2 /8} \ , \qquad\qquad
v(\rho) = {1 \over 4n} m_sm_v \rho^2 \ee to leading order in $1/n$,
and the transition to their vacuum values is controlled by a first
integral $\Psi( f,f',v,v') = {\rm const \ }$.  Large $n$ vortices
behave like a conglomerate of ``solid'' $n=1$ vortices. The area
scales as $n$, so the radius goes like $\sqrt n L_0$, where $L_0 = 2
(\sqrt{m_sm_v})^{-1}$. The transition region between the core and
asymptotic values of the fields is of the same width as for $n=1$
vortices Fig. \ref{fvgraphs} shows the functions $f_{NO}, \ v_{NO}$
for $n=50$, $\beta = 0.5$  (note that for $\beta>1$ these multiply
winding solutions are unstable to separation into $n=1$ vortices which
repel one another).

\bigskip $\bullet$ Energy considerations:

The energy per unit length of such configurations (static and
z-independent) is therefore 
\be {E} = \int d^2x \[ |D_m\Phi|^2 + \half B^2 + \lambda
(\Phidag\Phi - {\eta^2\over 2})^2 \] \ee
where  $ m,n = 1,2$ and 
$B=\d_mY_n - \d_nY_m$ 
is the $z$-component of the magnetic field.

In order to have solutions with finite energy per unit length we must
demand that, as $\rho\to\infty$, 
$D_\mu \Phi$, $|\Phi|^2 - {\eta^2 / 2} $ 
and $Y_{mn} $ all go to zero faster than 
$1/\rho$.

The vacuum manifold (\ref{vacman}) 
is a circle and strings form when the asymptotic field configuration
of the scalar field winds around this circle.  The important point
here is that there is no way to extend a winding configuration inwards
from $\rho = \infty$ to the entire $xy$ plane continuously while
remaining in the vacuum manifold. Continuity of the scalar field
implies that it must have a zero somewhere in the $xy$ plane. This
happens even if the $xy$ plane is deformed, and at all
times, and in three dimensions one finds a
continuous line of zeroes which signal the position of the string 
(a sheet in spacetime). Note that the string can have no ends;
it is either infinitely long or a closed loop.

The zeroes of the scalar field are forced by the non-zero topological
degree of the map 
\be 
\eqalign{
S^1 \quad &\to \quad {\cal V} \cr \varphi \ 
\quad &\to \quad \Phi(\rho = \infty, \varphi) \ \ ,
\cr}
\ee
usually called the {\it winding number} of the vortex; 
the resulting vortices are called topological because they
are labelled by non-trivial elements of the first homotopy group of
the vacuum manifold 
(where non-trivial means ``other than the identity element''). Thus,
$\pi_1( {\cal V}) = \pi_1(S^1) \neq 1$, is a necessary condition for 
the existence of topological vortices.
Vortices whose asymptotic scalar field configurations are associated
with the identity element of $\pi_1({\cal V})$ are called non-topological.
In particular, if 
${\cal V}$ is simply connected, i.e. $\pi_1 ({\cal V}) = 1$, 
one can only have non-topological vortices.

A few comments are needed at this point:

\bigskip$\bullet$ Quantization of magnetic flux: 

Recall that $B$ is  the $z$-component of the magnetic field. 
The magnetic flux $F_Y$ through the $xy$-plane is therefore
\be 
F_Y \equiv \int d^2x B = \int_{\rho=\infty}
{\vec Y}_\infty \cdot {\vec dl} = \int_0^{2\pi}  \d_\varphi \chi
d\varphi = {2\pi n\over q} 
\ee 
and is quantized in units of $2\pi / q$. This is due to the fact that 
$\Phi(\rho=\infty, \varphi) = \eta e^{iq\chi (\varphi)}/\sqrt{2}$, 
$D_\varphi \Phi = {\eta / \sqrt 2} 
[ i q\partial_\varphi \chi - iq Y_\varphi ] = 0$ 
and $\Phi$ must be singlevalued, thus $q [\chi(2\pi) - \chi(0)] = 2\pi n$. 
The integer $n$ is, again, the winding number of the vortex.

\bigskip$\bullet$ Magnetic pressure: 

In an Abelian theory, 
the condition
$\vec{\nabla} \cdot \vec {B} = 0$ implies that parallel magnetic field
lines repel. A two-dimensional scale transformation
${\vec x} \to \lambda {\vec x}$ where the magnetic field is reduced
accordingly to keep the magnetic flux constant, $B_\Lambda =
\Lambda^{-2} B( {{\vec x} / \Lambda})$, reduces the 
magnetic energy $\  \int d^2x~ B^2 /2 \ $ \ 
by
$\Lambda^2$. What this means is that a tube of magnetic lines of area
$S_0$ can lower its energy by a factor of $\Lambda^2$ by spreading
over an area $\Lambda^2 S_0$. 

Note that later we will consider non-Abelian gauge symmetries, for
which $\vec{\nabla} \cdot {\vec B}\neq 0$ and the energy can also be
lowered in a different way.  In this case, one can think of the gauge
fields as carrying a magnetic moment which couples to the ``magnetic''
field and, in the presence of a sufficiently intense magnetic field,
the energy can be lowered by the spontaneous creation of gauge bosons.
In the context of the electroweak model, this process is known as
W-condensation \cite{AmbOle90} and its relevance for electroweak
strings is explained in section
\ref{stability}.

\bigskip$\bullet$
Meissner effect and symmetry restoration:

In the Abelian Higgs model, as in a superconductor, it is
energetically costly for magnetic fields to coexist with scalar fields
in the broken symmetry phase. Superconductors exhibit the Meissner
effect (the expulsion of external magnetic fields), but as the sample
gets larger or the magnetic field more intense, symmetry restoration
becomes energetically favourable.  An example is the generation of
Abrikosov lattices of vortices in type II superconductors, when the
external magnetic field reaches a critical value.

The same phenomenon occurs in the Abelian Higgs model. In a region
where there is a concentration of magnetic flux, the coupling term $
q^2 A^2 \Phi^2$ in the energy will tend to force the value of the
scalar field towards zero (its value in the symmetric phase).  This
will be important to understand the formation of semilocal 
(and possibly electroweak) strings, where there is no topological
protection for the vortices, during a phase transition (see section
\ref{networks}).  The back reaction of the gauge
fields on the scalars depends on the strength of the coupling constant
$q$. When $q$ is large (in a manner that will be made precise in
section \ref{networks}) semilocal strings tend to form
regardless of the topology of the vacuum manifold.

\subsection{Stability of Nielsen-Olesen vortices}

Given a solution to the classical equations of motion, there are
typically two approaches to the question of stability. One is to
consider the stability with respect to infinitesimal perturbations of
the solution. If one can establish that no perturbation can lower the
energy, 
then the solution is called classically stable. Small
perturbations that do not alter the energy are called zero modes, and
signal the existence of a family of configurations with the same
energy as the solution whose stability we are investigating
(e.g. because of an underlying symmetry). If one can guess an
instability mode, this approach is very efficient in showing that a
solution is unstable (by finding the instability mode explicitly) but
it is usually much more cumbersome to prove stability; mathematically
the problem reduces to an eigenvalue problem and one often has to
resort to numerical methods. A stability analysis of this type for
Nielsen-Olesen vortices has only been carried out recently by
Goodband and Hindmarsh \cite{GooHin95a}.  An analysis of the stability of
semilocal and electroweak strings can be found in later sections.

A second approach, due to Bogomolnyi, consists in finding a lower
bound for the energy in each topological sector and proving that the
solution under consideration saturates this bound. This immediately
implies that the solution is stable, although it does not preclude
the existence of zero modes or even of other configurations with the
same energy to which the solution could tunnel semiclassically.  We
will now turn to Bogomolnyi's method in the case of Nielsen-Olesen
vortices.

\bigskip$\bullet$ Bogomolnyi limit and bounds

Consider the scalar gradients:
\be 
\eqalign{
({D_1\Phi})^\dagger D_1\Phi + ({D_2\Phi})^\dagger D_2\Phi 
&= [{(D_1+iD_2)\Phi}]^\dagger (D_1+iD_2) \Phi 
- i [({D_1\Phi})^\dagger  D_2 \Phi - ({D_2\Phi})^\dagger  D_1\Phi ] \cr
&=  |(D_1+iD_2)\Phi|^2 -i \[ \d_1
(\Phidag D_2\Phi) - \d_2 (\Phidag D_1 \Phi)\] + i \Phidag [D_1,
D_2]\Phi \ \ .
\cr}
\label{Bog0}
\ee

Note that the second term in the RHS of (\ref{Bog0}) is the curl of
the current $J_i = -i \Phidag D_i \Phi $, and that $\oint {\vec
J}\cdot{\vec dl}$ tends to zero as $\rho \to \infty$ for
configurations with finite energy per unit length (because $D_i\Phi$
must vanish faster than $1/\rho$).
Now use the identity
$[D_1, D_2] \Phi = -iq F_{12}
\Phi = -iqB\Phi$.
to rewrite the energy per unit length as follows:

\be 
\eqalign{E &= \int d^2 x \[ |(D_1 \pm iD_2 )\Phi|^2 + \half B^2 \pm
qB\Phidag\Phi +  \lambda \Biggl(\Phidag\Phi - {\eta^2 \over 2}\Biggr)^2 \] \cr
&=
\int d^2 x \[ |(D_1 \pm iD_2 )\Phi|^2 + 
\half \Biggl \{ B \pm q \Biggl(\Phidag\Phi -
{\eta^2 \over 2}\Biggr) \Biggr \} ^2 + 
(\lambda - \half q^2)\Biggl(\Phidag\Phi -
{\eta^2 \over 2}\Biggr)^2
\]\cr 
&\ \ \ \ \ \ \ \ \ \pm q{\eta^2\over 2} \int B d^2x \cr}
\label{Bog1}
\ee
The last integral  is the total magnetic flux,
and we saw earlier that it has to be an integral multiple of $2\pi/q$,
so we can write, introducing $\beta = 2\lambda/q^2$, 
\be E = 2\pi (\pm n)  {\eta^2 \over 2} +
\int \[ |(D_1 \pm iD_2 )\Phi|^2 + \half \Biggl[B \pm q \Biggl(\Phidag\Phi -
{\eta^2\over 2}\Biggr) \Biggr]^2 + \half q^2 (\beta -1)\Biggl(\Phidag\Phi -
{\eta^2\over 2}\Biggr)^2
\]
\label{Bog2}
\ee
where the plus or minus
signs are chosen so that the first term is positive, depending on the
sign of the magnetic flux.

Note that, if $\beta\geq 1$ the energy is bounded below by 
\be 
E \geq \langle
\Phidag\Phi \rangle q F_Y \ \ , 
\label{Bogbound}
\ee
 where $F_Y$ is the magnetic flux.
\footnote{
When $\beta= 1$, the masses of the scalar and the vector are equal,
and the Abelian Higgs model can be made supersymmetric. 
In general, bounds of the form (Energy) $\geq$
(constant) $\times$ (flux) are called Bogomolnyi bounds, and their
origin can be traced back to supersymmetry.}

If $\beta = 1$, there are
configurations that saturate this bound: those that satisfy the first
order {\it Bogomolnyi equations}
\be (D_1 \pm iD_2 )\Phi = 0 \ \ , \qquad\qquad  B \pm 
q (\Phidag\Phi - {\eta^2\over 2}) = 0
\qquad .  
\label{Bog3}\ee 
or, in terms of $f(\rho)$ and $v(\rho)$,
\be
f'(\rho) + (\pm n) {v(\rho)-1 \over \rho} f(\rho) \ = \ 0 \ \ , \qquad\qquad 
(\pm n)  v'(\rho) + {q^2\eta^2
\over 2}
\rho (f^2 (\rho) -1)
\ =
\ 0
\label{Bog4}
\ee

However, when $\beta >1$ there does not exist a static solution with
$E = \pi |n| \eta^2$ 
since requiring, {\it e.g.}, $ B + q (\Phidag\Phi - \eta^2/2) = 0 $
and $(\Phidag\Phi - \eta^2/2) = 0 $ simultaneously would imply $B=0$,
which is inconsistent with the condition on the total magnetic flux,
$\int B d^2x = 2\pi n /q \ $.  This has an effect on the stability of
higher winding vortices when $\beta >1$: if $n>1$ the solution
breaks into $n$ 
vortices each with a unit of magnetic flux \cite{Bog76}, 
which repel one another.

If $n=1$ there are stable static solutions, but with an energy higher
than the Bogomolnyi bound.  This is because the topology of the vacuum
manifold forces a zero of the Higgs field, and then competition between
magnetic and potential energy fixes the radius of the solution. 
The same argument shows that $n=1$ strings are stable  
for every value of $\beta$. One still has to worry about angular 
instabilities, but a careful analysis by \cite{GooHin95a} shows there 
are none.

The dynamics of multivortex solutions is governed by the fact that
when $\beta < 1$ vortices attract, but with $\beta >1$ they repel.
This can be understood heuristically from the competition between
magnetic pressure and the desire to minimise potential energy by
having symmetry restoration in as small an area as possible.  The
width of the scalar vortex depends on the inverse mass of the Higgs,
$l_s$, that of the magnetic flux tube depends on the inverse vector
boson mass, $l_v$.
If $\beta < 1$, have $m_v > m_s$ so $l_v < l_s$ (the radii of the
scalar and vector tubes). The scalar tubes see each other first - they
attract.  Whereas if $\beta > 1$, the
vector tubes see each other first - they repel. 
For $\beta = 1$ there is no net force between vortices, and there are
static multivortex solutions for any $n$. In the Abelian Higgs case
they were explicitly constructed by Taubes \cite {JafTau80} and their
scattering at low kinetic energies has been investigated using the
geodesic approximation of Manton \cite{Man82} by Ruback
\cite{Rub88} and, more recently, Samols \cite{Sam92}. 
For $\beta <1$ Goodband and Hindmarsh 
\cite{GooHin95a} have found bound states of two $n=1$ vortices oscillating
about their centre of mass.

\section{Semilocal strings}
\label{semilocal}

The semilocal model is obtained when we replace the complex scalar
field in the Abelian Higgs model 
by an $N$-component multiplet, while keeping only the overall
phase gauged. In this section we will concentrate on $N=2$ because of
its relationship to electroweak strings, but the
generalisation to higher $N$ is straightforward, and is discussed below.

\subsection{The model}

Consider a direct generalization of the Abelian
Higgs model where the complex scalar field is replaced by an
$SU(2)$ doublet $\Phi^T \ = \ ( \phi_1, \phi_2 )$.  The action is 
\be S =
 \int d^4 x \left [ |(\partial_\mu - iq Y_\mu )  \Phi |^2 \ - \
{1 \over 4} Y_{\mu \nu} Y^{\mu \nu} \ - \ {\lambda } \(\Phidag \Phi
- {\eta^2\over 2} \)^2 \right ] \ ,
\label{semilocalaction}
\ee
where  $Y_{\mu} $ is the
$U(1)$ gauge potential and $Y_{\mu
\nu}= \partial_{\mu}Y_{\nu} - \partial_{\nu}Y_{\mu}$ its field
strength. 
Note that this is just the scalar sector of the GSW 
model for $ g=0, \ g' = g_z = 2q$, \ {\it i.e.} for
$\sin^2\theta_w = 1$, and $W_\mu^a = 0$.

Let us take a close look at the symmetries. The action is invariant
under
$G = SU(2)_{global} \times U(1)_{local}$, with transformations 
\be
\Phi \rightarrow e^{i q \gamma (x)  }\Phi  = 
  \pmatrix {e^{i q \gamma(x)}  & \ 0  \cr & \cr
 \ 0 &e^{i q \gamma (x)} \cr } \pmatrix {\phi_1 \cr \cr \phi_2 \cr}  ,
\ \ \ \ \ \ \quad 
Y_{\mu} \rightarrow Y_{\mu} +  \partial_{\mu} \gamma (x) \ ,
\ee
under $U(1)_{local}$, and
\be
\Phi \rightarrow e^{i\alpha_a \tau ^a} \Phi  =  
\pmatrix{ {\rm cos }{\left(\alpha  \over 2 \right)}
+in_3{\rm sin}{\left(\alpha  \over 2 \right)}
&i(n_1 \! - \! i n_2){\rm sin}{\left( \alpha  \over 2 \right)}  \cr 
& \cr  
i(n_1 \! + \!i n_2){\rm  sin}{\left(\alpha  \over 2 \right)}
&{\rm cos}{\left( \alpha \over 2 \right)}
-in_3{\rm sin}{\left(\alpha  \over 2 \right)}  }
\pmatrix {\phi_1 \cr \cr \phi_2 \cr} ,
\ \  Y_\mu \to Y_\mu 
\ee
under $SU(2)_{global}$, where $\alpha = \sqrt {\alpha_1^2 +
\alpha_2^2 + \alpha_3^2} \in [0, 4\pi)$ 
is a positive constant and $n_a = \alpha_a / \alpha$ is a constant unit 
vector. Note that a shift of
the function $\gamma(x)$ by $2\pi/q$ leaves the transformations
unaffected.
The model actually has symmetry $ G =
[SU(2)_{global} \times U(1)_{local}] /Z_2$; the $Z_2$ identification
comes because the transformation with $(\alpha, \gamma)$ is identified
with that with $(\alpha + 2\pi, \gamma + \pi/q)$. 
Once $\Phi$ acquires a vacuum expectation
value, the symmetry breaks down to $H = U(1)$ exactly as in the
GSW model, except for the
fact that the unbroken $U(1)$ subgroup is now {\it global} (for
instance, if the VEV
of the Higgs is $\langle\Phi^T \rangle = \eta (0,1)/ {\sqrt 2}$, 
the unbroken global $U(1)$ is the subgroup with $n_1 = n_2
= 0$, $n_3=1$, $q\gamma = \alpha/2$). 
Thus, the symmetry breaking is $ [SU(2)_{global} \times
U(1)_{local}] /Z_2 \to U(1)_{global}$.

Note also that, for any {\it fixed} $\Phi_0$ a global phase change can
be achieved with either a global $U(1)_{\rm local} $ transformation
or a $SU(2)_{\rm global} $ transformation. {The
significance of this fact will become apparent in a moment}

Like in the GSW model, the vacuum manifold is the three sphere
\be 
{\cal V} \ = \ 
\{\Phi \in \CC^2 \ | \ \Phidag \Phi = {\eta^2 \over 2} \}
\ 
\cong
\ S^3 \ \ , 
\ee
which is simply connected, so there are no
topological string solutions. On the
other hand, if we only look at the {\it gauged} part of the symmetry,
the breaking looks like $U(1) \to 1$, identical to that  of the Abelian
Higgs model, and this suggests that we should have local strings.

After symmetry breaking, the particle content
is two Goldstone bosons, one scalar of mass $m_s =\sqrt{2\lambda}
\eta$ and a massive
vector boson of mass $m_v = q\eta$. In this section it will be
convenient to use rescaled units throughout; after the rescaling
(\ref{natural}), and dropping hats, we find
\be {q^2 S} = 
 \int d^4 x \left [ |(\partial_\mu - i Y_\mu )  \Phi |^2 \ - \
{1 \over 4} Y_{\mu \nu} Y^{\mu \nu} \ - \ {\beta \over 2 } \(\Phidag
\Phi - 1 \)^2 \right ] \ ,
\ee
and, as in the Abelian Higgs case, $\beta  = {m_s^2 /
m_v^2} = {2\lambda / q^2}$ is the only free parameter in the model.
The equations of motion 
\be
\eqalign{
D_\mu D^\mu\Phi + \beta(|\Phi|^2 - 1)\Phi &= 0 \cr
\d^\mu Y_{\mu\nu} &= - i\Phidag{\Dbw}_{\nu}\Phi
 \ \ .\cr}
\label{eomnatural}
\ee
are exactly the same as in the 
Abelian Higgs model but replacing the scalar field by the
$SU(2)$ doublet,
and complex conjugation by hermitian conjugation of $\Phi$. Therefore, 
any solution ${\hat \Phi}(x), \ {\hat Y}_\mu (x)$ of (\ref{eom}) 
(in rescaled units) extends trivially
to a solution
$\Phi_{sl} ({x}),\ (Y_\mu)_{sl} ({x})$ of the semilocal model if we
take
\be
\Phi_{sl} ({x}) = {\hat \Phi} ({x}) \Phi_0 \qquad \qquad
(Y_\mu)_{sl} ({x}) = {\hat Y}_\mu ({x})
\ee
with $\Phi_0$ a constant $SU(2)$ doublet of unit norm, $\Phidag_0
\Phi_0 = 1$. In particular, the Nielsen-Olesen string can be embedded
in the semilocal model in this way. The configuration
\be
\Phi = f_{NO}(\rho)e^{in\varphi} \Phi _0 , \ \ \  
\ \ \   Y = n v_{NO}(\rho) d{ \varphi}
\label{nslsol}
\ee
remains a solution of the semilocal model with winding number $n$
provided $f_{NO}$ and $ v_{NO}$ are the solutions to the
Nielsen-Olesen equations (\ref{NOeqs}). In this context, the constant
doublet $\Phi_0$ is sometimes called the `colour' of the string (do
not confuse with $SU(3)$ colour!). One important difference
with the Abelian Higgs model is that a scalar perturbation can remove
the zero of $\Phi$ at the center of the string,
thereby reducing the potential energy stored in the
core. 

Consider the energy per unit length, in these units, of a
static, 
cylindrically symmetric configuration along the $z$-axis:
\be { E \over (\eta^2/2)} = \int d^2 x \left [ {1 \over 4} (\partial _m
        Y_n - \partial _n Y_m )^2 + |(\partial_m - iY_m ) \Phi |^2 +
        {\beta \over 2} (\Phidag \Phi - 1 ) ^2 \right ] \
\label{energynatural}
\ee
Note, first of all, that any finite energy configuration must satisfy 
$$ (\partial _m - i Y_m )\phi_1 \rightarrow 0 \ \ ,\ \ \ \ \ \ \
(\partial _m - iY_m )\phi_2 \rightarrow 0 \ \ , \ \ \ \ \
\ \ \phionebar \phi_1 + \phitwobar \phi_2 \rightarrow 1 \ \
\ 
\
\ \ \ \ \ \  {\rm as }\ \ \ \rho
\rightarrow \infty$$
(As before, $m,n = 1,2$ and ($\rho , \varphi$) are polar coordinates on the plane
orthogonal to the string). This leaves the phases of $\phi_1$ and
$\phi_2$ undetermined at infinity and there can be solutions where both
phases change by integer multiples of $2 \pi$ as we go around the
string; however, there is only one $U(1)$ gauge field available to
compensate the gradients of $\phi_1$ and $\phi_2$, and this introduces
a correlation between the winding in both components: the
condition of finite energy requires that the phases of $\phi_1$
and
$\phi_2$ differ by, at most, a constant, as $\rho
\rightarrow \infty$.  Therefore, a finite energy string must tend
asymptotically to a maximal circle on $S^3$
\be
\Phi  \rightarrow e^{in \varphi} 
\pmatrix {a e^{iC}\cr  \cr \sqrt {1  - a^2} } \equiv 
e^{in \varphi} \Phi_0 \ \
\
\
\
\ \ \ \
\  Y \rightarrow  n d {\varphi} \ \ \ \ \ \ \ \ \ \ 
\bigl( {\rm or} \ \  {\vec Y} \to
{n \over \rho}  {\hat \varphi} \bigr )\  \ ,
\ee
where $0 \leq a \leq 1$ and $C$ are real constants, and determine
the `colour' of the string.  
A few comments are needed at this point.

\bigskip$\bullet$ Note that the choice of
$\Phi_0$ is arbitrary for an isolated string (any value of $\Phi_0$
can be rotated into any other without any cost in energy) but the
relative `colour' between two or more strings is fixed. That is, the
{\it relative} value of
$\Phi_0$ is significant whereas the {\it absolute} value is not.

\bigskip$\bullet$ The number $n$ is the winding number of the string
and, although it is not a topological invariant in the usual sense
(the vacuum manifold, $S^3$, is simply connected), it {\it is}
topologically conserved. The reason is that, even though any maximal
circle can be continuously contracted to a point on $S^3$, all the
intermediate configurations have infinite energy.  The space that
labels finite energy configurations is not the vacuum manifold but,
rather, the {\it gauge orbit} from any reference point $\Phi_0 \in{
\cal V}$, and this space $(G_{local} / H_{local})$, is not simply
connected: $\pi_1 (G_{local} / H_{local}) = \pi_1 (U(1) / 1) =
\ZZ$. Thus, configurations with different winding numbers are
separated by infinite energy barriers, but this information is not
contained in $\pi_1({\cal V})$
\footnote{The fact that the gauge orbits sit 
inside ${\cal V}
= G/H$ without giving rise to non-contractible loops can be traced
back to the previous remark that every point in the gauge orbit of
$\Phi_0$ can also be reached from $\Phi_0$ with a global
transformation.}.

\bigskip$\bullet$ On the other hand,  because $\pi_1({\cal V}) = 1$, 
the existence of a topologically conserved winding number does not
guarantee that winding configurations are non-dissipative 
either. In contrast with the Abelian Higgs model, a field
configuration with non-trivial winding number at $\rho =
\infty$ can be extended inwards for all $\rho$ without ever leaving the
vacuum manifold.  {Thus, the fact that $\pi_1 (G_{local} / H_{local)}
\neq 1 $ only means that finite energy field configurations fall into
inequivalent sectors, but it says nothing about the existence of
stable solutions within these sectors.

$\bullet$ Thus, we have a situation where
\be  \pi_1 ({\cal V}) = \pi_1 (G/H)= \pi_1(S^3) = 1  \qquad {\rm but} 
\qquad \pi_1 (G_{\rm local}/H_{\rm local})= \pi_1(S^1) = \ZZ \ . 
\label{semilocalconditionforstring}\ee
and the effect of the global symmetry is to eliminate the topological
reason for the existence of the strings. Notice that this subtlety
does not usually arise because these two spaces are the same in
theories where all symmetries are gauged (like GSW, Abelian Higgs,
etc.).  We will now show that, in the semilocal model, the stability
of the string depends on the dynamics and is controlled by the value
of the parameter $\beta = 2\lambda / q^2$. Heuristically we expect
large $\beta$ to mimic the situation with only global symmetries
(where the strings would be unstable) , whereas small $\beta$ resembles
the situation with only gauge symmetries (where we expect stable
strings).

\subsection {Stability}

Let us first prove that there are classically stable
strings in this model. We can show this  analytically for
$\beta = 1$ \cite{VacAch91}. Recall the expression of the energy per unit
length (\ref{energynatural}).  The analysis in the 
previous section  goes through when the complex field
 is replaced by the $SU(2)$ doublet, and we can rewrite
\be {E \over (\eta^2/2)}  = 2\pi |n| \ + \ \int d^2 x \Bigl[ 
|D_1
\Phi \pm i D_2 \Phi |^2 \ + \  {1\over 2 } (B \pm
(\Phi^{\dag} \Phi - 1 ) )^2 \ + 
\ {1 \over 2} (\beta - 1) (\Phi^{\dag} \Phi - 1 )^2 \Bigr] \ ,  \ee
choosing the upper or lower signs depending on the sign of
$n$. Since $n$ is fixed for finite energy configurations this shows
that, at least for $\beta = 1$, a configuration satisfying the
Bogomolnyi equations
\be (D_1 \pm i D_2) \Phi = O \ \ \ \ \ \ \ \ \ \ \ \ 
 B \pm (\Phi^{\dag} \Phi -1) = 0, \ \ 
\label{semilocalbogomolnyieq}
\ee
is a local minimum of the energy and, therefore, automatically stable
to infinitesimal perturbations. But these are the same
equations as in the Abelian Higgs model, therefore the semilocal
string (\ref{nslsol}) automatically saturates the Bogomolnyi bound
(for any `colour' $\Phi_0$). Thus, it is classically stable for $\beta = 1$.

This argument 
does not preclude zero modes or other
configurations degenerate in energy. Hindmarsh \cite{Hin92} showed that,
for $\beta = 1$ there are indeed such zero modes, described below in
(\ref{semilocalzeromodes}).

We have just proved that, for $\beta = 1$, semilocal
strings are {\it stable}. This is surprising because the vacuum manifold is
simply connected and a field configuration that winds at infinity may unwind
without any cost in {\it potential} energy \footnote{ In the
Nielsen-Olesen case a configuration with a non-trivial winding number
{\it must} go through zero somewhere for the field to be
continuous. But here, a configuration like $ \Phi^T ( \rho \! =\!
{\infty}) = \eta( 0, e^{i {\varphi}}) /\sqrt 2$ can 
gradually 
change to $\Phi^T (\rho \! = \!0) = \eta (1,0) \sqrt 2$ as we move
towards the centre of the ``string'' without ever leaving the vacuum
manifold.  This is usually called `unwinding' or `escaping in the
third dimension' by analogy with condensed matter systems like nematic
liquid crystals.}. The catch is that, because $\pi_1
(G_{local}/H_{local}) =
\pi_1 (U(1)) = \ZZ$ is non trivial, leaving the $U(1)$ gauge orbit is
still expensive in terms of {\it gradient} energy.

As we come in from infinity, the field has to choose between unwinding
or forming a semilocal string, that is, between acquiring mostly
gradient or mostly potential energy.  The choice depends on the
relative strength of these terms in the action, which is governed by
the value of $\beta$, and we expect the field to unwind for large
$\beta$, when the reduction in potential energy for going off the
vacuum manifold is high compared to the cost in gradient energy for
going off the $U(1)$ orbits. And vice versa.  Indeed, we will now
show that, for $\beta > 1$, the $n=1$ vortex is unstable to
perturbations in the direction orthogonal to $\Phi_0$ \cite{Hin92}
while, for $\beta <1$, it is stable.
For $\beta = 1$, some of the perturbed
configurations become degenerate in energy with the semilocal vortex
and this gives a (complex) one-parameter family of solutions with
the same energy and varying core radius \cite{Hin92}.

\subsubsection {The stability of strings with  $\beta >1$}

Hindmarsh has shown \cite{Hin92} that for $\beta >1$ the semilocal
string configuration with unit winding is unstable to perturbations
orthogonal to $\Phi_0$, {which make the magnetic flux spread
to infinity }. As pointed out by Preskill
\cite{Pre92}, this is remarkable because the total amount of flux 
measured at infinity remains quantized, but the flux is {\it not}
confined to a core of finite size (which we would have
expected to be of the order of the inverse vector mass).

The semilocal string solution with $n=1$ is, in rescaled  units, 
\be 
\Phi_{sl} = f_{NO} (\rho)  e^{i\varphi} \Phi_0 \ \ ,\qquad  \qquad \
{Y_{sl}}= v_{NO}(\rho) { d\varphi} \ \ .
\label{slsol}
\ee

However, as pointed out in \cite{Hin92}, this is not the most general
static one-vortex ansatz compatible with cylindrical
symmetry. Consider the ansatz
\be 
\Phi = f (\rho) e^{i\varphi} \Phi_0  + g(\rho) e^{i
m\varphi}\Phi_\perp \ \ , \qquad\qquad  {Y} = {v(\rho)} {d \varphi} \ \ ,
\label{perturbedsl}
\ee
with $|\Phi_0| = |\Phi_\perp| = 1$ and ${\overline \Phi_0} \Phi_\perp
= 0$. The orthogonality of $\Phi_0$ and $\Phi_\perp$ ensures that the
effect of a rotation can be removed from $\Phi$ by a suitable $SU(2)
\times U(1)$ transformation, therefore the configuration is cylindrically 
symmetric. 
For the configuration to have finite energy we
require the boundary conditions $ f(0) = g'(0) = v(0) = 0$ and $f \to
1, \ g\to 0, \ v \to 1$ as $\rho \to \infty$

 We know that if $g=0$ the energy is minimised by the semilocal
string configuration $f=f_{NO}, \ v=v_{NO}$, because the problem is
then identical to the Abelian Higgs case. The question is whether a
non-zero $g$ can lower the energy even further, in which case the
semilocal string would be unstable. The standard way to
find out is to consider a small perturbation of (\ref{slsol}) of the
form $g=
\phi(\rho) e^{i\omega t}$ and look for solutions of the equations of
motion where $g$ grows exponentially, that is, where $\omega^2 <0$. The
problem reduces to finding the 
negative eigenvalue solutions to the Schr\"odinger-type
equation
\be 
\[ -{1\over \rho} {d\over d\rho} 
\left( \rho {d\over d\rho} \right) + 
{(v(\rho)-m)^2\over \rho^2} + \beta (f(\rho)^2 -1) \] \psi(\rho) =
 \omega^2 \psi (\rho)\ee 

First of all, it turns out that it is sufficient to examine the $m=0$
case only. Note that, since $0\leq v(\rho) \leq 1$, for
$m>1$ the second term is everywhere larger than for $m=1$, 
so if we one can show that all
eigenvalues are positive for $m=1$ then so are the eigenvalues for
$m>1$. But for $m=1$ the problem is identical to the
analogous one for instabilities in $f$ in the Abelian Higgs model, and
we know there are no instabilities in that case.
Therefore it is sufficient
to check the stability of the solution to perturbations with $m=0$
(negative values of $m$ also give higher eigenvalues than $m=0$.)

If $m=0$, the above ansatz yields
\be
{E \over (\eta^2 / 2)} = 2\pi \int_0^\infty \rho \left[ (f')^2 + (g')^2 + {1
\over 2\rho^2} (v')^2 + {(1 - v)^2 \over \rho^2} f^2 + {v^2 \over \rho^2} g^2 +
\half \beta (f^2 + g^2 - 1)^2 \right] d\rho
\ee
for the (rescaled) energy functional (\ref{energynatural}).
Notice that a non-zero $g$ at $\rho = 0$ (where $f \neq 1$) reduces the
potential energy but increases the gradient energy for small values of
$\rho$.  If $\beta$ is large, this can be energetically favourable
(conversely, for very small $\beta$, the cost in gradient  energy
due to a non-zero $g$ could outweigh any reduction in
potential energy). Indeed, Hindmarsh showed that there are no
minimum-energy vortices of finite core radius when $\beta >1$ by
constructing a one-parameter family of configurations whose energy
tends to the Bogomolnyi bound as the parameter $\rho_0$ is increased:
\be
 f(\rho) = {\rho \over \rho_0} \[ 1 + {\rho^2 \over \rho_0^2} \]^{-1/2} \qquad
 g(\rho) = \[ 1 + {\rho^2 \over \rho_0^2} \]^{-1/2}		\qquad
 v(\rho) = {\rho^2 \over \rho_0^2} \[ 1 + {\rho^2 \over \rho_0^2}
\]^{-1} 
\ee
The energy per unit length of these configurations is 
${E} = \pi\eta^2(1 + 1/3\rho_0^2)$ which, as $\rho_0 \to
\infty$, tends to the Bogomolnyi bound. This shows that any stable
solution must {\it saturate} the Bogomolnyi bound, but this is
impossible because, when $\beta > 1$, saturation would require $B=0$
everywhere, which is incompatible with the total magnetic flux being
$2\pi /q$ (see the comment after eq. (\ref{Bog4}). While
this does not preclude the possibility of a metastable solution,
numerical studies have found no evidence for it
\cite{Hin92,AchKuiPerVac92}.  All indications are that, for $\beta >1$, the
semilocal string is unstable towards developing a condensate in its
core which then spreads to infinity.

\begin{figure}[tbp]
\caption{\label{unstablestring} A two-dimensional simulation of the 
evolution of a perturbed isolated semilocal string with $\beta >1$,
from
\cite{AchKuiPerVac92}. The plot shows the (rescaled) energy density per unit
length in the plane perpendicular to the string.  $\beta = 1.1$ The
initial conditions include a large destabilizing perturbation in the
core, $\Phi^T(t=0) = (1, f_{NO}(\rho) e^{i\varphi})$,  which is seen to
destroy the string.}
\smallskip
\epsfxsize = \hsize \epsfbox{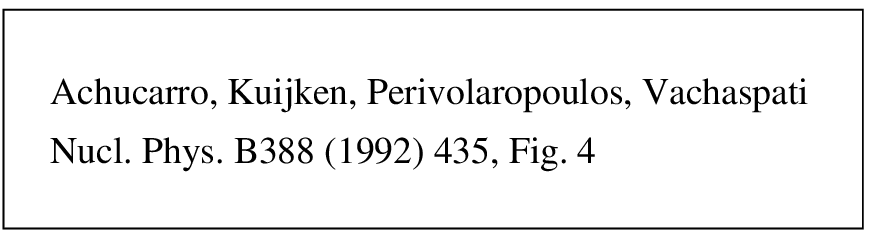}
\vskip 1truecm
\end{figure}

\begin{figure}[tbp]
\caption{\label{stablestring} The evolution of a string with $\beta <1$. 
The initial configuration is the same as in Fig. \ref{unstablestring} but
now, after a few oscillations, the configuration relaxes into a
semilocal string, $\Phi^T = (0, f_{NO}(\rho) e^{i\varphi})$. $\beta = 0.9$}
\smallskip
\epsfxsize = \hsize \epsfbox{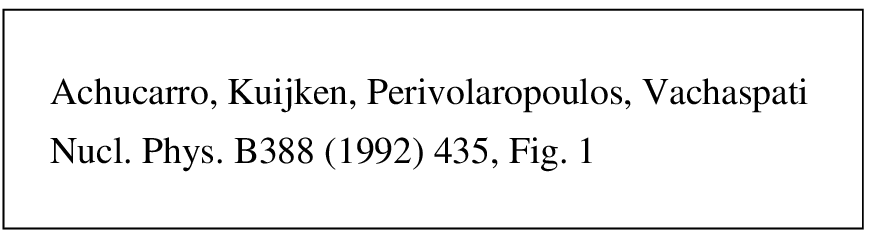}
\vskip 1 truecm
\end{figure}

Thus, the semilocal model with $\beta >1$ is a system where magnetic
flux is quantized, the vector boson is massive and yet there is no
confinement of magnetic flux \footnote{Preskill has emphasized
that the ``mixing'' of global and local generators is a necessary
condition for this behaviour, that is, there must be a generator of
$H$ which is a non-trivial linear combination of generators of $G_{\rm
global}$ and $G_{\rm local}$ \cite{Pre92}.}.

\subsubsection {The stability of strings with   $\beta <1$}
 
Semilocal strings with $\beta < 1$ are stable
to small perturbations. Numerical analysis of the eigenvalue equations
\cite{Hin92,Hin93} shows no negative
eigenvalues, and numerical simulations of the solutions themselves
indicate that they are stable to $z$-independent perturbations
\cite{AchKuiPerVac92,AchBorLid98}, including those with angular dependence. 
Note that the stability to $z$-dependent perturbations is automatic,
as they necessarily have higher energy. These results are confirmed by
studies of electroweak string stability \cite{GooHin95b,AchGreHarKui94} 
taken in the limit $\theta_w \to \pi/2$.

\subsubsection {$\beta = 1$ zero modes and skyrmions}
\label{semilocalzeromodes}

Substituting the ansatz (\ref{perturbedsl}) into the (rescaled)
Bogomolnyi equations for $n=1$ gives :
\be
\eqalign{ f'(\rho) + {v(\rho) -1\over \rho} f(\rho) &= 0 \cr
 g'(\rho)  + {v(\rho) \over \rho} g(\rho) &= 0 \cr
v'(\rho) + \rho(f^2(\rho) + g^2(\rho) - 1) &= 0\cr}
\label{semilocalbogomolnyi}
\ee

When $\beta = 1$ we showed earlier that the semilocal string
$f=f_{NO}, \ g=0, v=v_{NO}$ saturates the Bogomolnyi bound, so it is
necessarily stable (since it is a minimum of the energy). There may
exist, however, other solutions satisfying the same boundary conditions 
and with the same energy. Hindmarsh showed that this is indeed
the case by noticing that the eigenvalue equation has a
zero-eigenvalue solution \cite{Hin92} 
\be 
\psi = \psi_0 \exp \[ -\int_0^\rho  d{\hat\rho} 
 {{v({\hat\rho})} \over {\hat\rho}} \] \ \ , \qquad \psi_0 = {\rm const} \ ,
\ee
which signals a degeneracy in the solutions to the Bogomolnyi
equations. 
(Note that the `colour' at infinity, $\Phi_0$,
is fixed, so this is not a zero mode associated with the 
global $SU(2)$
transformations; its dynamics have been studied in \cite{Lee92}.)  

It can be shown that the zero mode exists for any value of $g$,
not just $g = 0$; the Bogomolnyi equations
(\ref{semilocalbogomolnyi}) are not independent since,
\be
g(\rho) = q_0 {f(\rho) \over \rho} 
\ee 
is a solution of the second equation for any (complex) constant
$q_0$. Solving the other two equations leads to the most 
general solution with winding number one and centred at $\rho=0$. It is
labelled by the complex parameter $q_0$, which fixes the size and
orientation of the vortex:

\be
     \left( \phi_1 \atop \phi_2 \right) ={1\over \sqrt{\rho^2+|q_0|^2}}
      \left( q_0 \atop \rho{\rm e}^{i\varphi} \right) \exp \left\{ {1\over
      2}\, u(\rho;|q_0|) \right\} \> ,
\label{skyrmsol}
\ee
where $u = \ln |\Phi|^2 $ is the solution to 
\be
\nabla^2 u + 2(1 - {\rm e}^u ) = \nabla^2 \ln (\rho^2 + |q_0|^2) \, ,
\qquad\qquad u \to 0 \quad  {\rm as} \quad \rho \to \infty \ \ .
\ee

If $q_0 \ne 0$, the asymptotic behaviour of these solutions is very
different from that of the Nielsen-Olesen vortex; the Higgs field is
non-zero at $\rho=0$ and approaches its asymptotic values like
$O(\rho^{-2})$. Moreover, the magnetic field tends to zero as $B \sim
2|q_0|^2 \rho^{-4}$, so the width of the flux tube is not as 
well-defined as in the $q_0 = 0$ case when $B$ falls off exponentially.
These $q_0 \neq 0$ solutions have been dubbed `skyrmions'.    
In the limit $|q_0| \to 0$, 
one recovers the semilocal string solution (\ref{slsol}), with $u
= \ln (f^2_{NO})$, the Higgs vanishing at $\rho=0$ and approaching the
vacuum exponentially fast.  On the other hand, when $|q_0| \gg 1$,
$u\approx 0$ the scalar field is in vacuum everywhere and the solution
approximates a $\CC P^1$ lump
\cite{Hin92, LeeSam93}. 
Thus, in some sense, the `skyrmions' interpolate between
vortices and $\CC P^1$ lumps.

\subsubsection{Skyrmion dynamics}

We have just seen that, for $\beta = 1$ the semilocal vortex
configuration is degenerate in energy with a whole family of
configurations where the magnetic flux is spread over an arbitrarily
large area. It is interesting to consider the dynamics of these
`skyrmions' when $\beta \neq 1$ \cite{Hin93,BenBuc93}: large skyrmions tend
to contract if $\beta <1$ and to expand if $\beta>1$. The timescale
for the collapse of a large skyrmion increases quadratically with its
size \cite{Hin93}. Thus large skyrmions collapse very slowly. 

Benson and Bucher \cite{BenBuc93} derived the energy spectrum of
delocalized `skyrmion' configurations in 2+1 dimensions as a function
of their size. More precisely, they defined an `antisize' 
$\chi = E_{\rm magnetic}/E_{\rm total}$ as the ratio of the magnetic energy
$\int d^2 x {B}^2/2$ to the total energy (\ref{energynatural}). Note that
when the flux lines are concentrated, magnetic energy is high compared
to the other contributions, and vice versa. Thus, $\chi \to 0$
corresponds to the limit in which the magnetic flux lines are spread
over an infinitely large area, which explains the name `antisize'.

For large skyrmions - those with  
$\chi \leq \beta/(1+\beta)$ - they concluded that the
minimum energy configuration among all delocalized configurations with
antisize $\chi$ satisfies
\be
E(\beta, \chi) \ = \ 2\pi {\eta^2 \over 2} ~ {\beta \over \beta - \chi
(\beta - 1)}\ee (if $\chi > \beta / (1 + \beta)$ the analysis does not
apply). Therefore, energy decreases monotonically with decreasing 
$\chi$ for
$\beta >1$ and increases monotonically for $\beta <1$, confirming that 
delocalized configurations tend to grow in size if $\beta >1$ and
shrink if $\beta <1$. 

This behaviour is observed in numerical simulations
\cite{AchBorLid97}. Benson and Bucher \cite{BenBuc93} have pointed out that in a
cosmological setting the expansion of the Universe could drag the
large skyrmions along with it and stop their collapse. The simulations
in flat space are at least consistent with this, in that they show that
delocalised configurations tend to live longer when artificial
viscosity is increased, but a full numerical simulation of the
evolution of semilocal string networks has not yet been performed and
is possibly  the only way to answer these questions reliably.

Finally, we stress that the magnetic flux of a skyrmion does not
change when it expands or contracts (the winding number is conserved)
but this does not say anything about how localized the flux is.  In
contrast with the Abelian Higgs case, the size of a skyrmion can be
made arbitrarily large with a finite amount of energy.

\subsection{Semilocal string interactions} 

\subsubsection{Multivortex solutions, $\beta = 1$, same colour}

Multi-vortex solutions in 2+1 dimensions corresponding to parallel
semilocal strings with the same colour have been 
constructed by Gibbons, Ruiz-Ruiz, Ortiz and Samols
\cite{GibOrtRuiSam92} for the critical case $\beta = 1$. Their analysis
closely follows that of \cite{JafTau80} in the case of the Abelian Higgs
model, and starts by showing that, as in that case, the full set of
solutions to the (second order) equations of motion can be obtained by
analysing the solutions to the (first order) Bogomolnyi equations.

In the Abelian Higgs model, solutions with winding number $n$ are
labelled by $n$ unordered points on the plane (those where the scalar
field vanishes) which, for large separations, are identified with the
positions of the vortices. In the semilocal model, 
the solutions have other degrees of freedom, besides position,
describing their size and orientation.

Assuming without loss of generality that the winding number $n$ is
positive, and working in temporal gauge $Y_0 = 0$,
any solution with winding number $n$ is specified
(up to symmetry transformations) by two holomorphic polynomials
\begin{eqnarray*}
     P_n(z) && =\prod_{r=1}^n (z-z_r)
     \\
     &&\equiv  z^n + p_{n-1} z^{n-1} + \ldots + p_1 z + p_0
\end{eqnarray*}
and
\be
     Q_n(z) \qquad \equiv q_{n-1} z^{n-1} + \ldots + q_1 z + q_0
\ee
where $z = x+iy$ is a complex coordinate on the $xy$ plane. 
The solution for the Higgs fields is, up to gauge
transformations,
\be
     \left( \phi_1 \atop \phi_2 \right) ={{\rm e}^{{1\over 2} u (z,
\overline{z})}\over
\sqrt{|P_n|^2+|Q_n|^2}} \left(Q_n \atop P_n \right)
\ee
where the function $u(z, \overline{z}) = \ln (|\phi_1|^2 + |\phi_2|^2)$
must satisfy 
\be
\nabla^2 u + 2( 1 - {\rm e}^u ) = \nabla^2 \ln (|P_n|^2 + |Q_n|^2) \, ,
\ee
and tend to 0 as $|z| \to \infty$.  Although its form is not known
explicitly, ref. \cite{GibOrtRuiSam92} proved the existence of a unique
solution to this equation for every choice of $P_n$ and $Q_n$
(if $P_n$ and $Q_n$ have
a common root then $\exp [u/2]$ has a zero there, so the
expression for the Higgs field is  everywhere well-defined).
The gauge field 
can then be read off from the Bogomolny equations
(\ref{semilocalbogomolnyieq}).   
This generalises (\ref{skyrmsol}) to arbitrary $n$. 
The coefficients of $P_n(z)$, $Q_n(z)$ parametrise the moduli space,
$\CC^{2n}$.

The Nielsen-Olesen vortex has $Q_n=0$.
If $P_n \neq 0$, then in regions where $|Q_n| << |P_n|$ one finds
\be
|\phi_1| \sim 1 - \half \left| {Q_n \over P_n} \right|^2 \ \ , \qquad
|\phi_2| \sim \left| {Q_n\over P_n} \right| \ \ , \qquad v \sim 1 -
\left| {Q_n \over P_n} \right|^2
\ee
indicating that the scalar fields fall off
as a power law, as opposed to the usual exponential fall off found in
NO vortices. The same is true of the magnetic field.

The low energy scattering of semilocal vortices and skyrmions with
$\beta = 1$ was studied in \cite{LeeSam93} in the geodesic approximation
of \cite{Man82}. The behaviour of these solitons was found to be
analogous to that of $\CC P^1$ lumps but without the singularities,
which are smoothed out in the core.

\subsubsection{Interaction of parallel strings, $\beta <1$, 
different colours}

Ref. \cite{AchKuiPerVac92} carried out a numerical study in two dimensions of
the interaction between stable ($\beta<1$) strings with different
``colour'' with non-overlapping cores. It was found that the strings
tend to radiate away their colour difference in the form of Goldstone
bosons, and there is little or no interaction observed. The position
of the strings remains the same during the whole evolution while the
fields tend to minimize the initial relative $SU(2)$ phase (see
figure \ref{interaction}).

Thus, we expect interactions betwen infinitely long  semilocal
strings with different colours  to be  essentially the same as for
Nielsen-Olesen strings. 
This expectation is confirmed by numerical simulations of two- and
three-dimensional semilocal string networks \cite{AchBorLid97,AchBorLid98},
discussed in \ref{networks}.

\begin{figure}[tbp]
\epsfxsize = \hsize \epsfbox{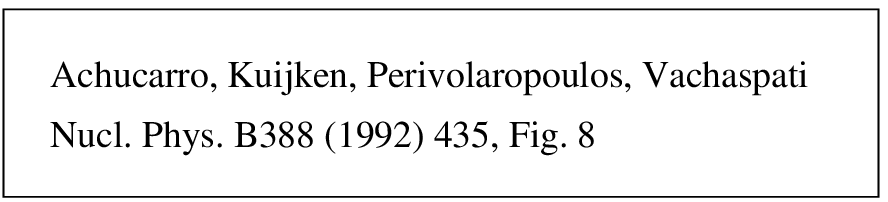}
\smallskip
\caption{\label{interaction} A numerical simulation of the 
interaction between two parallel semilocal strings with different
`colour', from Ref. \cite{AchKuiPerVac92}. The initial configuration has one
string with $\Phi_1^T = (0, f(\rho_1) e^{i\varphi_1} )$  and the other
with $\Phi_2^T = (if(\rho_2) e^{i\varphi_2}, 0)$, where  $(\rho_i,
\varphi_i)$ are polar coordinates centred at the cores of each string.
The energy density of the string pair is plotted in the plane
perpendicular to the strings. The colour difference is radiated away
in the form of Goldstone bosons, and the strings cores remain at their
initial positions.  $\beta = 0.5$. }
\vskip 1truecm
\end{figure}

\subsection {Dynamics of string ends}

Note that, in contrast with  Nielsen-Olesen strings,
there is no topological reason  that forces a semilocal  string to continue
indefinitely or form a closed loop.  Semilocal strings  can
end in a ``cloud'' of energy, which behaves like 
a global monopole \cite{Hin92}.

Indeed, consider the following asymptotic configuration for the Higgs field:
\be
\Phi = {\eta \over \sqrt 2} \(\begin{array}{c}
                       \cos\half\theta \\ 
			\sin\half\theta\, e^{i\varphi}
          \end{array}\)
\ee
which is ill-defined at $\theta = \pi$ and at $r=0$. We can make the
configuration regular by introducing profile functions such that the
Higgs field vanishes at those points:
\be
\Phi = {\eta \over \sqrt 2} \(\begin{array}{c}
                       h_1(r,\theta)\cos\half\theta \\
                       h_2(r,\theta)\sin\half\theta\, e^{i\varphi} 
         \end{array}\)
\ee
where $h_1$ and $h_2$ vanish at $r=0$ and $h_2(r,\pi)=0$. 
This configuration describes a string in the $z<0$
axis ending in a monopole at $z=0$.

At large distances, $r>>1$, the Higgs field is everywhere in vacuum
(except at $\theta \approx \pi$) and we find $\Phidag {\vec \tau} \Phi \sim
\vec x$, just like for a Hedgehog in $O(3)$ models.
On the other hand, the configuration for the gauge fields resembles
that of a semi-infinite solenoid; the string supplies U(1) flux which
spreads out from $z=0$.

This is the $\theta_w \to \pi/2$ limit of a configuration first
discussed by Nambu \cite{Nam77} in the context of the GSW model  -see
section \ref{zoo} - but here the energy of the monopole is
linearly divergent because there are not enough gauge fields to cancel
the angular gradients of the scalar field.

Angular gradients provide an important clue to understand the dynamics
of string ends.  If $\beta < 1$, numerical simulations show that
string segments grow to join nearby segments or to form loops (see
figures \ref{loopformation} and \ref{semi4}) \cite{AchBorLid98}. This
confirms analytical estimates in refs. \cite{GibOrtRuiSam92,Hin93}. In
other cases the string segment collapses under its own tension, with
the monopole and antimonopole at the ends annihilating each other.

\begin{figure}[t!]
\centering
\leavevmode\epsfysize=7.5cm \epsfbox{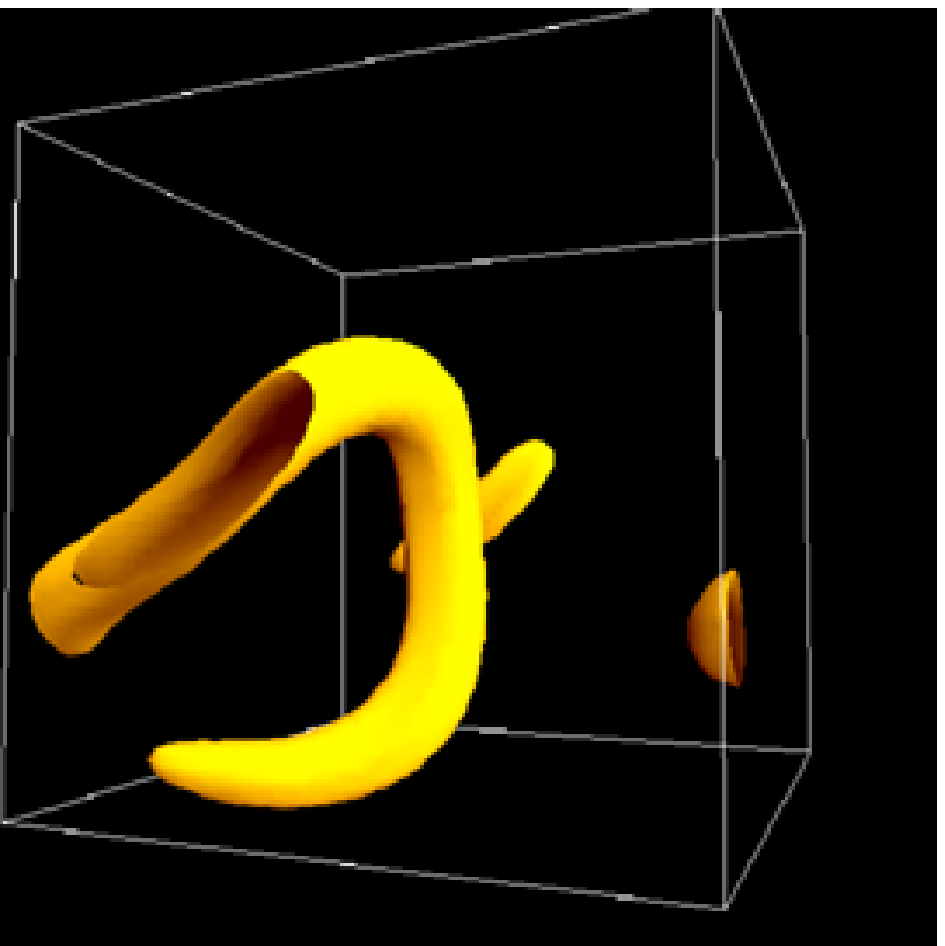}\\ % Colour
\vspace*{5pt}
\leavevmode\epsfysize=7.5cm \epsfbox{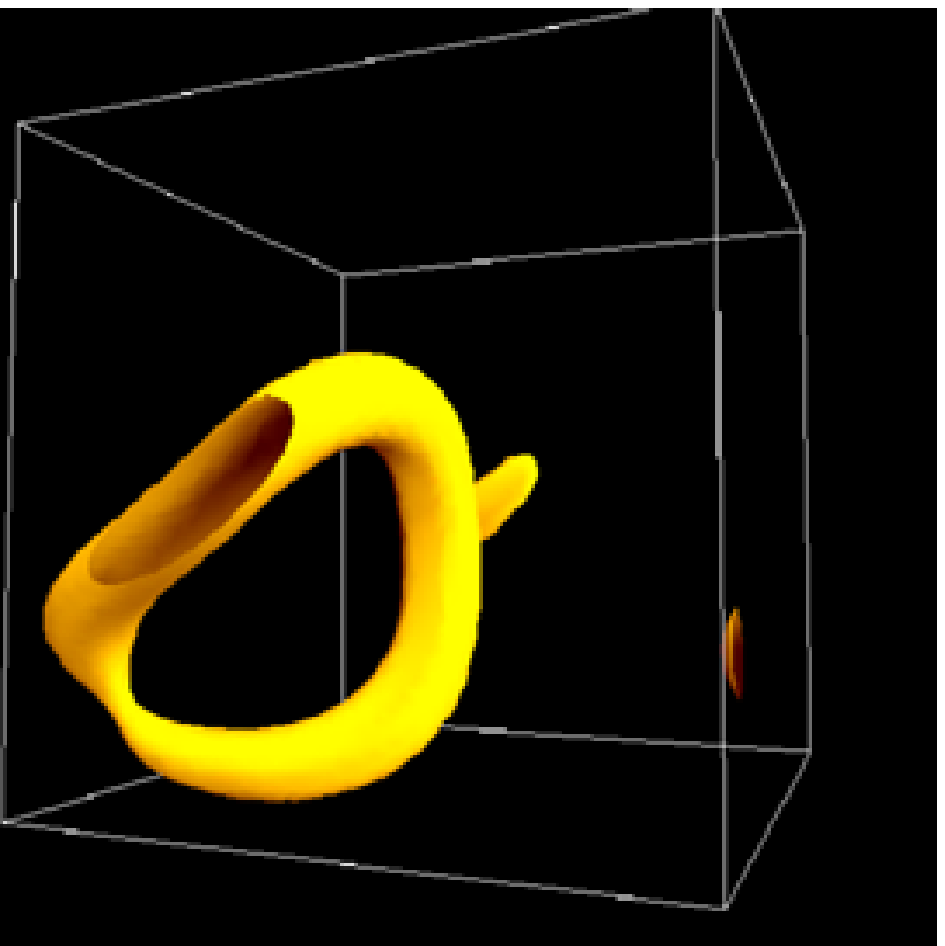}\\ % Colour
\vspace*{5pt}
\caption[looopformation]{\label{loopformation} 
Loop formation from semilocal string segments.  The figure shows two
snapshots, at $t = 70$ and $t = 80$, of a $64^3$ numerical simulation
of a network of semilocal strings with $\beta = 0.05$ from
Ref.\cite{AchBorLid98}, where the ends of an open segment of string join up
to form a closed loop (see section \ref{networks} for a discussion of
the simulations). Subsequently the loops seem to behave like those of
topological cosmic string, contracting and disappearing.}
\vskip 1 truecm
\end{figure}

\begin{figure}[t!]
\centering
\leavevmode\epsfysize=7.5cm \epsfbox{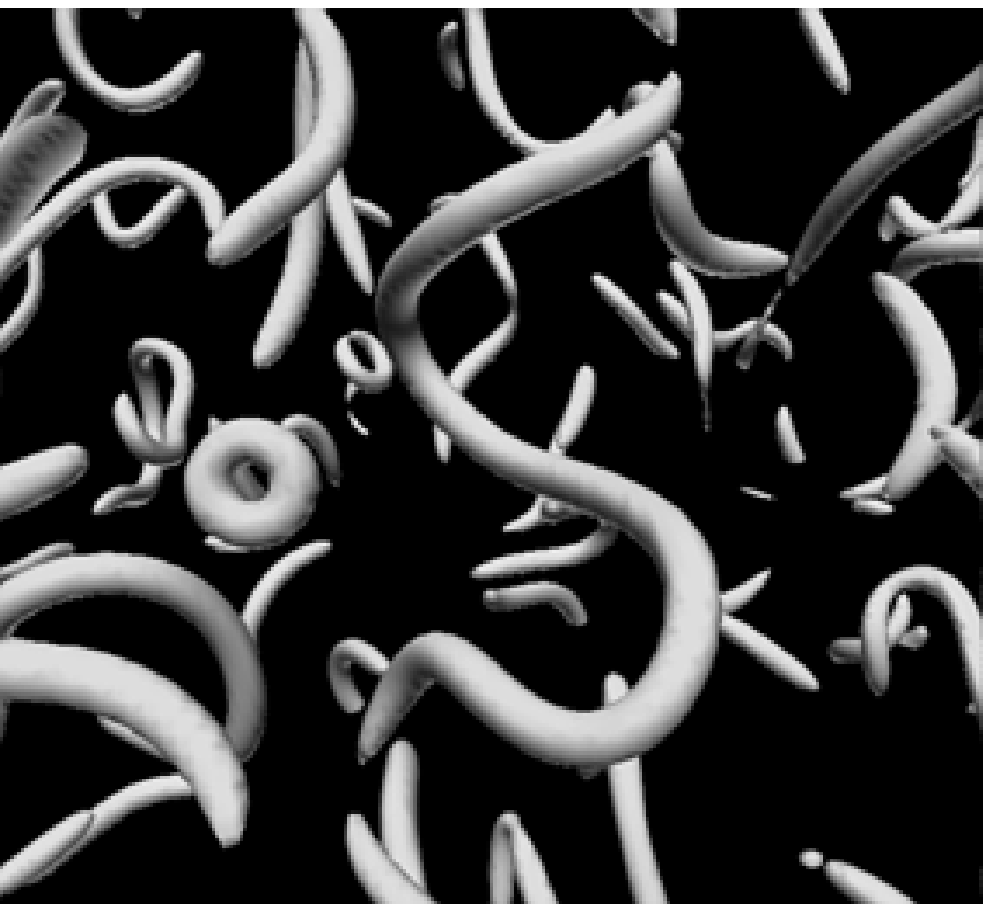}\\ % B&W
\vspace*{5pt}
\leavevmode\epsfysize=7.5cm \epsfbox{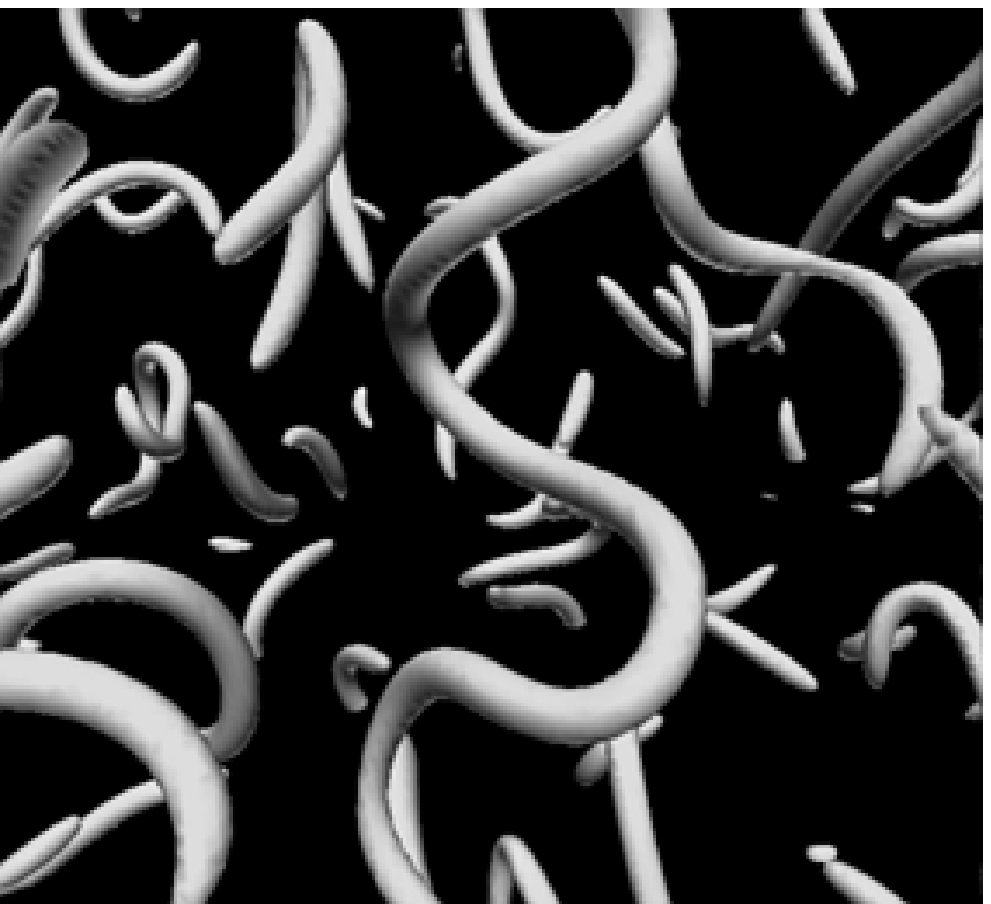}\\ % B&W
\vspace*{5pt}
\caption[semi4]{\label{semi4}
{\it  The growth of string segments to form longer strings}. The figure
shows two snapshots, at time $t=60$ and $t=70$ of a large $256^3$
numerical simulation of a network of semilocal strings with $\beta =
0.05$ from Ref. \cite{AchBorLid98}. Note several joinings of string
segments, e.g.~two separate joinings on the long central string, and
the disappearance of some loops. The different apparent thickness of
strings is entirely an effect of perspective. The
simulation was performed on the Cray T3E at the National Energy
Research Scientific Computing Center (NERSC).  See section
\ref{networks} for a discussion.}
\vskip 1 truecm
\end{figure}

\subsection{Numerical simulations of semilocal string networks}
\label{networks}

As the early Universe expanded and cooled to become what we know today
it is very likely that it  went through a number of phase transitions where
topological (and possibly non-topological) defects are expected to
have formed according to the Kibble
mechanism \cite{Kib76,Zur85,reviews}. Although the cosmological evidence
for the existence of such defects remains unclear
\cite{cosm}, there is plenty of experimental evidence from
condensed matter systems that networks of defects do form in symmetry
breaking phase transitions
\cite{NATO}, the most
recent confirmation coming from the Lancaster-Grenoble-Helsinki
experiments in vortex formation in superfluid Helium
\cite{nature}. 
An important question is whether semilocal (and electroweak)
strings are stable enough to form in a phase transition.

We defer discussion of the electroweak case to section
\ref{cosmologicalapplications}. 
Here we want to review recent numerical simulations of
the formation and evolution of a network of $\beta<1$ semilocal strings
\cite{AchBorLid97,AchBorLid98,AchBorLid98b} which show that such 
strings should indeed
form in appreciable numbers in a phase transition.  The results
suggest that, even if no vortices are formed immediately after
$\Phi$ has acquired a non-zero vacuum expectation value,
the interaction between the gauge fields and the scalar fields 
is such that vortex formation
does eventually occur
simply because it is energetically favourable for the random
distribution of magnetic fields present after the phase transition
to become concentrated in regions where the Higgs field has a value 
close to that of the symmetric phase. 
%\footnote{At the time
%this review went to press, colour images and movies of the
%three-dimensional simulations could be found on the WWW at \\ {\small
%http://cfpa.berkeley.edu/$\sim$borrill/defects/semilocal.html}.  Also,
%a 100 frame {\tt mpeg} movie (0.5Mb) of a two-dimensional simulation
%was available at \\ {\small
%http://star-www.cpes.susx.ac.uk/people/arl\underline{~}recent.html}.}.

Even though they do not account for the expansion of the Universe,
these simulations represent a first step towards understanding
semilocal string formation in cosmological phase transitions and they
have already provided very interesting insights into the dynamical
evolution of such a network.

\subsubsection{Description of the simulations}

>From a technical point of view, the numerical simulation of a
network of semilocal strings has additional complications over
that of $U(1)$ topological strings. 
Because there are not enough gauge degrees of
freedom to cancel all of the scalar field gradients, the existence of
string cores depends crucially on the way the fields (scalar and
gauge) interact.  Another problem, generic to all non-topological
strings, is that the winding number is not well defined for
configurations where the scalar is away from a maximal circle in the
vacuum manifold, and this makes the identification of strings much more
difficult than in the case of topological strings.

The strategy proposed in \cite{AchBorLid97} to circumvent 
these problems is to follow the evolution of the 
gauge field strength in numerical simulations, 
since the field strength provides
a gauge invariant indicator for the presence of vortices.
The initial conditions are obtained by an extension of the
Vachaspati-Vilenkin algorithm
\cite{VacVil84} appropriate to non-topological defects, plus a short
period of dynamical evolution including a dissipation term (numerical
viscosity) to aid the relaxation of configurations in the `basin of
attraction' of the semilocal string.  

As with any new algorithm, it is essential to check that it reproduces
previously known results accurately, and this has been done in
\cite{AchBorLid97}. Note that setting $\phi_2=0$ in the semilocal model
obtains the Abelian Higgs model, thus comparison with topological
strings is straightforward, and it is used repeatedly as a test case,
both to check the simulation techniques and to minimise systematic
errors when quoting formation rates. In particular, the proposed
technique is tested in a two-dimensional toy model (representing
parallel strings) in three different ways: a) restriction to the
Abelian Higgs model gives good 
agreement with analytic and
numerical estimates for cosmic strings in \cite{VacVil84}; b) the results
are robust under varying initial conditions and numerical viscosities
(see Figure \ref{semi2}), and c) they are in good agreement
with previous analytic and numerical estimates for semilocal string
formation in
\cite{AchKuiPerVac92,Hin93}.

The results are summarized in Fig.
\ref{semi6}. We refer the reader to 
refs. \cite{AchBorLid97,AchBorLid98,AchBorLid98b} 
for details;  however, a few comments are needed to understand those
figures.

$\bullet$ The study takes place in flat spacetime. Temporal gauge and
rescaled units (\ref{natural}) are chosen.  Gauss' law, which here is
a constraint derived from the gauge choice $Y_0=0$, is used to test
the stability of the code. 

$\bullet$ Space is discretized into a lattice with periodic boundary
conditions.  The equations of motion (\ref{eomnatural}) are solved
numerically using a standard staggered leapfrog method; however, to
reduce its relaxation time an {\it ad hoc} dissipation term was added
to each equation ($\eta
\dot{\Phi}$ and $\eta
\dot{Y}_i$ respectively).  A range of strengths of dissipation was tested, 
and it did not
significantly affect the number densities obtained.
The simulations  displayed in this section all have  
have $\eta = 0.5$.

$\bullet$ The number density
of defects is estimated by an extension of the Vachaspati-Vilenkin algorithm
\cite{VacVil84} by first  generating a random
initial configuration for the scalar fields drawn  from
the vacuum manifold, which is not discretised, and then finding  the
gauge field configuration that minimizes the energy associated
with (covariant) gradients\footnote{In fact, it turns out that the
energy-minimization condition is redundant, since the early
stages of dynamical evolution carry out this role anyway.}.
If space is  a grid of dimension $N^3$, 
the correlation length is chosen
to be some number $p$ of grid points 
($p=16$ in \cite {AchBorLid97,AchBorLid98}; 
the size of the lattice is either  $N=64$ or $N=256$.)
To obtain a reasonably smooth configuration
for the scalar fields, one throws down random vacuum values on a
$(N/p)^3$ subgrid; the
scalar field is then interpolated onto the full grid by
bisection. Strings are always
identified with the location of magnetic flux tubes. 

For cosmic strings, the two-dimensional toy model accurately 
reproduces the formation rates of \cite{VacVil84}.
For semilocal strings, on the other hand, the initial configurations
generated in this way have a complicated flux structure with extrema
of different values (top panel of Fig. \ref{semi1}), and it is far
from clear which of these, if any, might evolve to form semilocal
vortices; in order to resolve this ambiguity, the 
initial configurations are
evolved forward in time.  As anticipated, in the unstable regime
$\beta > 1$ the flux quickly dissipates leaving no strings. By
contrast, in the stable regime $\beta < 1$ stringlike features emerge
when configurations in the ``basin of attraction''of the semilocal
string relax unambiguously into vortices (bottom panel of
Fig. \ref{semi1}).

Since the initial conditions are somewhat artificial, the results
were checked against various other choices of initial conditions, in
particular different initial conditions for the gauge field and also
thermal initial conditions for the scalar field (see Fig. \ref{semi2}).
All the initial conditions in \cite{AchBorLid97,AchBorLid98} had zero initial
velocities for the fields. Initial conditions with non-zero field momenta
have not yet been investigated.

\begin{figure*}
\centering
\leavevmode\epsfysize=10.9cm \epsfbox{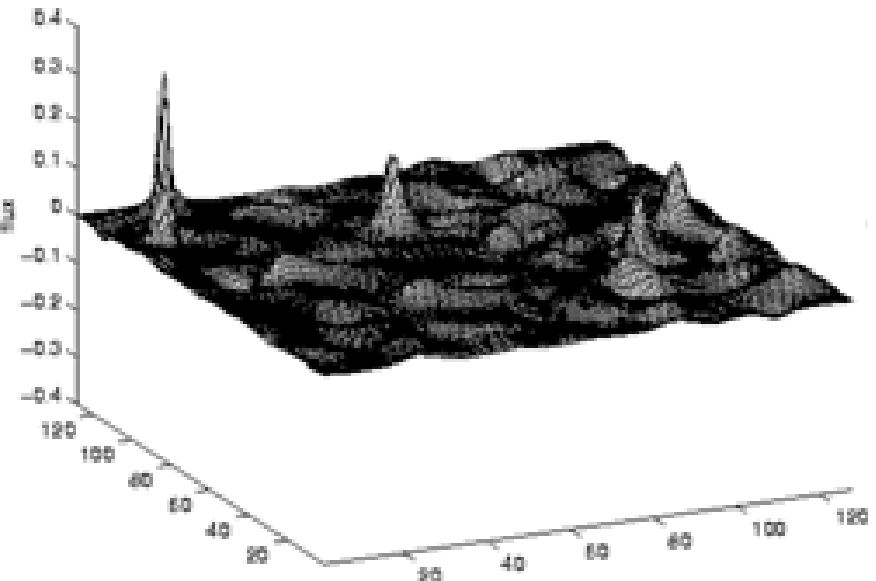}\\
\leavevmode\epsfysize=10.9cm \epsfbox{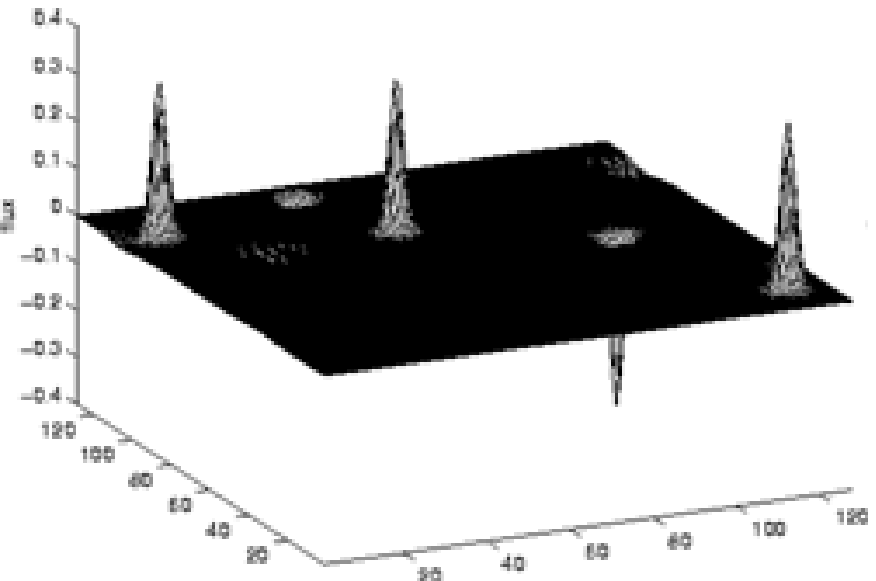}\\
\caption[semi1]{\label{semi1} The flux tube structure in a
  two-dimensional semilocal string simulation with $\beta = 0.05$,
  from Ref \cite{AchBorLid97}. The upper panel ($t=0$) shows the initial
  condition after the process described in the text. The lower panel
  shows the configuration resolved into five flux tubes by a short
  period of dynamical evolution ($t=100$). These flux tubes are
  semilocal vortices.}
\vskip 1 truecm
\end{figure*}

\subsubsection {Results and discussion}

These simulations give very important information on the dynamics and
evolution of a network of semilocal strings. In particular, they
confirm our discussion in the previous subsection of the behaviour of
the ends of string segments, and of strings with different
colours. String segments are seen to grow in order to join nearby ones
or form closed loops, and very short segments are also observed to
collapse and disappear. The colour degrees of freedom do not seem to
introduce any new forces between strings. Because the strings tend to
grow or form closed loops, time evolution makes the network resemble
more and more a network of topological strings (NO vortices) but with
lower number densities\footnote{However, one important point is that
no intersection events were observed in the semilocal string
simulations, so the rate of reconnection has not been determined.}

Note that the correlation length in the simulations is constrained to
be larger than the size of the vortex cores, to avoid overlaps. This
results in a minimal value of the parameter $\beta$ of around 0.05 (if
$\beta$ is lowered further, the scalar string cores become too wide to
fit into a correlation volume, in contradiction with the vacuum values
assumed in a Vachaspati-Vilenkin algorithm).  Figure \ref{semi6} shows
the results for seven different values of $\beta$ by taking several
initial configurations on a $64^3$ grid smoothed over every $16$
grid-points.  As expected, for $\beta < 1$ the formation rate depends
on $\beta$, tending to zero as $\beta$ tends to 1. The
ratio of semilocal string density to cosmic string density in an
Abelian Higgs model for the same value of $\beta$ is {less
than} but of order one.
{For} the lowest value of $\beta$ simulated ($\beta
= 0.05$), the 
semilocal
string density is about one third {of} that of cosmic strings.

\begin{figure}
\centering
\leavevmode\epsfysize=6cm \epsfbox{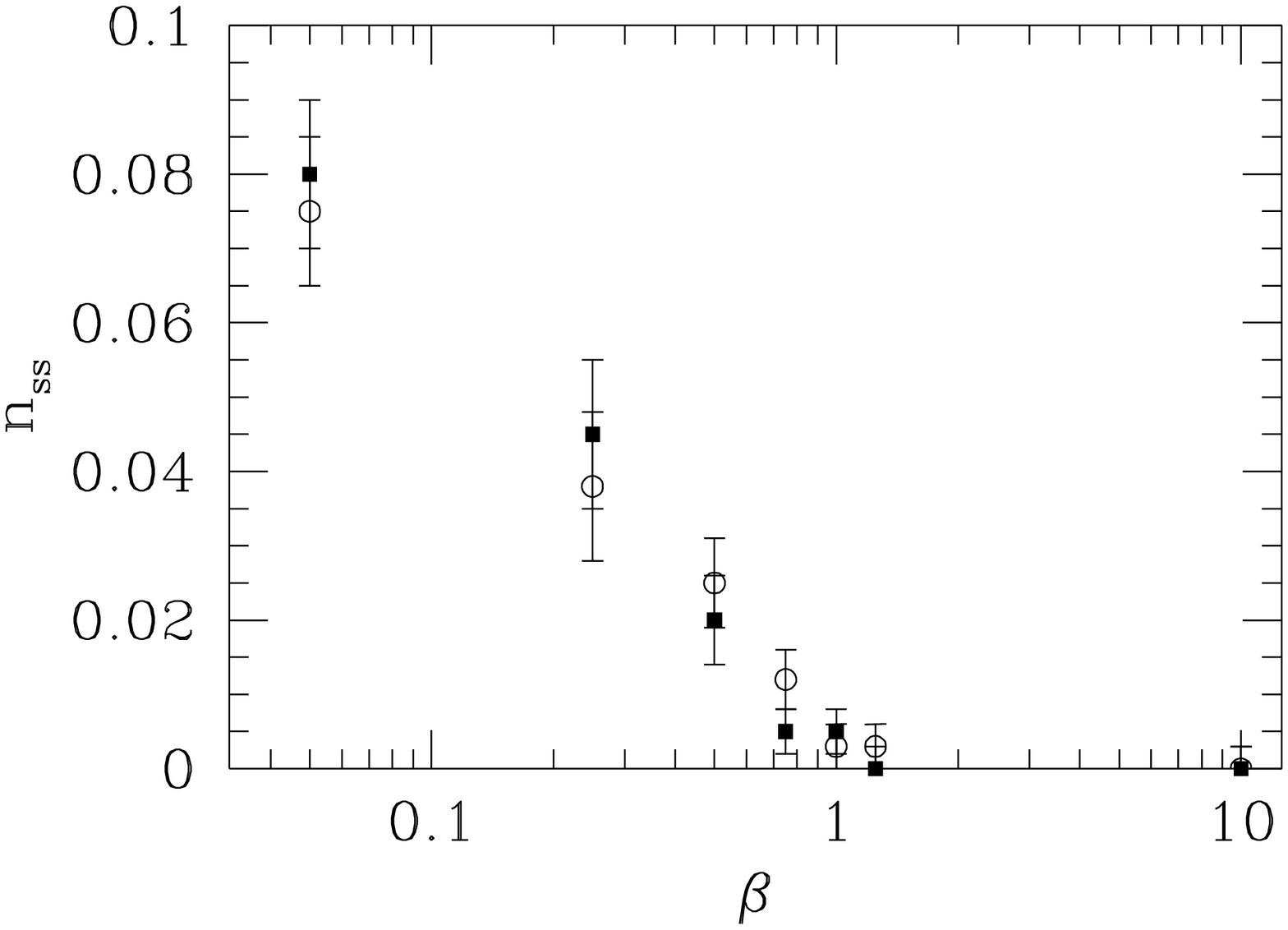}\\
\caption[semi2]{\label{semi2} A test of the sensitivity of the 
results to the choice of initial conditions in a two-dimensional
simulation with the algorithm proposed in section \ref{networks}. The
plot shows the number of semilocal strings formed per initial
two-dimensional correlation volume. Each point is an average over ten
simulations.  Squares indicate that the vacuum initial conditions
described in the text were used, while open circles indicate that
non-vacuum (thermal) initial conditions were used.  Both sets of
initial conditions are seen to give comparable results. Statistical
results are derived from a large suite of simulations (700 in all)
carried out on a $64^3$ grid (from
Ref. \cite{AchBorLid97}) }
\vskip 1 truecm
\end{figure}

\begin{figure}[t]
\centering
\leavevmode\epsfysize=6cm \epsfbox{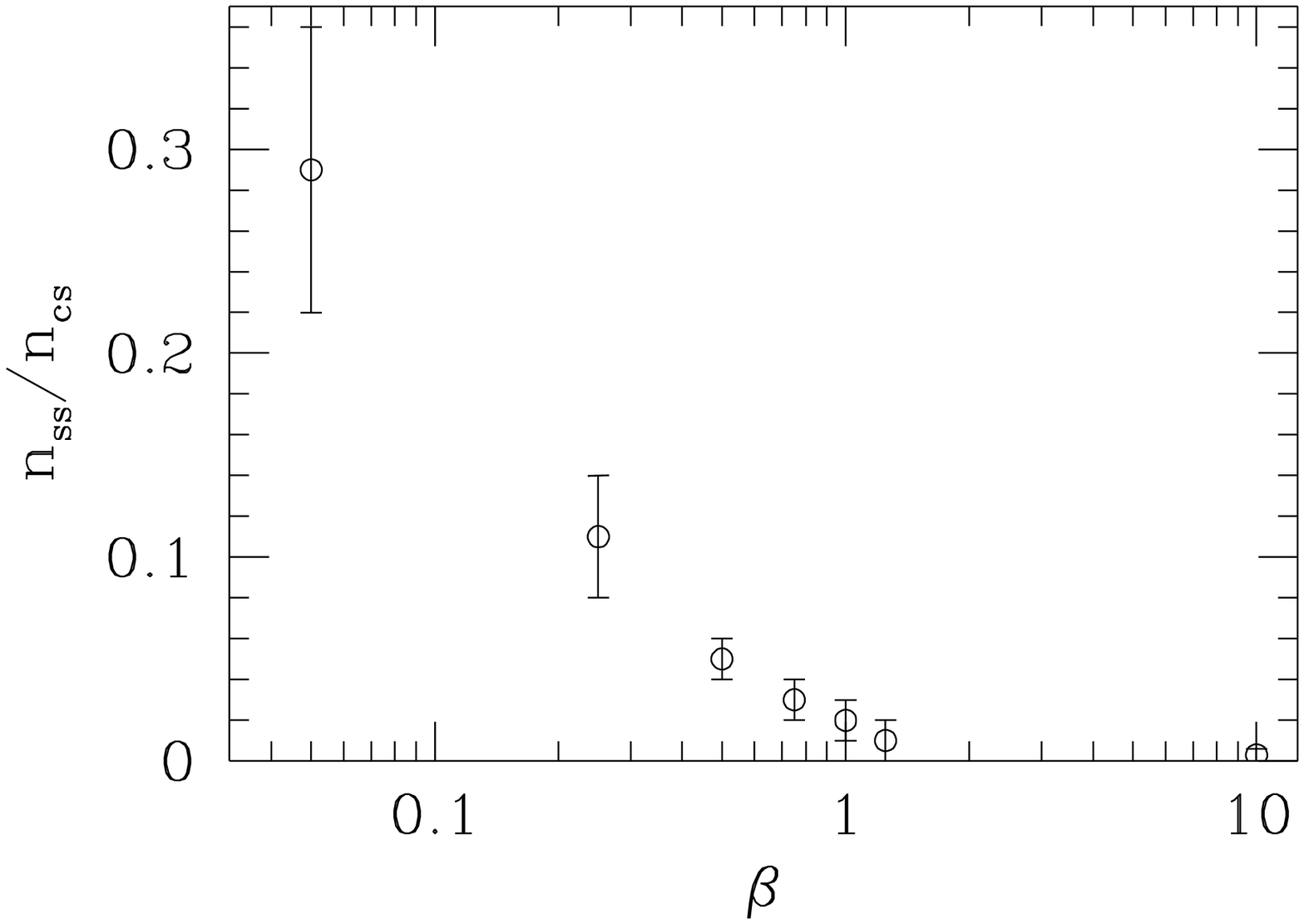}\\
\caption[semi6]{\label{semi6} The ratio of lengths of semilocal and cosmic
strings as a function of the stability parameter $\beta$, from
\cite{AchBorLid98}.}
\vskip 1 truecm
\end{figure}

One final word of caution about the possible cosmological implications
of these simulations. We mentioned above that numerical viscosity was
introduced to aid the relaxation of configurations close to the
semilocal string. In an expanding Universe the expansion rate would
provide some viscosity, though $\eta$ would typically not be
constant. This may have an important effect on the production of
strings. Indeed, note the different numbers of upward and downward
pointing flux tubes in Fig. \ref{semi1}, despite the zero net flux boundary
condition. The missing flux resides in the smaller `nodules', made
long-lived by the numerical viscosity; these are none other than the
`skyrmions' described in section \ref{semilocal}. As was explained
there, the natural tendency of skyrmions when $\beta<1$ is to collapse
into strings, but the timescale for collapse increases quadratically
with their size and Benson and Bucher \cite{BenBuc93} have argued that the
effect of the expansion could stop the collapse of large skyrmions
almost completely. On the other hand one expects skyrmions to be
formed with all possible sizes, so the effect of the expansion on the
number density of strings remains an open question.  Another important
issue that has not yet been addressed is whether these semilocal
networks show scaling behaviour, and whether reconnections are as rare
as the above simulations suggest. Both would have important
implications for cosmology.  However, the answer to these and other
questions may have to wait until full numerical simulations are
available.

\subsection{Generalisations and final comments}

i) Charged semilocal vortices

The semilocal string solution described earlier in this
section is {\it strongly} static and
z-independent, by which we mean that $D_t(\Phi) = D_z(\Phi) = 0$. It
is possible to relax these conditions while still keeping the
Lagrangian and the energy independent of $z$ 
The idea is that, as we move along the
$z$-direction, the fields move along the orbit of the global
symmetries; in other words, Goldstone bosons are excited.

Abraham has shown that it is possible to construct semilocal vortices
with finite energy per unit length carrying a global charge
\cite{Abr93} in the Bogomolnyi limit $\beta = 1$ \footnote{By contrast,
charged solutions with $D_0(\Phi) \neq 0$ in the Abelian Higgs model
have infinite energy per unit length \cite{JulZee75}.}. They
satisfy a Bogomolnyi-type bound and are therefore stable.
Perivolaropoulos \cite{Per94} has
constructed spinning vortices (however these have infinite
energy per unit length). 

ii) {Semilocal models  with 
$SU(N)_{\rm global} \times U(1)_{\rm local}$ symmetry}

The generalization of semilocal strings to so-called Extended Abelian
Higgs models with an N-component multiplet of scalars whose overall
phase is gauged is straightforward
\cite{VacAch91,Hin92}, and has been analysed in detail in 
\cite{Hin93,GibOrtRuiSam92}.
The strings are stable (unstable) for $\beta < 1$ ($\beta >1$) and for
$\beta = 1$ they are degenerate in energy with skyrmionic
configurations labelled by an $N-1$ complex vector. For winding $n$,
and widely separated vortices, the $Nn$ complex parameters that
characterize the configurations can be thought of as the $n$ positions
in $\RR^2 \sim \CC$ and the $(N-1)n$ `orientations'.

iii) Semilocal monopoles and generalized semilocality

We have seen that semilocal strings have very special properties
arising from the fact that $\pi_1(G/H) = 0$ but $\pi_1(G_{\rm
local}/H_{\rm local}) \neq 0$.  An immediate question is whether it is
possible to construct other non-topological defects such that
\be 
\pi_k(G/H) = 0 \qquad\qquad {\rm but} \qquad\qquad  \pi_k(G_{\rm
local}/H_{\rm local}) \neq 0 \ \ .
\label{semilocalcondition}
\ee 
This possibility would be particularly interesting in the case of
monopoles, $k=2$, since they might retain some of the features of
global monopoles, in particular a higher annihilation rate in the
early Universe.  Surprisingly, the answer seems to be negative.
Within a very natural set of assumptions, it was shown in \cite{VacAch91}
that the condition (\ref{semilocalcondition}) can only be satisfied if
the gauge group $G_{\rm local}$ is Abelian, and therefore one cannot
have semilocal monopoles (nor any other defects satisfying
conditions (\ref{semilocalcondition}) with $k>1$). 

However, Preskill has remarked that it is possible to define a wider
concept of semilocality \cite{Pre92} by considering the larger
approximate symmetry $G_{\rm approx}$ which is obtained in the limit
where gauge couplings are set to zero. The symmetry $G_{\rm approx} $
is partially broken to the exact symmetry $ G \sim G_{\rm local}
\times G_{global}$ (modulo discrete transformations) when the gauge
couplings are turned on It is then possible to have generalized
semilocal monopoles associated with non-contractible spheres in
$G_{\rm local} /H_{\rm local}$ which are contractible in the
approximate vacuum manifold $G_{\rm approx}/H_{\rm approx}$ even
though they are still non-contractible in the exact vacuum manifold
$G/H$. 

Another obvious possibility is to have topological monopoles with
``colour'', by which we mean extra global degrees of freedom, if the
symmetry $G \sim G_{\rm global} \times G_{\rm local}$ 
is such that the
gauge orbits are non-contractible two-spheres, $\pi_2 (G_{\rm
local}/H_{\rm local}) \neq 1$.  Given that there are no semilocal
monopoles \cite{VacAch91}, these monopoles must have $\pi_2 (G/H) \neq 1$,
so they are topologically stable, and they have additional global
degrees of freedom.

iv) {Semilocal defects and Hopf fibrations}

In the semilocal model, the action of the gauge group fibres the vacuum
manifold $S^3$ as a non-trivial bundle over $S^2 \sim \CC P^1$, the
Hopf bundle. 
The fact that this bundle is non-trivial is at the root of conditions
(\ref{semilocalconditionforstring}), 
and is ultimately the reason why the topological
criterion for the existence of strings fails.  
In view of this,
Hindmarsh \cite{Hin93} has proposed an alternative definition of a
semilocal defect: it is a defect in a theory whose vacuum manifold is
a non-trivial bundle with fibre $G_{\rm local}/H_{\rm local}$.

Extended Abelian Higgs models \cite{Hin93} are similarly related to the
fibrations of the odd-dimensional spheres $S^{2N-1}$ with fibre $S^1$
and base space $\CC P^{N-1}$. 
A natural question to ask is if the remaining Hopf fibrations of
spheres can also be realised in a field theoretic model. This question
was answered affirmatively in \cite{HinHolKepVac93} for the $S^7 \ {S^3 \atop
\rightarrow}\  S^4$
fibration in a quaternionic model. Other non-trivial bundles were
also implemented in this paper, but to date the field theory
realisation of the $S^{15} \ {S^7 \atop \rightarrow} \ 
S^8$ Hopf bundle remains an open problem.

v) Monopoles and textures in the semilocal model:

Since the gauge field is Abelian, ${\rm div} \vec B =0$,
and isolated
magnetic monopoles are necessarily singular in semilocal models.  
The only way to make the singularity disappear is by embedding the
theory in a larger non-Abelian theory which provides a regular core,
or by putting the singularity behind an event horizon \cite{GibOrtRuiSam92}.
One important question that has not yet been addressed is if the
scalar gradients in these spherical monopoles make them unstable to
angular collapse into a flux tube.  A related system where this
happens is in $O(3)$ global monopoles where the spherically symmetric
configuration is unstable. In the semilocal case, it is possible
that the pressure from the magnetic field
might prevent the instability towards angular collapse.

Finally, note that, because $\pi_3(S^3) = \ZZ$, there is also the
possibility of textures in the semilocal model
(\ref{semilocalaction}). In contrast with purely scalar $O(4)$ models,
their collapse seems to be stopped by the pressure from the magnetic
field \cite{Hin93}. Of course they can still unwind by tunnelling.

vi) We should point out that systems related to the semilocal model
have been studied in condensed matter. In \cite{BurKop87}, the system
was an unconventional superconductor where the role of the global
SU(2) group was played by the spin rotation group. In \cite{Vol84} the
hypothetical case of an ``electrically charged'' A-phase of $^3 \rm
He$, {\it i.e.} a superconductor with the properties of $^3 \rm He$-A,
was considered (see section \ref{lightening} for a brief discussion of
the A and B phases of $^3 \rm He$). In this case the global group was
SO(3), the group of orbital rotations. Both papers discussed
continuous vortices in such superconductors, which correspond to the
``skyrmions'' discussed here.

\section{Electroweak strings}
\label{ewsection}

In this section we introduce electroweak strings. There are two kinds:
one, more precisely known as the Z-string, carries Z-magnetic flux,
and is the type that was discussed by Nambu and that becomes stable as
it approaches the semilocal limit.  It is associated with the subgroup
generated by
$$
T_Z = n^aT^a - \sin^2\theta_w Q \ .
$$

There are other strings in the GSW model that carry $SU(2)$ magnetic
flux, called $W$-strings. 
 There is a one-parameter family of
W strings which are all gauge equivalent to one another, and they are
all unstable. They are generated by a linear combination of the
SU(2) generators $T^+$ and $T^-$. These will be discussed in more
detail in the next section.

\subsection{The Z string}

Modulo gauge transformations, the configuration describing a 
straight, infinitely long Z-string along the $z$-axis is
\cite{Vac92}:
\be
\eqalign{
\Phi &= {\eta \over {\sqrt{2}}} f_{NO}(\rho) e^{i\varphi} 
\pmatrix{0\cr 1\cr} \ , \cr
\ \ \ 
Z &= - {2 \over g_z}  {{v_{NO}(\rho)}} d\varphi
\quad{\rm or, \ in \ vector \ notation, \ } {\vec Z} = 
 \( -{2 \over g_z} \){{v_{NO}(\rho)} \over {\rho}} {\hat \varphi}   \cr
A_\mu &= W_\mu^\pm = 0 \cr}
\label{Zstring}
\ee
where $f$ and $v$ are the Nielsen-Olesen profiles that solve the
equations (\ref{NOeqs}).  It is straightforward to show that this is a
solution of the bosonic equations of motion (alternatively,
one can show that it is an extremum of the energy \cite{Vac92}).
Equations (\ref{Zstring}) describe a string with unit winding. The
solutions with higher winding number can be constructed in an
analogous way, but note that the winding number is not a topological
invariant. The unstable string can decay 
by unwinding until it reaches the vacuum sector.

The solution (\ref{Zstring}) reduces to the semilocal string in the limit
$\sin^2\theta_w = 1$, and therefore it is classically stable for
$\beta < 1$ and unstable for $\beta > 1$ (see section
\ref{stability}), where $\beta$ is now the ratio between the Higgs
mass, $\sqrt 2 \lambda \eta$ and the Z-boson mass $g_z \eta /2$, thus
\be \beta = { 8 \lambda \over g_z^2 }\ee

The Z-string
configuration is axially symmetric, as it is invariant under the
action of the generalised angular momentum operator 
\be M_z = L_z + S_z + I_z\ee
where $L_z$, $S_z$ and $I_z$ are the orbital, spin and isospin parts,
respectively, defined in section \ref{chernsimons}.

The Z-string carries a Z-magnetic flux
\be F_Z = {4\pi \over g_z }\ee
thus particles whose Z charge is not an integer multiple of $g_z / 2$ will
have Aharonov-Bohm interactions with the string (see section
\ref{scattering}).
The Z-string can terminate on magnetic monopoles (such configurations
are discussed in section \ref{zoo}). When a string
terminates, 
the discrete Aharonov-Bohm 
interaction can be smoothly deformed to the trivial interaction.
The smoothness is provided by the presence of the magnetic flux
of the monopole.

\bigskip$\bullet$ Note that,
in the background given by (\ref{Zstring}), the covariant derivative
becomes
\be
{\bf d}_\mu \equiv D_\mu|_{\rm Z-string} = 
\d_\mu + i{g_z \over 2}  \[ -2(T^3 - Q\sin^2 \theta_w) \] Z_\mu
\label{smalldmu}
\ee
in particular,  left and right fermion fields couple to
$Z_\mu$ with different strengths, since the effective Z-charge
\be
{\bf q}  = -2(T^3 - Q\sin^2 \theta_w)
\label{zcharge}
\ee 
has different values, $q_R = q_L \pm 1$. (Note that $\bf q$
is proportional to the string generator $T_z$, defined in equation
(\ref{QandTzgenerators}). The proportionality factor has been
introduced for later convenience). This will be important in section
\ref{scattering}. Note also that, for the Higgs field,
\be 
{\bf q}  = {\rm diag} (-\cos 2\theta_w, 1)
\ee

\bigskip
$\bullet$ Ambj{\o}rn and Olesen \cite{AmbOle89} and, more recently, Bimonte
and Lozano \cite{BimLoz94a} have derived Bogomolnyi-type bounds for
periodic configurations in the GSW model.  They consider static
configurations such that physical observables are periodic in the
$xy$-plane and cylindrically symmetric in each cell. If $A$ is the
area of the basic cell, they find that the energy (per unit length)
satisfies
\be 
\cases{
 {E} \geq ({1 / 2e^2}) m_W^2(2 g'F_Y - m_W^2 A) &if $m_H \geq m_Z$ \cr
&\cr
 {E} \geq ({1 /2e^2}) ({m_W m_H / m_Z})^2 (2 g'F_Y - m_W^2 A)
&if $m_H < m_Z$ \cr }
\label{BiLobound}
\ee 
where $F_Y$ is the magnetic flux of the hypercharge field through the
cell.  Note that the top line of (\ref{BiLobound}) 
reduces to the familiar $E \geq \langle
\Phidag\Phi \rangle q F_Y$ for the Abelian Higgs and semilocal case in
the $g \to 0$ limit (with $q = g'/2$). In the non-Abelian case the
bound involves an area term and therefore does not admit a topological
interpretation.

In the Bogomolnyi limit, $m_H = m_Z$, the bound is saturated for
configurations satisfying the first order Bogomolnyi equations
\be
\eqalign{
D_1+iD_2 \Phi &= 0 \cr Y_{12} + {g'\over 2} \left( \Phidag\Phi -
{\eta^2 \over 2\sin^2\theta_w} \right) &= 0 \cr 
W_{12}^a + {g \over 2}\Phidag \tau^a \Phi &= 0 \cr}
\ee
A solution to these equations describing a lattice of Z-strings was
constructed in \cite{BimLoz94a}. Other periodic configurations with
symmetry restoration had been previously found in the presence of an
external magnetic field in \cite{AmbOle89}.

\section{The zoo of electroweak defects}
\label{zoo}

The electroweak Z-string is one member in the zoo of electroweak
defects. Other members include the electroweak monopole, dyon and the
W-string. The latter fall in
the class of ``embedded defects'' and this viewpoint provides a simple
way to characterize them. The electroweak sphaleron is also related to
the electroweak defects.

\subsection{Electroweak monopoles}
\label{ewmonopoles}

To understand the existence of magnetic monopoles in the GSW 
model, recall the following sequence of facts:
\begin{itemize}
\item The Z-string does not have a topological origin and 
hence it is possible for it to terminate. 
\item As the hypercharge component of the Z-field in the string 
is divergenceless it cannot terminate.
Therefore it must continue from within the string to beyond the terminus. 
\item However, beyond the terminus, the Higgs is in its vacuum and the 
hypercharge magnetic field is massive. Then, if the massive 
hypercharge flux was to continue beyond the string, it would
cost an infinite amount of energy and this is not possible. 
\item The only means by which the hypercharge field can continue beyond 
the terminus is in combination with the SU(2) fields such that it 
forms the massless electromagnetic magnetic field. 
\end{itemize}
So the terminus of the Z-string is the location
of a source of electromagnetic magnetic field, that is, a
magnetic monopole \cite{Nam77}.
We now make this argument more quantitative.

Assume that we have a semi-infinite Z-string along the $-z$ axis
with terminus at the origin (see Fig. \ref{fluxbalance}). Let us denote 
the A- and Z- magnetic fluxes through a spatial surface by $F_A$ and 
$F_Z$. These are given in terms of the W- and Y- fluxes by taking surface 
integrals of the field strengths 
(see eqs. (\ref{amunu}), (\ref{zmunu})). Therefore 
\be
F_Z = \cos\theta_w F_n - \sin\theta_w F_Y \ ,
\ \ \ \
F_A = \sin\theta_w F_n + \cos\theta_w F_Y \ ,
\label{fzfa}
\ee
where we have denoted the SU(2) flux (parallel to $n^a$ in group
space) by $F_n$ and the hypercharge flux by $F_Y$.

\begin{figure}[tbp]
\caption{\label{fluxbalance} The outgoing hypercharge flux of the
monopole passing through the surface ${\Sigma -S}$ should equal the 
incoming hypercharge flux through the Z-string.
}
\vskip 1truecm
\epsfxsize = \hsize \epsfbox{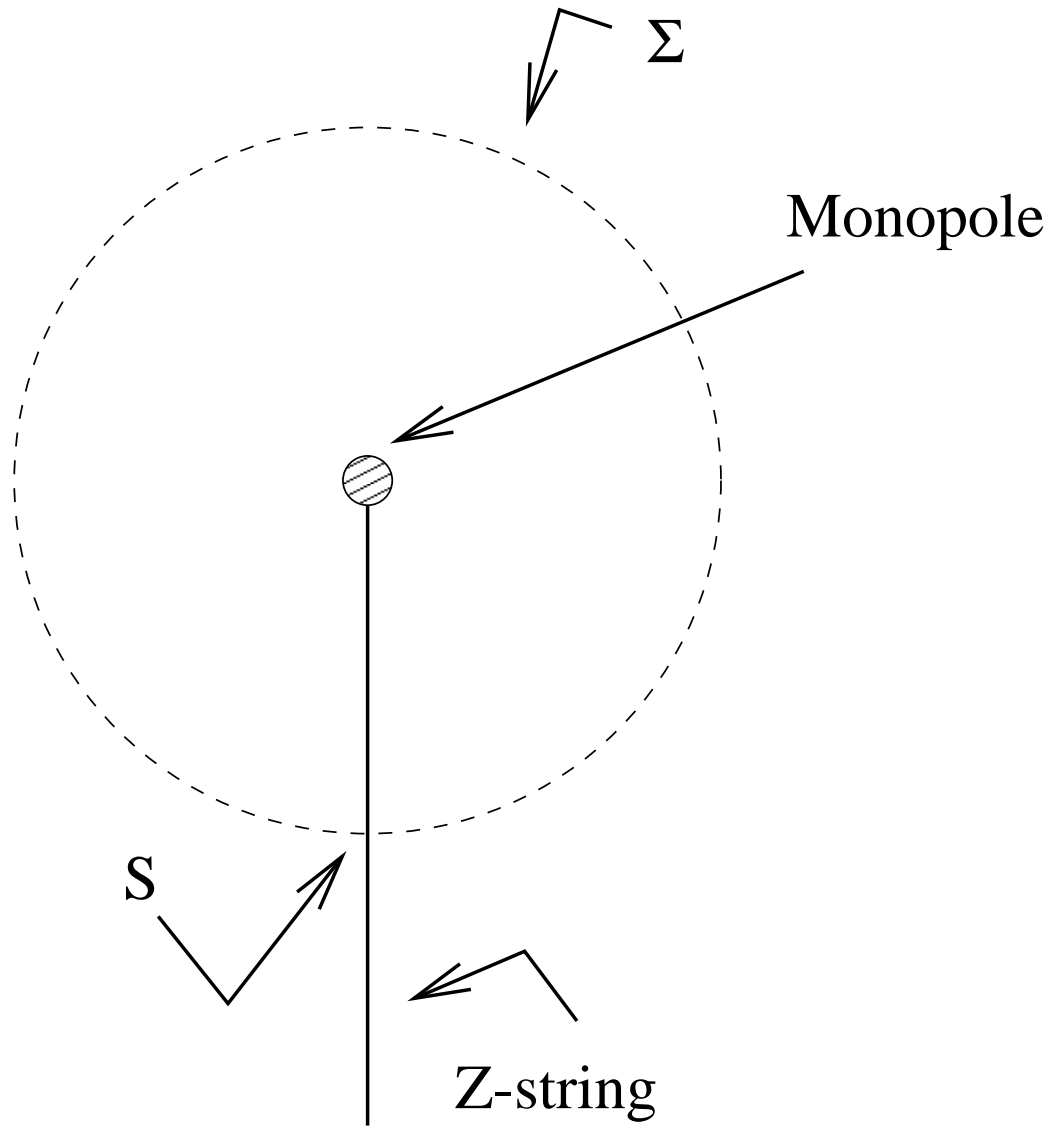}
\end{figure}

Now consider a large sphere $\Sigma$ centered on the string terminus. 
The field configuration is such that 
there is only A-flux through $\Sigma$ except near
the South pole ($S$) of $\Sigma$, where there is only a $Z$ magnetic 
flux. Hence, 
\be
F_Z |_{\Sigma -S} = 0 \ , \ \  F_A |_S = 0  \ .
\label{fzfaonsurfaces}
\ee
Together with (\ref{fzfa}) this gives,
\be
F_n |_{\Sigma -S} = \tan\theta_w F_Y |_{\Sigma -S} \ , \ \ \
F_n |_S = - \cot \theta_w F_Y |_S \ .
\label{fnfyfnfy}
\ee

The hypercharge flux must be conserved as it is divergenceless. So
\be
F_Y |_{\Sigma -S} = - F_Y |_S \equiv F_Y \ ,
\label{fy}
\ee
and, inserting this and (\ref{fnfyfnfy}) in (\ref{fzfa}) yields
\be
F_A |_{\Sigma -S} = {{F_Y} \over {\cos\theta_w}} \ , \ \ \
F_Z |_S = {{F_Y} \over {\sin\theta_w}} \ .
\label{fafyfzfy}
\ee

Now the flux in the $Z-$string along the $-z$ axis is quantized in
units of $4\pi /g_z$ (recall $g_z =
e/\cos\theta_w\sin\theta_w$ gives the coupling of the 
Z boson  to the Higgs field). Therefore, for the unit winding string,
\be
F_Z |_S = {{4\pi} \over {g_z}} \ .
\label{fzs}
\ee
Then (\ref{fafyfzfy}) yields,
\be
F_Y = {{4\pi} \over {g_z}} \sin\theta_w \ , \ \ \
F_A |_{\Sigma -S}= {{4\pi} \over {g_z}} \tan\theta_w
                  = {{4\pi} \over {e}} \sin^2 \theta_w \
\label{fyfa}
\ee
Hence the terminus of the string has net A-flux emanating 
from it and hence it is a magnetic monopole.

The electromagnetic flux of the electroweak monopole appears to
violate the Dirac quantization condition. However this is not true
since one must also take the Z-string into account when deriving the
quantization condition relevant to the electroweak monopole This
becomes clearer when we work out the magnetic flux for the $SU(2)$
fields.  Using (\ref{fnfyfnfy}) with (\ref{fyfa}), the net non-Abelian
flux is: 
\be
F_n = F_n |_S + F_n |_{\Sigma -S} = {{4\pi} \over {g}}
\label{fn}
\ee
just as we would expect for a 't Hooft-Polyakov 
monopole \cite{tHoPol74}. That is,
the Dirac quantization condition works perfectly well for the SU(2)
field and the monopole charge is quantized in units of $4\pi /g$.
Another way of looking at (\ref{fn}) is to say that the electroweak
monopole is a genuine SU(2) monopole in which there is a net emanating
$U(1)_n \subset SU(2)$ flux. The 
structure of the theory, however, only permits a linear
combination of this flux and hypercharge flux to be long range and so
there is a string attached to the monopole. But this string 
is made of Z field which is orthogonal to the electromagnetic field
and so the string does not surreptitiously return the monopole 
electromagnetic flux. 
Also, the magnetic charge on the monopole is conserved and
electroweak monopoles can only disappear by annihilating with
antimonopoles.

It is useful to have an explicit expression describing the asymptotic
field of the electroweak monopole and string. Nambu's monopole-string
configuration, denoted by $({\bar \Phi}, {\bar W}_\mu ^a, {\bar Y}_\mu
)$, is 
\be
{\bar \Phi} = {{\eta} \over {\sqrt{2}}}
          \pmatrix{ \cos(\theta /2) \cr \sin(\theta /2) e^{i\varphi }} \
\label{barphi}
\ee
where, $\theta$ and $\varphi$ are spherical coordinates centred on the
monopole, and the gauge field configuration is,
\be
g {\bar W}_\mu ^a = - \epsilon^{abc} n^b \partial_\mu n^c + i \cos^2 \theta_w
   n^a ({\bar \Phi}^{\dag} ~ \partial_\mu {\bar \Phi} - 
        \partial_\mu {\bar \Phi}^{\dag} ~{\bar \Phi} )
\label{gwbarmua}
\ee
\be
g' {\bar Y}_\mu = - i \sin^2 \theta_w
 ({\bar \Phi}^{\dag} ~ \partial_\mu {\bar \Phi} - 
   \partial_\mu {\bar \Phi}^{\dag} ~{\bar \Phi} ) \
\label{gpbarymu}
\ee
where, $n^a$ is given in eq. (\ref{nadef}). 

Note that there is no electroweak configuration that represents a
magnetic monopole surrounded by vacuum.  

\subsection{Electroweak dyons}
\label{ewdyons}

Given that the electroweak monopole exists, it is natural to ask if 
dyonic configurations exist as well. We now write down dyonic 
configurations that solve the asymptotic field equations \cite{Vac95}.  
The existence of such configurations is implicit in Nambu's original paper
in the guise of what he called ``external'' potentials \cite{Nam77}.
Essentially, the dyon solution is an electroweak monopole together
with a particular external potential.

The ansatz that describes an electroweak dyon connected by a
semi-infinite $Z$ string is: 
\be
\Phi = {\bar \Phi}
\label{phibarphi}
\ee
\be
W ^a = {\bar W} ^a - dt {{ n^a {\dot \zeta}} \over
                                              \cos\theta_w}
\label{wadyon}
\ee
\be
Y = {\bar Y} - dt {{\dot \zeta} \over \sin\theta_w}
\label{ydyon}
\ee
where, $\zeta= \zeta (t, \vec x )$, 
overdots denote partial time derivatives and barred
fields have been defined in the previous subsection.

We now need to insert this ansatz into the field equations and to
find the equation satisfied by $\zeta$. Some algebra leads to
\be
\partial ^i \partial_i {\dot \zeta} = 0 \ , \ \ \ 
\partial _t \partial ^i {\dot \zeta} =0
\label{dotzeta}
\ee
which can be solved by separating variables,
\be
\zeta = \xi (t) f(\vec x ) \ .
\label{zetaxif}
\ee
This leads to
\be
{\ddot \xi} = 0 \ , \ \ \ \nabla ^2 f = 0 \ .
\label{ddotxi}
\ee
The particular solution that we will be interested in is the
solution that gives a dyon. Hence, we take:
\be
\xi = \xi_0 t \ , \ \ \
 f(r) = - {{q \sin\theta_w \cos\theta_w} \over {4\pi \xi_0}} {1 \over r} \ ,
\label{xisoln}
\ee
where, $\xi_0$ and $q$ are constants. Now, using (\ref{xisoln}), together 
with (\ref{wadyon}), (\ref{ydyon}) and (\ref{zetaxif}),
we get the dyon electric field:
\be
{\vec E}_A = {q \over {4\pi}} {{\vec r} \over {r^3}} \ .
\label{ecoulomb}
\ee

For a long segment of string, the monopole and the antimonopole at the
ends are well separated and we can repeat the above analysis for both
of them independently. Therefore, the electric charge on the
antimonopole at one end of a Z-string segment is uncorrelated with the
charge on the monopole at the other end of the string. This means that
we can have dyons of arbitrary electric charge at either end of the
string. The situation will change with the inclusion of fermions since
these can carry currents along the string and transport charge from
monopole to antimonopole. 

This completes our construction of the dyon-string system in the GSW
model. As of now, the charge $q$ on the dyon is arbitrary. Quantum
mechanics implies that the electric charge must be quantized. If we
include a $\theta$ term in the electroweak action (but no fermions):
\be
S_\theta = {{g^2 \theta} \over {32 \pi^2}} \int d^4 x
            W_{\mu \nu}^a {\tilde W}^{\mu \nu ^a} \
\label{thetaaction}
\ee
where
\be
{\tilde W}^{\mu \nu ^a} = {1\over 2} \epsilon^{\mu \nu \lambda \sigma}
                             W_{\mu \nu}^a \ ,
\ee
then the charge quantization condition becomes
\be
q = \left ( n + {\theta \over {2\pi}} \right ) e \ .
\label{dyonquant}
\ee
This agrees with the standard result for dyons \cite{Wit79}.

In the GSW model with fermions, it is known that the $\theta$ term
can be eliminated by a rotation of the fermionic fields. This argument can
be turned around to argue that the CP violation in the mass matrix of the 
fermions will lead to an effective $\theta$ term and so the 
electroweak monopoles will indeed have a fractional charge with $\theta$
being related to the CP violation in the mass matrix. The precise
value of the fractional electric charge on electroweak monopoles has not
yet been calculated and remains an open problem.

It should be mentioned that, even though the electric charge on an
electroweak dyon can be fractional as in (\ref{dyonquant}), the total
electric charge on the dyon-string system is always integral because
the CP violating fractional charge on the monopole is equal and
opposite to that on the antimonopole.  

\subsection{Embedded defects and W-strings:}
\label{embeddeddefects}

A very simple way of understanding the existence of electroweak string
solutions is in terms of embedded defects. While this method does not
shed any light on the stability of the electroweak string, it does
provide a scheme for finding other solutions.

The idea is that the electroweak symmetry group contains several
$U(1)$ subgroups which break completely when the electroweak symmetry
breaks. Corresponding to each such breaking, one might have a
string solution. A more complete analysis tells us when such a
solution can exist 
\cite{VacBar92,BarVacBuc94,DavLep95,LepDav95}.

Consider the general symmetry breaking
\be
G \rightarrow H
\label{gtoh}
\ee
Suppose $G_{emb}$ is a subgroup of $G$ which, in this process, breaks down 
to $G_{emb} \cap H$. Then we ask the question: when are topological defects
in the symmetry breaking
\be
G_{emb} \rightarrow G_{emb} \cap H
\label{gembtogembinth}
\ee
also solutions in the full theory? An answer to this question requires
separating the gauge fields into those that transform within the $G_{emb}$
subgroup and those that do not. Similarly, the Higgs field components
are separated into those that lie in the embedded vector space of scalar
fields and those that do not. Then, it is possible to write down general 
conditions under which solutions can be embedded 
\cite{BarVacBuc94,DavLep95}.
Here we shall not describe these conditions but remark that the Z-string
is due to the embedded symmetry breaking
\be
U(1)_Z \rightarrow 1
\label{u1zto1}
\ee
where the $U(1)_Z$ is generated by $T_Z$, defined in
eq. (\ref{QandTzgenerators}).
Now, there are other $U(1)$'s that can be embedded in the GSW model 
which lie entirely in the $SU(2)$ factor. For example, we can choose 
$U(1)_1$ which is generated by $T^1$ (one of the off-diagonal 
generators of $SU(2)$). 
Since we have
\be
U(1)_1 \rightarrow 1
\label{u11to1}
\ee
when the electroweak symmetry breaks, there is the possibility of another
string solution in the GSW model. Indeed, it is easily checked
that this string can be embedded in the GSW model and the solution
is called a W-string. By considering a one parameter family of $U(1)$ 
subgroups generated by
\be
T_\zeta = \cos(\zeta ) ~T^1 + \sin(\zeta ) ~T^2
\label{tgenerator}
\ee
we can generate a one parameter family of W-strings 
\be
\Phi = {\eta \over {\sqrt{2}}}
f_{NO} (\rho ) \pmatrix{ \cos\varphi \cr ie^{-i\zeta} \sin\varphi}
\label{phiforwstring}
\ee
\be
W^1 = -{2 \over g} \cos\zeta ~{v_{NO}(\rho )} d\varphi ~ ,
\ \ 
W^2 = -{ 2\over g} \sin\zeta ~{v_{NO} (\rho )} d\varphi ~ ,
\label{wforwstring}
\ee
and all other fields vanish. Although the string solutions are gauge
equivalent for different values of $\zeta$, the parameter does take on
physical meaning when considering multi-string configurations in which
the value of $\zeta$ is different for different strings
\cite{BarVacBuc94}.

Note that the generator
(\ref{tgenerator}) can be obtained from $T^1$ by the action of the
unbroken (electromagnetic) group,
\be
T_\zeta = e^{i\zeta Q} T^1 e^{-i\zeta Q}
\label{tzetadef}
\ee
With this in mind, Lepora {\it et al} \cite{LepDav95, LepKib99a} 
have
classified embedded vortices. The idea is that, for a general symmetry
breaking $G \to H$,the Lie algebra of $G$, $\cal G$, decomposes
naturally into a direct sum of the space $\cal H$ of generators of the
unbroken subgroup $H$ (the ones associated with massless gauge bosons)
and the space $\cal M$ of generators associated with massive gauge
bosons: ${\cal G} = {\cal H} + {\cal M}$. The action of $H$ on the
subspace $\cal M$ further decomposes $\cal M$ into irreducible
subspaces. The classification of embedded vortices is based
on this decomposition, as we now explain. 

Recall (eq. \ref{gaugeorbit}) that finite energy vortices are
associated with gauge orbits on the vacuum manifold\footnote{The gauge
orbits are geodesics of a squashed metric on the vacuum manifold which
is different from the isotropic metric relevant to the scalar sector
\cite{LepKib99b}.}. Choosing a base point $\Phi_0$ in the vacuum
manifold, each embedded vortex can be associated to a Lie algebra
generator which is tangent to the gauge orbit describing the
asymptotic scalar field configuration of the vortex.  The unbroken
subgroup $H$ at $\Phi_0$ ``rotates'' the various gauge orbits among
themselves as in eq. (\ref{tzetadef}). Thus, the action of $H$ splits
the space of gauge orbits into irreducible subspaces.

Except for critical values of the coupling constants (which could lead
to so-called {\it combination vortices}), it can be shown 
\cite{BarVacBuc94,LepDav95} that embedded vortices have to lie entirely in
one of these irreducible subspaces. If the subspaces have dimension
greater than one, then there may be a {\it family} of gauge equivalent
vortices.

In the GSW model, for instance, the Lie Algebra decomposes into ${\cal
H} + {\cal M}_1 + {\cal M}_2 \ $ where $\cal H$ is spanned by the charge
$Q$, ${\cal M}_1$ is a one-dimensional subspace spanned by $T_Z$
(corresponding to the Z-string) and ${\cal M}_2 $ is a two-dimensional
subspace comprising all W-string generators $T_\zeta$.

Both the W- and the Z-string are embedded string solutions in the GSW 
model. What makes the Z-string more interesting is its unexpected stability
properties. It can be shown \cite{LepDav95} that only those vortices lying
in one-dimensional subspaces can have a stable semilocal limit. Thus,
embedded vortices belonging to a family are always unstable.

Another important difference is that the Z-string is known to
terminate on magnetic monopoles but this is not true of the
W-string. The W-string can terminate without any emanating
electromagnetic fields since it is entirely within the $SU(2)$ sector
of the GSW model.

It is straightforward to embed domain walls in the GSW model.  There
are no embedded monopoles in the GSW model since there is no SU(2)
subgroup that is broken to U(1).

\section{Electroweak strings in extensions of the GSW model}
\label{ewstringinextensions}

Electroweak strings have been discussed in various extensions of
the GSW model. We describe some of this work below. We do not, however,
discuss extensions in which topological strings are produced at the
electroweak scale \cite{DvaSen94,BimLoz94b}.

\subsection{Two Higgs model}
\label{twohiggs}

As discussed in Sec. \ref{embeddeddefects}, the Z-string is an embedded 
string in the GSW model. The general conditions that enable the embedding 
are valid even with a more complicated Higgs structure. Here we will 
consider the two Higgs doublet model which is inspired by supersymmetric 
extensions of the GSW model.

In a two Higgs doublet model, the Higgs structure of the GSW 
model is doubled so that we have scalars $\Phi_1$ and $\Phi_2$ and the 
scalar potential is \cite{Kas93} 
\begin{eqnarray}
V(\Phi_1,\Phi_2) & = & 
\lambda_1(\Phi_1^{\dagger}\Phi_1 - \frac{\nu_1^2}{2})^2 + 
\lambda_2(\Phi_2^{\dagger}\Phi_2 - \frac{\nu_2^2}{2})^2 +  
\lambda_3[(\Phi_1^{\dagger}\Phi_1 - \frac{\nu_1^2}{2})  + 
          (\Phi_2^{\dagger}\Phi_2 - \frac{\nu_2^2}{2})]^2 
                                              \nonumber \\ 
                 & + & 
\lambda_4[(\Phi_1^{\dagger}\Phi_1)(\Phi_2^{\dagger}\Phi_2) 
        - (\Phi_1^{\dagger}\Phi_2)(\Phi_2^{\dagger}\Phi_1)] +
\lambda_5[{\rm Re}(\Phi_1^{\dagger}\Phi_2) 
        - \frac{\nu_1\nu_2}{2}\cos\xi]^2 
                                              \nonumber \\
                 & + & 
\lambda_6[{\rm Im}(\Phi_1^{\dagger}\Phi_2) 
        - \frac{\nu_1\nu_2}{2}\sin\xi]^2 \  
                                              \nonumber \\
                 & + & 
\lambda_7[{\rm Re}(\Phi_1^{\dagger}\Phi_2) 
        - \frac{\nu_1\nu_2}{2}\cos\xi]~[{\rm Im}(\Phi_1^{\dagger}\Phi_2) 
        - \frac{\nu_1\nu_2}{2}\sin\xi] \ \ .
\label{twohiggspotential}
\end{eqnarray}
Here $\nu_1$ and $\nu_2$ are the respective VEVs
of the two doublets, $\lambda_i$ are coupling constants and 
the parameter $\xi$ is a
phase 

In polar coordinates, the solution for the two Higgs Z-string is:
\be
\Phi _1 = \nu_1 f_1 (\rho ) e^{i\varphi} \pmatrix{0\cr 1\cr}
\label{phi1for2higgs}
\ee
\be
\Phi_2 = \nu_2 f_2 (\rho ) e^{i\varphi } \pmatrix{0\cr 1\cr}
\label{phi2for2higgs}
\ee
\be
{\vec Z} = -{2 \over {g_z}} {{v(\rho )}\over {\rho}} {\hat \varphi}
\label{afor2higgs}
\ee
with the profile functions
satisfying differential equations similar to the Abelian-Higgs
case. These have been studied in Ref. \cite{EarJam93} where the
stability has also been analyzed (also see \cite{Tom97}).  

\subsection{Adjoint Higgs model}
\label{adjointhiggs}

The GSW model with an additional $SU(2)$ field
in the adjoint representation, $\vec \chi$,
is what we shall refer to as the
``adjoint Higgs model''. The impact of the adjoint field on
electroweak defects was considered in Ref.
\cite{KepVac96}.

The bosonic sector of the adjoint Higgs model is:
\begin{equation}
L = T_{ew} + |(\partial_\mu +ig \epsilon^a W_\mu ^a ) {\vec \chi}|^2
      - V(\Phi , {\vec \chi}) + L_f\
\label{adjointmodellagrangian}
\end{equation}
where, $T_{ew}$ is the gradient part of the bosonic sector of the
electroweak Lagrangian, $L_f$ is the fermionic part of the
Lagrangian, $\epsilon^a_{ij} = \epsilon_{aij}$ ($a,i,j = 1,2,3$)
and,
\begin{equation}
V(\Phi , {\vec \chi}) =
- \mu_2^2 \Phi^{\dag}\Phi - \mu_3^2 {\vec \chi}^2
+ \lambda_2 (\Phi^{\dag}\Phi)^2
+\lambda_3 {\vec \chi}^4 + a {\vec \chi}^2 \Phi^{\dag}\Phi
+ b {\vec \chi} \cdot \Phi^{\dag} {\vec \tau} \Phi \; .
\label{adjointmodelpotential}
\end{equation}
If we impose an additional $Z_2$ symmetry on the Lagrangian under $
\Phi \rightarrow +\Phi \ , \ \ \
{\vec \chi} \rightarrow -{\vec \chi} \ .  $ the symmetry is
$([SU(2)_L\times U(1)_Y]/Z_2) \times Z_2$ and we must set $b=0$. In
what follows, we shall only consider this case and henceforth ignore
the last (cubic) term in the potential.  In this case, an additional
simplification is that the leptons and quarks do not couple to $\vec
\chi$ and so $L_f$ is identical to the fermionic Lagrangian of the GSW
model.  (If the $Z_2$ symmetry is absent, the cubic term in the
potential is allowed but is constrained to be small by experiment.)

In a cosmological context, as the universe cools down from high
temperatures, if the parameters lie in a certain range \cite{KepVac96}
there will first be a phase transition in which the adjoint
field gets a VEV.  The VEV of the adjoint will break the $SU(2)$
factor of the high temperature symmetry group to $U(1)$. If 
the VEV of ${\vec \chi}$ is along the $(0,0,1)$ direction,
the generator of this $U(1)$ will be $T^3$ and we will denote
the unbroken subgroup as $U(1)_3$. So the symmetry breaking
pattern at this stage is
\be
([SU(2) \times U(1)_Y]/Z_2) \times Z_2 \rightarrow 
                     ([U(1)_3 \times U(1)_Y]/Z_2) \times Z_2
\label{adjointstage1}
\ee
and topological magnetic monopoles will be produced with pure $U(1)_3$
flux 

At a lower temperature, the doublet field will also get a VEV with the
effect,
\be
([U(1)_3 \times U(1)_Y]/Z_2) \times Z_2 \rightarrow U(1)_{em} \ .
\label{adjointstage2}
\ee
where, as usual, the electromagnetic charge operator is
\be
Q =  T^3 + {Y\over 2}  
\label{qt3y}
\ee
The electromagnetic component (A) 
from the monopoles is massless but the orthogonal part (Z) of the flux
is massive and gets confined to a string. This is the Z-string. In
addition, the breaking of the $Z_2$ factor gives domain walls.

In the second stage of symmetry breaking, the Z-string is topological
and hence is stable. The presence of magnetic monopoles from the
earlier symmetry breaking means that the Z-strings can break by
terminating on monopoles. But, as the monopoles form at a higher
energy scale, their mass is much larger than the energy scale at which
strings form and which sets the scale for the tension in the
string. So the string can only break by instanton processes.  

At a yet lower temperature, the VEV of the adjoint turns off. This 
makes no difference to the symmetry structure of the model (apart from
restoring the $Z_2$ symmetry and eliminating the domain walls) and
hence no significant difference to the monopoles connected by
strings. However, it does affect the stability of
the strings since the monopoles are no longer topological.

\section{Stability of electroweak strings}
\label{stability}

\subsection{Heuristic stability analysis}
\label{heuristic}

As described in \cite{Vac92}, the
Z-string goes over into the semilocal string in the 
limit $\theta_w \rightarrow \pi /2$ and hence the stability 
of the Z-string should match on continuously to that of the
semilocal string. Therefore we expect that Z-strings
should be stable if $\theta_w$ is close to $\pi/2$ and
$m_H \le m_Z$.

The stability analysis to certain subsets of perturbations
can be carried out much more easily than to the completely
general perturbations. The subset includes perturbations in 
the Higgs field and W-fields separately. Such analyses may
be found in \cite{Vac92,Vac93,BarVacBuc94} and \cite{Perk93}.

\noindent (i) {\it Higgs field perturbations:}
Perturbations in the Higgs field alone have maximum destabilizing 
effect for $\theta_w = \pi /4$ \cite{BarVacBuc94} and, in this case, 
it is easy to see that the Z-string is unstable. Consider the one 
parameter family of field configurations
\be
\Phi (\vec x ; \xi ) = \cos\xi ~\Phi_0 (\cos\xi ~{\vec x}) +
                       \sin\xi ~\Phi_\perp
\label{phiconfigpiover4}
\ee
\be
Z_j (\vec x ; \xi ) = \cos\xi ~Z_{(0)j} (\cos\xi ~\vec x )
\label{zconfigpiover4}
\ee
where, the string solution is denoted by the $0$ subscript,  
$\xi \in [0,\pi /2]$ and 
\be
\Phi_\perp = {\eta \over {\sqrt{2}}}\pmatrix{1\cr 0\cr} \ .
\label{phiperp}
\ee
For $\xi=0$ the field configuration is the unperturbed Z-string 
while for $\xi =\pi/2$ it is the vacuum. The energy per unit length
of this field configuration can be evaluated and is found to be:
\be
E(\xi ) = \cos^2\xi ~ E(\xi =0) \ .  
\label{energywithxi}
\ee
Hence the energy per
unit length of the string is a
monotonically decreasing function of $\xi$ and so the string is
unstable to decay into the vacuum.

\noindent (ii) {\it Incontractible two spheres:} James \cite{Jam94}, 
and, Klinkhamer and Olesen \cite{KliOle94} have constructed the Z- and
W-string solutions by considering incontractible two spheres in the
space of electroweak field configurations in two spatial
dimensions. The idea was introduced by Taubes \cite{Tau82} and was
used by Manton to construct the sphaleron
\cite{Man83,KliMan84}.  The procedure (known as the ``minimax'' procedure) is
to construct a set of field configurations that are 
labelled by some parameters $\mu_i$. If this set is incontractible
in the space of field configurations, then there exist (subject to
certain assumptions \cite{Man83}) values of the parameters for which
the field configuration extremizes the energy functional. For example,
Klinkhamer and Olesen \cite{KliOle94} give the following construction
for the Z-string in terms of a two parameter ($\mu , \nu$) family of
field configurations 
\begin{eqnarray}
\pi /2 \le [ \mu \nu ] \le \pi \, : && W=0 \ , \ \ \ Y=0 \nonumber \\
&&\Phi = ( 1-\{ 1-h(\rho )\}  \sin [ \mu \nu ] ) {\eta \over {\sqrt{2}}}
           \pmatrix{0\cr 1\cr} 
\label{fkpowyphi1}
\end{eqnarray}
\begin{eqnarray}
0 \le [ \mu \nu ] \le \pi /2 \, : && W= -f(\rho ) G^a T^a \ , \ \ \ 
                               Y= f(\rho ) \sin^2\theta_w F^3  \nonumber \\
&&\Phi = h(\rho )  {\eta \over {\sqrt{2}}}
          \Omega U  \pmatrix{0\cr 1\cr} 
\label{fkpowyphi2}
\end{eqnarray}
where, $W$ and $Y$ are Lie algebra
valued 1-forms ({\it e.g.} $W = W_\mu ^a T^a dx^\mu$), 
$[\mu \nu] \equiv {\rm max}(|\mu |, |\nu|)$, 
\be
F^a T^a = 2i U^{-1} dU \ ,
\label{fataua}
\ee
\be
G^a T^a  = \Omega U 
           [ F^1 T^1 + F^2 T^2 + \cos^2\theta_w F^3 T^3 ] 
                     U^{-1}  \Omega^{-1} \ ,
\label{gataua}
\ee
\be
U(\mu , \nu , \varphi ) = -i \sin\mu \tau_1 -i \cos\mu \sin\nu  
\tau_2 -i \cos\mu \cos\nu \sin\varphi \tau_3 + 
\cos\mu \cos\nu \cos\varphi {\bf 1} \ ,
\label{umatrix}
\ee
\be
\Omega = U(\mu , \nu , \varphi = 0 )^{-1} \ ,
\label{omegadefn}
\ee
and the functions $f(\rho )$ and $h(\rho )$ satisfy the boundary conditions
\be
f(0) = 0 = h(0) \ , \ \ \ f(\infty )=1=h(\infty ) \ .
\label{fkpofhbcs}
\ee
This set of field configurations labelled by the
parameters $\mu , \nu \in [-\pi , \pi ]$ defines an incontractible two
sphere in the space of field configurations.  This is seen by
considering the fields as if they were defined on the three-sphere 
on which the coordinates are $\varphi$, $\mu$ and $\nu$ and then 
showing that the
field configurations define a topologically non-trivial mapping from
this $S^3$ to the vacuum manifold which is also an $S^3$. Then the
minimax procedure says that there is an extremum of the energy at some
value of the parameters. By inserting the field configurations into
the energy functional, it can be checked that the extremum occurs at
$\mu =0=\nu$, when the configuration coincides with that of the
Z-string. Furthermore, for $\theta_w \le \pi /4$, the extremum is a
maximum and hence the Z-string is unstable.  

A very similar analysis has been done \cite{Jam94,KliOle94}
for the W-string confirming the result \cite{BarVacBuc94} that 
it is always unstable.

\noindent (iii) {\it W-condensation:}
There is also a well-known \cite{AmbOle90} instability to 
perturbations in the W-fields alone called ``W-condensation''. 
Application of this instability to the Z-string may be found in 
\cite{Perk93,Vac93,Vac94,AchGreHarKui94}. 
A heuristic argument goes as follows.

The energy of a mass $m$, charge $e$ and spin $s$ particle in a
uniform magnetic field $\vec B$ along the z-axis is given by:
\be
E^2 = p_z^2 +m^2 +(2n+1) e B -2e {\vec B}\cdot {\vec s}
\label{energyspininb}
\ee
where $n=0,1,2,...$ labels the Landau levels and $p_z$ is the
momentum along the z-axis. Now, if $s=1$, the right-hand
side can be negative for $p_z =0$, $n=0$ provided
\be
B > {{m^2}\over e} ~ .
\label{bcritical}
\ee
This signals an instability towards the spontaneous creation
of spin one particles in sufficiently strong magnetic fields
\cite{AmbOle90}.

In our case, the magnetic field is a Z-magnetic field and this couples
to the spin one W-particles. If the string thickness is larger than
the Compton wavelength of the W-particles, the Z-magnetic field may be
considered uniform. Also, the relevant charge in this case is the
Z-charge of the W-bosons and is $g_Z\cos^2\theta_w$.  The constraint
that the string be thick so that the Z-magnetic field appears uniform
and that the charge not be too small means that $\theta_w$ should be
small. Hence the instability towards W-condensation applies for small
$\theta_w$. This analysis can be performed more quantitatively
\cite{Perk93} with the result that there is a relatively hard bound
$\sin^2\theta_w > 0.8$ for the string to be stable to
W-condensation. 

\subsection{Detailed stability analysis}
\label{detailedstability}

To analyse the
stability of electroweak strings, we perturb the string solution,
extract the quadratic dependence of the energy on the perturbations
and then determine if the energy can be lowered by the perturbations
by solving a Schr\"odinger equation. The
analysis is quite tedious
\cite{JamPerVac92,AchGreHarKui94,GooHin95b,MacTor94,MacTor95} and here
we will only outline the main steps. We use the vector notation in
this section for simplicity.

The general perturbations of the Z-string are 
\be
(\phi_\perp,\phi_\parallel, \delta {\vec Z}, {\vec W}^{\bar a}, 
{\vec A})
\label{pertlist}
\ee
where, ${\bar a}=1,2$, $\phi_\perp$ and 
$\phi_\parallel$ are scalar field fluctuations defined by
\be
\Phi = \pmatrix{\phi_\perp\cr \phi_{NO} + \phi_\parallel\cr} \ ,
\label{perturbedphi}
\ee
$\delta {\vec Z}$ is defined by
\be
{\vec Z} = {\vec Z}_{NO} + \delta {\vec Z} \ .
\label{perturbedz}
\ee
(The subscript $NO$ means that the field is the unperturbed
Nielsen-Olesen solution for the string as described in Sec. \ref{NO}.)
The fields ${\vec W}^{\bar a}, {\vec A}$ are perturbations since 
the unperturbed values of these fields vanish in the Z-string.

The perturbations can depend on the $z-$coordinate and
the $z-$components of the vector fields can also be non-zero. However,
since the vortex solution has translational invariance along the
$z-$direction, it is easy to see that it is sufficient to consider $z$
independent perturbations and to ignore the $z-$components of the
gauge fields.  This follows from the expression for the energy
resulting from the Lagrangian in eq. (\ref{GSWb}) where the relevant
$z-$dependent terms in the integrand are:
\be
          {1\over 2} G_{i3} ^a G_{i3} ^a + {1\over 4} F_{Bi3} F_{Bi3}
          + (D_3 \Phi ) ^{\dag} (D_3 \Phi )
\label{zdepenergy}
\ee
and explicitly provide a positive contribution to the energy.
Hence we drop all reference to the $z-$coordinate with the understanding
that the energy is actually the energy {\it per unit length} of the string.
 
Now we calculate the energy of the perturbed configuration,
discarding terms of cubic and higher
order in the infinitesimal perturbations. We find,
\begin{equation}
E = ( E_{NO} + \delta E_{NO} ) + E_\perp + E_c + E_W
\label{perturbede1}
\end{equation}
where,
$E_{NO}$ is the energy of the Nielsen-Olesen string and $\delta E_{NO}$ is
the energy variation due to the perturbations $\phi_\parallel$ and 
$\delta {\vec Z}$.
The term $E_\perp$ is due to the perturbation $\phi_\perp$ in the upper
component of the Higgs field, $E_c$ is the cross-term between perturbations
in the Higgs and gauge fields, while $E_W$ is the contribution from 
perturbing the gauge fields alone:
\begin{equation}
E_{\perp} = \int d^2 x \left [
       |{\bar d}_j \phi_\perp |^2 + 
         \lambda \eta^2 ( f^2 - 1 ) | \phi_\perp |^2
                         \right ] \  ,
\label{e1}
\end{equation}
\begin{equation}
E_c = i {{g_z}\over 2}  \cos\theta_w \int d^2 x \left [
           \Phi^{\dag} T^{\bar a} {\bf d}_j \Phi
   - ( {\bf d}_j \Phi ) ^{\dag} T^{\bar a} \Phi \right ] 
                    W_j ^{\bar a} \ , 
\label{ec} 
\end{equation} 
with ${\bf d}_j$ defined in \ref{smalldmu}, 
\begin{eqnarray} E_W = &
           \int d^2 x \biggl [
            \gamma {\vec W} ^1 \times {\vec W} ^2 \cdot
              \vec \nabla \times \vec Z
+ {1\over 2} |\vec \nabla \times {\vec W} ^1
                  + \gamma {\vec W} ^2 \times \vec Z | ^2
\nonumber \\
&+
{1\over 2} |\vec \nabla \times {\vec W} ^2
                  + \gamma \vec Z \times {\vec W} ^1 | ^2
+ {1\over 4} g^2 f^2 ( {\vec W } ^{\bar a} ) ^2
+ {1\over 2} ( \vec \nabla \times \vec A )^2
                     \biggr ] \ , 
\label{ew}
\end{eqnarray}
where $\gamma \equiv g \cos\theta_w$, 
\be
{\bar d} _j \equiv \partial _j - i{g_z \over 2} \cos(2\theta_w ) Z_j 
\label{barddefn}
\ee
and, the $f$ and $\vec Z$ fields in the above equations are the
unperturbed fields of the string.

The two instabilities discussed in the previous subsection can also be
seen in eq. (\ref{perturbede1}). First consider perturbations in the
Higgs field alone. Then only $E_\perp$ is relevant.  For $\theta_w
=\pi/4$, ${\bar d} _j = \partial_j$, and $E_\perp$ is the energy of a
particle described by the wavefunction $\phi_\perp$ in a purely negative
potential in two dimensions since $f^2 \le 1$ everywhere. It is known 
that a purely negative potential
in two dimensions always has a bound state.  \footnote{For some
potentials though, the wavefunction of the bound state may have
singular (though integrable) behaviour at the origin and such bound
states would be inadmissible for us since we require that the
perturbations be small. This turns out not to be the case for the
potential in eq. (\ref{e1}).}  Hence the energy can be lowered by at
least one perturbation mode and so the string is unstable when
$\theta_w = \pi/4$.  The instability towards W-condensation can be
seen in $E_W$. The term with $\gamma$ can be negative and its strength
is largest for small $\theta_w$. Hence W-condensation is most relevant
for small $\theta_w$. 

Returning to the full stability analysis, we first note that the 
perturbations of the fields that make up the string do not couple to the 
other available perturbations {\it i.e.} the perturbations in the
fields $f$ and $v$ only occur inside the variation $\delta E_{NO}$. 
Now, since we know that the Nielsen-Olesen
string with unit winding number is stable to perturbations for any values of
the parameters then necessarily, $\delta E _{NO} \ge 0$ and the perturbations
$\phi_\parallel$ and $\delta {\vec Z}$ cannot destabilize the vortex. Then, 
we are justified in ignoring these perturbations and setting 
$\delta E _{NO} = 0$.
Also we note that the ${\vec A}$ boson only appears in the last term of eq.
(\ref{ew}) and is manifestly positive. So we can set ${\vec A}$ to zero.
 
The remaining perturbations can be expanded in modes:
\be
\phi _\perp = \chi (\rho ) e^{im\varphi}
\label{phi1modes}
\ee
for the $m^{th}$ mode where $m$ is any integer. For the gauge fields
we have,
\be
{\vec W}^1 = 
\left [
\left \{ 
{\bar f}_{1} (\rho ) \cos(n\varphi ) + f_{1} (\rho )
\sin(n\varphi ) 
\right \}
{\hat e} _\rho +
{1 \over \rho} 
\left \{
- {\bar h}_{1} (\rho )\sin(n\varphi ) + 
h_{1} (\rho )\cos(n \varphi ) 
\right \}
{\hat e} _\varphi
\right ]
\label{w1modes}
\ee
\be
{\vec W}^2 = 
\left [
\left \{ - {\bar f}_{2} (\rho ) \sin(n\varphi ) + 
f_{2} (\rho )\cos(n\varphi ) 
\right \} {\hat e} _\rho + {1 \over \rho }
\left \{
           {\bar h}_{2} (\rho )\cos(n\varphi ) + 
h_{2} (\rho ) \sin(n \varphi ) \right \}
{\hat e} _\varphi
\right ]
\label{w2modes}
\ee
for the $n^{th}$ mode where $n$ is a non-negative integer.

The most unstable mode is the one with $m=0$ and $n=1$. This is because
these have the lowest gradient energy and are the only perturbations
that can be non-vanishing at $\rho =0$. Further analysis shows that the
string is most unstable to the $h_1 + h_2$ mode. Hence, we can ignore
$f_i$, $h_1-h_2$ and the barred variables. A considerable amount of
algebra then yields:
\be
\delta E[ \chi , \xi_+ ] = 2\pi \int d\rho \  \rho ~
              ( \chi , \xi _+ ) {\bf O} \pmatrix{\chi\cr \xi_+\cr}
\label{deltaeofchiandxi}
\ee
where, {\bf O} is a $2\times 2$ matrix differential operator and
\be
\xi_+ = {{h_1+h_2} \over 2} \ .
\label{xiplusdefn}
\ee

Before proceeding further, note that a gauge transformation on the 
fields does not change the energy. However, we have not fixed the gauge 
in the preceding analysis and hence it is possible that some of the 
remaining perturbations, $(\chi, \xi_+)$, might correspond to
gauge degrees of freedom and may not affect the energy.
So we now identify the combination of perturbations
$\chi$ and $\xi_+$ that are pure gauge transformations of the string
configuration. 

The $SU(2)$ gauge transformation, $\exp (ig\psi )$, of an electroweak
field configuration leads to first order changes in the fields of the
form
\be
\delta\Phi=ig\psi\Phi_0 ~ ,{\ {\delta W}_i=-iD^{(0)}_{i}\psi} \  ,
\label{gaugeonfields}
\ee
where 
$W_i = W_i^a T^a$, $\psi =\psi^a T^a$, 
and the $0$ index denotes the unperturbed field and covariant
derivative. In our analysis above, we
have fixed the form of the unperturbed string and so we should
restrict ourselves to only those gauge transformations that leave the
Z-string configuration unchanged. (For example, $\delta \Phi$ should
only contain an upper component and no lower component.) This
constrains $\psi$ to take the form
\be
\psi=s(\rho )\pmatrix{0&ie^{-i\varphi }\cr -ie^{i\varphi }&0\cr}
\label{psiform}
\ee
where $s(\rho )$ is any smooth function. This means that perturbations
given by 
\begin{equation}
\pmatrix{\chi (\rho) \cr\xi_+ (\rho)}\ =
\ s(\rho )\pmatrix{-g\eta f(\rho)/\sqrt{2}\cr 
                     2(1-2 \cos^2\theta_w v(\rho ))\cr}
\label{puregaugemode}
\end{equation}
are pure gauge perturbations that do not affect the string configuration.
Therefore, such perturbations cannot contribute to the energy variation
and must be annihilated by ${\bf O}$. Then, in the two-dimensional
space of $(\chi , \xi_+ )$ perturbations, we can choose a basis in
which one direction is pure gauge and is given by (\ref{puregaugemode}) 
and the other orthogonal direction is the direction of physical perturbations.
The physical mode can now be written as,
\be
\zeta (\rho )\ =\ (1-2 \cos^2\theta_w v(\rho ))\chi (\rho )\ +
                    \ {g\eta f(\rho )\over {2\sqrt{2}}}\xi_+ (\rho ) ~ .
\label{zetadefn}
\ee
So now the energy functional reduces to one depending only on 
$\zeta (\rho )$:
\be
\delta E[ \zeta ] = 2\pi \int d\rho ~  \rho ~ 
              \zeta  {\overline O} \zeta
\label{deltaefinal}
\ee
where ${\overline O}$ is the differential operator
\be
{\overline O} = - {1 \over \rho} {d \over {d\rho }} \left (
             {\rho \over {P_+}} {d \over {d\rho }} \right ) + U(\rho )
\label{overlineo}
\ee
and
\be
U(\rho ) = {{{f'}^2} \over {P_+ f^2}} + 
             {{4 S_+} \over {g^2 \eta^2 \rho^2 f^2}} +
          {1 \over \rho } {d \over {d\rho }} \biggl (
                     {{\rho f'} \over {P_+ f}} \biggr ) \  ,
\label{schrodingerudefn}
\ee
where
\be
P_{+} = (1 - 2 \cos^2\theta_w v )^2 + {g^2 \eta^2 \rho^2 f^2\over 4}
\label{pplusdefn}
\ee
\be
S_+ (\rho ) =  {g^2 \eta^2 f^2\over 4} - 
               {{4 \cos^4\theta_w {v'}^2} \over {P_+ (\rho )}}+ 
    \rho {{d \  } \over {d\rho }} \left [
   2 \cos^2\theta_w {{v'} \over \rho } 
        {{(1 - 2 \cos^2\theta_w v )} \over {P_+ (\rho )}}
                          \right ] \  .
\label{splusdefn}
\ee

The question of Z-string stability reduces to asking if there are negative 
eigenvalues $\omega$ of the Schr\"odinger equation,
\begin{equation} 
{\overline O} \zeta =\omega\zeta \  . 
\label{eigenvalueproblem}
\end{equation}
The eigenfunction $\zeta (\rho )$ must also satisfy the boundary conditions 
$\zeta (\rho =0) = 1$ and $\zeta ' (0) =0$ where prime denotes differentiation
with respect to $\rho$. In this way the stability analysis reduces to a single
Schr\"odinger equation which can be solved numerically. 

The results of the stability analysis are shown in
Fig. (\ref{stabilityregion}) as a plot in parameter space $(m_H/m_Z,
\sin^2 \theta_w)$, demarcating regions where the Z-string is unstable
(that is, where negative $\omega$ exist) and stable
(negative $\omega$ do not exist).  It is evident that the
experimentally unconstrained values: $\sin^2 \theta_w=0.23$ and $m_H /
m_Z > 0.9$ lie entirely inside the unstable sector. Hence the Z-string
in the GSW model is unstable.
 
\begin{figure}[tbp]
\caption{\label{stabilityregion} The Z-string is stable in the triangular
shaded region of parameter space. At $\sin^2\theta_w =0.5$, the string
has a scaling instability. The experimentally allowed parameters are
also shown. 
}
\vskip 1truecm
\epsfxsize = \hsize \epsfbox{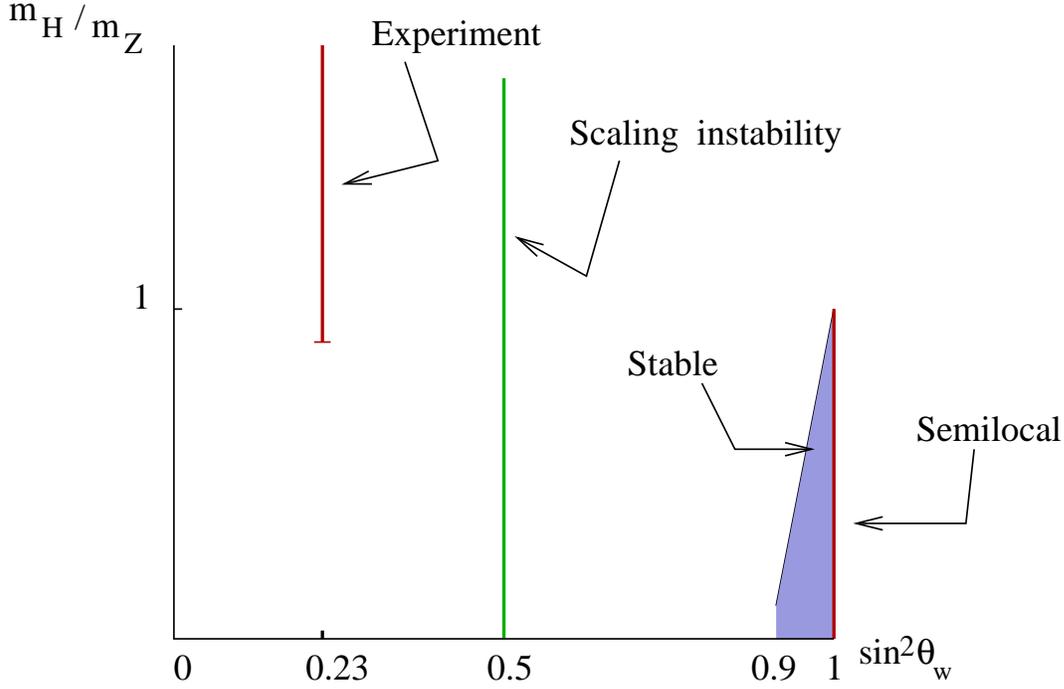}
\end{figure}

The stability analysis of the Z-string described above leaves open
the possibility that the string might be stable in some special circumstances
such as, the presence of extra scalar fields, or a magnetic field background,
or fermions. We now describe some circumstances in which the Z-string
stability has been analyzed.

\subsection{Z-string stability continued}
\label{stabilitycontd}

The stability of Z-strings has been studied in various other
circumstances: 

(i) {\it Thermal effects:}
In \cite{HolHsuVacWat92} 
the authors examined thermal effects on Z-string 
stability using the high temperature effective potential
and found slight modifications to the stability. The conclusion
is that Z-strings in the GSW
model are unstable at high temperatures
as well. In the same paper, left-right symmetric models were studied
and it was found that these could contain stable strings that are similar
to the Z-string.

(ii) {\it Extra scalar fields:} It is natural to wonder if the
presence of extra scalar fields in the model can help provide
stability.  In \cite{EarJam93} the stability was examined in the
physically motivated two Higgs doublet model with little advantage. In
\cite{VacWat93} it was shown that an extra (globally) charged scalar 
field could enhance stability. The extra complex scalar field, $\psi$, 
is coupled to the electroweak Higgs by a term $|\psi|^2 \Phi^\dag \Phi$ 
and hence the charges have lower energy on the string where 
$\Phi^\dag \Phi \sim 0$ than outside the string where $\Phi$ has a 
non-zero VEV. 
the background of a string and hence a sufficient amount of charge 
can stabilize the string. This is exactly as in the case of non-topological 
solitons or Q-balls \cite{Ros68,FriLeeSir76,Col85}. However, scalar
global charges attract and this can cause an instability of the
charge distribution along the string \cite{CopKolLee88,VacWat93}.
For realistic parameters, stable Z-strings do not seem likely even
in the presence of extra scalar fields.

(iii) {\it Adjoint scalar field:}
A possible variant of the above scheme is that an $SU(2)$ adjoint
can be included in the GSW model as described in Sec. (\ref{adjointhiggs}).
Now, since the Z-string is topological within the second symmetry
breaking stage in eq. (\ref{adjointstage2}), it is stable. However, to be 
consistent with current 
experimental data the VEV of the $SU(2)$ adjoint must vanish at a lower
energy scale. At this stage the Z-string becomes unstable. Hence, in this
scheme, there could be an epoch in the early universe where Z-strings 
would be stable.

(iii) {\it External magnetic field:}
An interesting possibility was studied by Garriga and Montes \cite{GarMon95} 
when they considered the stability of the Z-string placed in an external 
electromagnetic magnetic field of field strength $B$ 
parallel to the string. First, note that $B$ 
should be less than $B_c = m_W^2/e$, otherwise the vacuum outside the string 
is unstable to W-condensation \cite{AmbOle90}. 
Then they found that the Z-string could be stable if
$B > \sqrt{\beta} B_c$, where $\beta = m_H^2/m_Z^2$ should be less than 1
for stability of the ambient vacuum.
The region of stability for a few values of the magnetic field (given
by $K=g_z B/2m_Z^2$) is sketched in Fig. \ref{stabilityregionb}.
For a certain range of
$K \sim 0.85$, stable Z-strings in the GSW model
are still just possible. 

\begin{figure}[tbp]
\caption{\label{stabilityregionb} The triangular regions depict the
parameter range for which the electroweak vacuum and the Z-string are
both stable in the presence of a uniform external magnetic field whose
strength is proportional to $K$. For a range of magnetic field
($K \sim 0.85$), stable strings are possible even with 
the experimentally constrained parameter values.
}
\vskip 1truecm
\epsfxsize = \hsize \epsfbox{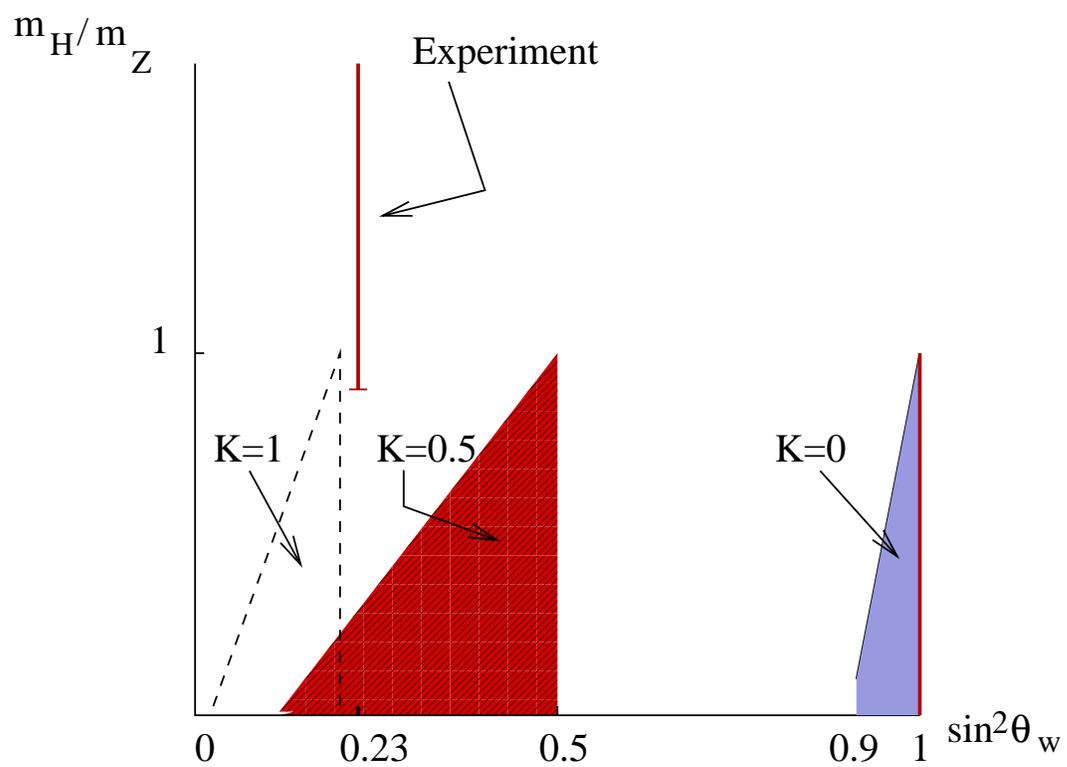}
\end{figure}

A way to understand the enhanced stability of
the Z-string in a magnetic field is to realize that the W-condensation 
instability is due to the interaction of 
$W^3_ \mu = \sin \theta_w A_\mu + \cos \theta_w Z_\mu$ and 
$W^\pm_\mu$. The Z-string itself has a Z magnetic flux. Then the external
electromagnetic flux can serve to lower the net $W^3$ flux. This reduces the
efficiency of W-condensation and makes the string more stable. Another
viewpoint can be arrived at if we picture the Z-string instability to be 
one in which the string breaks due to the production of a monopole- 
antimonopole pair on the string. If the external magnetic field 
is oriented in a direction that prevents the nucleated magnetic monopoles 
from accelerating away from each other, it will suppress the
monopole pair production process, leading to a stabilization of the 
string for sufficiently strong magnetic fields.

(iv) {\it Fermions:} The effect of fermions on the stability of the
Z-string has been considered in 
Refs. \cite{EarPer94,Nac95,KonNac95,LiuVac96}.
Naculich \cite{Nac95} found that
fermions actually make the Z-string unstable. In
\cite{LiuVac96} it was argued that this effect of fermions is quite
general and also applies to situations where the strings form at a low
energy scale due to topological reasons but can terminate on very
massive monopoles formed at a very high energy scale.  This most
likely indicates that the Z-string solution itself should be different
from the Nielsen-Olesen solution when fermions are included. We shall
describe these results in greater detail in
Sec. \ref{superconductivity} after discussing fermion zero modes on
strings.  

Z-strings have also been considered in the presence of a cold bath of
fermions \cite{BimLoz95}. The effect of the fermions is to induce an
effective Chern-Simons term in the action which then leads to a long
range magnetic field around the string.

\subsection{Semiclassical stability}
\label{semiclassicalstability}

Preskill and Vilenkin \cite{PreVil92} have calculated the decay
rate of electroweak strings in the region of parameter space where
they are classically stable. The instability is due to quantum tunneling
and is calculated by finding the semiclassical rate of nucleation of
monopole-antimonopole pairs on electroweak strings. The bounce action
is found to be
\be
S \sim {{4\pi^2}\over {g^2}} {{a_\infty} \over {a_s}}
\label{bounceaction}
\ee where, the strings are classically stable if the ratio of
parameters ${{a_\infty} / {a_s}}$ is larger than 1. ($a_\infty /a_s$
is the ratio of energy in the magnetic flux when it is spread over an
infinite area to that if it is confined within the string.) The
semiclassical decay probability of the string per unit length per unit
time is proportional to $\exp[-S]$.

The decay rate gets suppressed as we approach the semilocal string 
($g \rightarrow 0$) thus the semilocal string is also stable semiclassically.

\section{Superconductivity of electroweak strings}
\label{superconductivity}

\subsection{Fermion zero modes on the Z-string}
\label{zeromodes}

Here we shall consider the fermionic sector of the 
GSW model
in the fixed background of the unit winding Z-string for which the 
solution is given in eq. (\ref{Zstring}).
The Dirac equations for a single family of leptons and quarks 
are obtained from the Lagrangian in Sec. \ref{fermionicsector}. 
These have been solved in the background of a straight Z string in 
\cite{EarPer94,GarVac95,MorOakQui95}. 
The analysis is similar to that for $U(1)$ strings \cite{JacRos81} 
since the Z-string is an embedded $U(1)$ string in the 
GSW model (see Sec. \ref{embeddeddefects}). A discussion of the fermion zero modes in connection with index theorems can be found in \cite{KliRup97,Kli98}

In polar coordinates with the Z-string along the $z$-axis, a
convenient representation for the $\gamma$ matrices is:
\begin{equation}
\gamma^\rho = \pmatrix{0&e^{-i\varphi }&0&0\cr
                    -e^{i\varphi }&0&0&0\cr
                    0&0&0& -e^{-i\varphi }\cr
                    0&0&e^{i\varphi}&0\cr}\ , \ \ \ 
\gamma^\varphi = \pmatrix{0&-ie^{-i\varphi}&0&0\cr
                         -ie^{i\varphi}&0&0&0\cr
                         0&0&0&ie^{-i\varphi}\cr
                         0&0&ie^{i\varphi}&0\cr}\ , \label{gamma1}
\end{equation} 
\begin{equation} 
\gamma^t = \pmatrix{\tau^3&0\cr
0&-\tau^3\cr}\ , \ \ \ \gamma^z = \pmatrix{0&{\bf 1}\cr -{\bf 1}&0\cr}\ ,
\ \ \ \gamma^5 = \pmatrix{0&{\bf 1}\cr {\bf 1}&0\cr}\ .  
\label{gamma2}
\end{equation} 
(Note that the derivative $\gamma^\mu \partial_\mu$ is given
by $\gamma^t\partial_t + \gamma^\rho \partial_\rho
+\gamma^\phi \partial_\phi /\rho +\gamma^z\partial_z$.)
Then the electron has a zero mode solution
\begin{equation} e_L = \pmatrix{1\cr 0\cr -1\cr 0\cr} \psi_1 (\rho ) \ , \
\ \ e_R = \pmatrix{0\cr 1\cr 0\cr 1\cr} i \psi_4(\rho )  \label{eler}
\end{equation} where, 
\begin{equation} 
\psi_1 ' + {{qv} \over \rho} \psi_1 =
-h{\eta \over \sqrt{2}} f \psi_4 \label{psi1eom} \end{equation}
\begin{equation} 
\psi_4 ' - {{(q-1)v} \over \rho } \psi_4 = -h {\eta \over
\sqrt{2}}
 f \psi_1 \ .
\label{psi4eom}
\end{equation}
In these equations $q$ is the eigenvalue of the operator ${\bf q}$
defined in eq. (\ref{zcharge}) and denotes the Z-charge of the various
left-handed fermions. (For the electron, $q =
\cos(2\theta_w)$.) The boundary conditions are that $\psi_1$ and
$\psi_4$ should vanish asymptotically.  This means that there is only
one arbitrary constant of integration in the solution to
eqns. (\ref{psi1eom}) and (\ref{psi4eom}). This may be taken to be a
normalization of $\psi_1$ and $\psi_4$.

For the $d$ quark, the solution is the same
as in eqns. (\ref{eler}), (\ref{psi1eom}) and (\ref{psi4eom}) except that 
$q=1 -  (2/3)\sin^2\theta_w$. For the $u$ quark
the solution is:
\begin{equation}
u_L = \pmatrix{0\cr 1\cr 0\cr -1\cr} \psi_2 (\rho ) \ , \ \ \
u_R = \pmatrix{1\cr 0\cr 1\cr 0\cr} i \psi_3(\rho )
\label{ulur}
\end{equation}
where,
\begin{equation}
\psi_2 ' - {{qv} \over \rho } \psi_2 = -G_u {\eta \over \sqrt{2}}
f \psi_3
\label{psi2eom}
\end{equation}
\begin{equation}
\psi_3 ' + {{(q+1)v} \over \rho } \psi_3 = -G_u {\eta \over \sqrt{2}}
f \psi_2
\label{psi3eom}
\end{equation}
where, $q = -1 + (4/3) \sin^2\theta_w$. Note that 
(\ref{psi1eom}), (\ref{psi4eom}) are related to (\ref{psi2eom}),
(\ref{psi3eom}) by $q \rightarrow -q$.

The right-hand sides of the neutrino Dirac equations (corresponding to
eqns. (\ref{psi1eom}) and (\ref{psi4eom})) vanish since the neutrino
is massless. The solutions can be found explicitly in terms of the
string profile equations in the case when the Higgs boson mass ($m_H =
\sqrt{2\lambda }\eta$) equals the Z boson mass ($m_Z = g_z\eta /2$)
\cite{GarVac95}. 
Recall that the string equations in the $m_H=m_Z$ case are
\cite{Bog76}:
\begin{equation}
f' = {f \over \rho} (1-v)
\label{bogo1}
\end{equation}
\begin{equation}
v' =  {{m_Z^2} \over 2} \rho (1-f^2 )  
\label{bogo2}
\end{equation}
yielding the useful relation:
\begin{equation}
\int d\rho {v \over \rho} = {\rm ln} \biggl ( {m_Z \rho \over f} \biggr )
\label{vr}
\end{equation}
where we have included a factor of $m_Z$ to make the argument of the 
logarithm dimensionless.
Now the zero mode profile functions for the massless fermions are:
\begin{equation}
\psi _1 = c_1 m_Z^{3/2} \biggl ( {{m_Z \rho} \over f} \biggr )^{-q}
\  ,  \ \ \
\psi _4 = c_4 m_Z^{3/2} \biggl ( {{m_Z \rho} \over f} \biggr )^{q-1}
\label{nupsi1psi4}
\end{equation}
where, $c_1$ and $c_4$ are independent constants that can be
chosen to normalize the left- and right-handed fermion states
and the spinors are given in (\ref{eler}). The boundary condition 
that the left-handed fermion wavefunction should vanish at infinity 
is only satisfied if $q > 0$. Hence (\ref{nupsi1psi4}) can only give 
a valid solution for $q > 0$ for the left-handed fermion. If we also
require normalizability, we need $q > 1$. (Note that there is no 
singularity at $\rho =0$ because $f \propto \rho $ when $\rho \sim 0$.) 
If we 
have a left-handed fermion with $q \le -1$, the correct equations to 
use are the equations corresponding to the up quark equations given in
(\ref{psi2eom}) and (\ref{psi3eom}) and these are solved
by letting $q \rightarrow -q$ in (\ref{nupsi1psi4}). In this
case, the spinors are given in (\ref{ulur}).

For the electroweak neutrino, the right-handed component is absent 
and $q = -1$. This means that the neutrino has the same
spinor structure as the left-handed up quark and the solution
is that in (\ref{nupsi1psi4}) with $q$ replaced by $+1$.
Therefore the wave function falls off as $1/\rho$ 
and the state is strictly not normalizable - the normalization
integral diverges logarithmically. However, depending on the
physical situation, one could be justified in imposing
a cut-off. For example, when considering closed loops of string,
the cutoff is given by the radius of the loop.

Next we give the explicit solutions to the Dirac equations in 
(\ref{psi1eom}) and (\ref{psi4eom}) in the
case when the fermion mass ($m_f = h\eta /\sqrt{2}$) is equal to the scalar
mass which is also equal to the vector mass. This so-called
``super-Bogomolnyi'' limit is not realized in the GSW model
but may be of interest in other situations (for example, in 
supersymmetric models). Then, if the charge
on the left-handed fermion vanishes ($q=0$), the solution can be
verified to be: 
\begin{equation}
\psi _1 (\rho ) = N m_Z ^{3/2} (1-f(\rho )^2)
\label{3.5a}
\end{equation}
\begin{equation}
\psi _4 (\rho ) = 2 N m_Z^{1/2}  {f(\rho ) \over \rho} (1-v(\rho ))  
\label{3.5b}
\end{equation}
where $N$ is a dimensionless normalization factor.
For the same set of parameters, the solution for the up quark equations 
can be written by using the transformation $q \rightarrow -q$ in the above
solutions.  Further, this solution can also be derived using supersymmetry 
arguments \cite{VecFer77,DavDavTro97}.

The left-handed fermion wave-functions found above can be multiplied
by a phase factor ${\rm exp}[i(E_p t - pz)$ and the resulting 
wave-function will still solve the Dirac equations provided
\begin{equation}
E_p = \epsilon_i p
\label{ep}
\end{equation}
where, $i$ labels the fermions, and,
\begin{equation}
\epsilon_\nu = +1 = \epsilon_u \ , \ \ \ 
\epsilon_e   = -1 = \epsilon_d \ .
\label{epsi1}
\end{equation}

In other words, $\nu_L$ and $u$ travel parallel to the string flux
while $e$ and $d$ travel anti-parallel to the string flux.  

\begin{figure}[tbp]
\caption{\label{zeromodedirections} The direction of propagation of quark
and lepton zero modes on the Z-string.
}
\vskip 1 truecm
\epsfxsize = \hsize \epsfbox{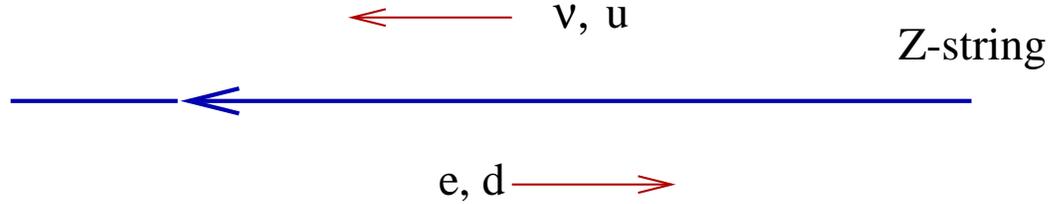}
\vskip 0.5 truecm
\end{figure}

We should mention that the picture of quarks travelling 
along the Z-string may be inaccurate since QCD effects
have been totally ignored.  At the present time it is not known if the
strong forces of QCD will confine the quarks on the string into mesons
and baryons (for example, pions and protons). Further, the
electromagnetic interactions of the particles on the string might lead
to bound states of electrons and protons on the string. This would
imply a picture where hydrogen (and other) atoms are the fundamental
entities that live on the string.

\subsection{Stability of Z-string with fermion zero modes}
\label{stabilitywithfermions}

In Fig. \ref{pertonfermions} we show the effect that perturbations of
order $\epsilon$ in the Z-string fields have on the fermion (u and d
quarks) 
zero modes.
The zero momentum modes acquire an $O(\epsilon )$ mass while the
non-zero momentum modes get an $O(\epsilon^2 )$ mass. For the
perturbation analysis to make sense, we require that the u and d quark
zero momentum modes are either both filled or both empty.  In that
case, the $O(\epsilon )$ terms in the variation in the energy will
cancel and we will be left with something that is $O(\epsilon^2 )$.
In fact,
\be
\Delta E = - {{\epsilon^2}\over 2} \vert m_1 \vert ^2 L 
                 \sum_{k=1}^N {1 \over k}
\ee
where $m_1$ is a matrix element having to do with the interactions of
the u and d quarks, $L \rightarrow \infty$ is the length of the string
on which periodic boundary conditions have been imposed, and $N
\rightarrow \infty$ is a cut-off on the energy levels which
are labeled by $k$. 
The crucial
piece of this formula is the minus sign which shows that the energy of
the string is lowered due to perturbations \cite{Nac95}.

In Ref. \cite{LiuVac96} it was argued that an identical calculation
could be done for any classically stable string that could terminate
on (supermassive) magnetic monopoles. However, in the low energy
theory, the strings are effectively topological and hence, it seems
unlikely that fermions can lead to an instability. This suggests that
the bosonic string configuration gets modified by the fermions and
the stability analysis around the Nielsen-Olesen solution may be
inappropriate. 

%In addition, the stability analysis with fermions only considers the
%zero modes and ignores the infinitely many massive fermion modes.  At
%the present time, there appears to be no feasible analysis that can
%treat the full spectrum of fermions in the string background. Hence
%the effect of fermions on the stability of the Z-string remains a
%vital open problem. 

So far, the stability analysis with fermions presented here only
considered the zero modes and ignored the infinitely many massive
fermion modes. Very recently, Groves and Perkins \cite{GroPer99}
have analysed the full spectrum of massless and massive fermionic 
modes in the background of the electroweak string. They then
calculate the effect of the
Dirac sea on the stability of electroweak strings by calculating 
the renormalised energy shift of the Dirac sea when a Z-string is
perturbed by introducing a non-zero upper component to the Higgs
doublet.  This energy shift is negative and so destabilises the
string, but it is small, leading them to conclude that if positive
energy fermionic states are populated, it is conceivable that the
total fermionic contribution could be to stabilise the string. This
work is still in progress. In the meantime, the stability of Z--strings
remains an open question.

%At
%the present time, there appears to be no feasible analysis that can
%treat the full spectrum of fermions in the string background. Hence
%the effect of fermions on the stability of the Z-string remains a
%vital open problem. 

\begin{figure}[tbp]
\caption{\label{pertonfermions} The effect of perturbations of the
Z-string on fermion zero modes.
}
\vskip 1 truecm
\epsfxsize = \hsize \epsfbox{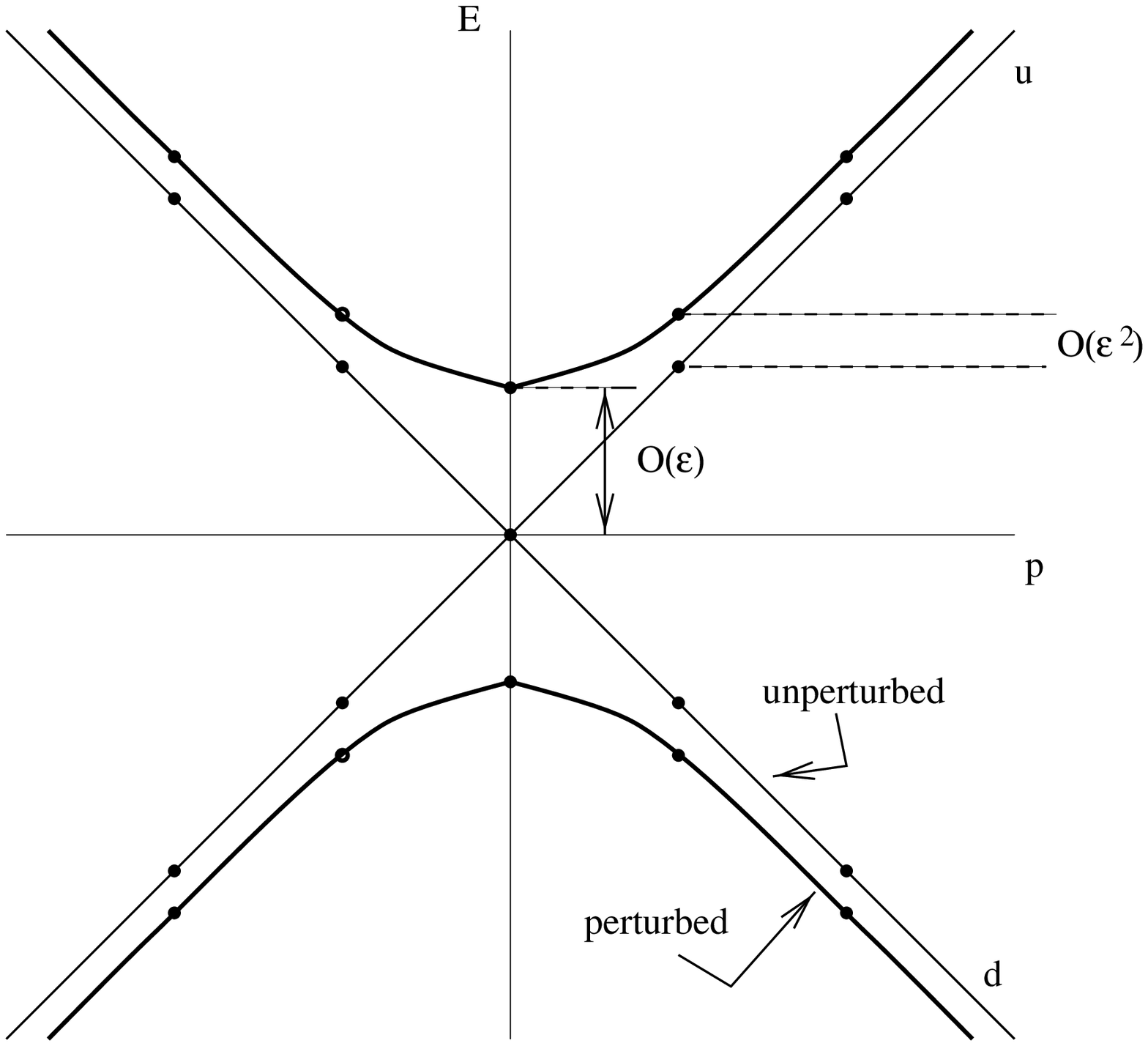}
\end{figure}

\subsection{Scattering of fermions off electroweak strings}
\label{scattering}

The elastic scattering of fermions off semilocal 
and electroweak strings has
been considered in \cite{Gan93,DavMarGan94,Lo95}. 

The main feature of the cross section is that
the scattering violates helicity \cite{Gan93}.
It is straightforward to  show that the helicity operator $\Sigma \cdot
\Pi$, where $\Sigma^i = \epsilon^{ijk} \gamma^i\gamma^j$ is the spin 
operator and $\Pi^i$ are the canonical momenta,  does not commute with
the hamiltonian. If $\Phi^T = (\phi^+,
\phi^0)$, the commutator is proportional to $(D\phi^0)$ terms. 
Consider for a moment the usual representation of Dirac matrices,
\be
\gamma^0 = \pmatrix{0 \ \ \  \  1 \cr 1 \ \ \  \  0} \qquad\qquad
\gamma^i = \pmatrix{0 \ \ -\tau^i \cr \tau^i \ \ \  \ 0} \qquad\qquad
\gamma_5 = \pmatrix{1 \ \  \ \  \ 0 \cr 0 \ \ -1} 
\ee
Then, for an incoming electron, one finds 
\be
	[{\rm H},{ \Sigma}\cdot{\Pi}]=i h \left(
	\begin{array}{cc} 0 & \tau^{j}(D^{j}\phi^{0})^{\dagger}\\
	\tau^{j}D^{j}\phi^{0} & 0\\ \end{array} \right),
\ee
where $h$ is the Yukawa coupling and $(D_j\phi^0)$ is given in
eq. (\ref{smalldmu}).
Therefore helicity-violating processes can take place in the
 core of the string. 

A preliminary calculation by Ganoulis in ref. \cite{Gan93} showed that,
for an incoming plane wave, the dominant mode of scattering gives
identical cross sections for positive and negative helicity scattered
states. More precisely, for an incoming electron plane wave of
momentum $k$, energy $\omega$ and positive helicity it was found that,
to leading order,
\be
{ d\sigma \over dk}\biggr |_\pm \sim 
{1 \over k} \left( \omega - k \over 2\omega \right)^2 \sin^2 (\pi q_R)
\ee
where $\omega^2 = k^2 + m_e^2$, $q_R$ is the
Z-charge of the right fermion field, given in eq. (\ref{zcharge})
(recall that right and left fermion fields have different Z-charges,
$q_R = q_L \pm 1$).

A more detailed calculation was done by Davis, Martin and Ganoulis
\cite{DavMarGan94}, and later extended by Lo \cite{Lo95}, using a 
`top hat' model
\be 
f(r) = \cases{ 0 &$r<R$ \cr {\eta / \sqrt{2}} &$r>R$ \cr}
\qquad\qquad\qquad
  v(r) = \cases{ 0 &\ \ $r<R$ \cr {2 / g_z} &\ \ $r>R$
\cr} \ \ , 
\ee
which is expected to be a reasonable approximation since the
scattering cross section in the case of cosmic strings has been shown
to be insensitive to the core 
model \cite{PerPerDavBraMat91}.  Note that there
is a discontinuous jump in the fermion mass and string flux; however
the wave functions are matched so that they are continuous at $r=R$.
Note
that the left and right fields decouple in the core of the string, so
helicity violating processes are concentrated at $r=R$. 

The authors of \cite{DavMarGan94,Lo95} 
confirmed that, in the massive case,
there are helicity-conserving and helicity-flip scattering cross
sections of equal magnitude. The latter goes to zero in the massless
limit (in that case, the left and right fields decouple, and no
helicity violation is possible), suggesting that helicity violation
may be stronger at low energies. For ``fractional string flux'' ({\it
i.e.} for fractional $q$) the cross section is of a modified
Aharonov-Bohm form, and independent of string radius. For integer $q$
it is of Everett form \cite{Eve81} (the strong interaction cross section
is suppressed by a logarithmic term).  

Another interesting feature has to do with the amplification of the
fermionic wave function in the core of the string.  
Lo \cite{Lo95} has remarked that there is a regime in
which the scattering cross section for electroweak strings is much
less sensitive to the fermion charge (that is, to $\sin^2 \theta_w$)
than for cosmic strings.  In contrast with, {\it e.g.}, baryon number
violating processes, which show maximal enhancement only for discrete
values of the fractional flux, the helicity violating cross section
for electroweak strings in the regime $k \sim m, \ kR << 1$ shows a
plateau for $0< \sin^2 \theta_w < 1/2$ where amplification is maximal
and the cross section becomes of order $m_f^{-1}$. This can be traced
back to the asymmetry between left and right fields; while the
wave function amplification 
is a universal feature, different components of the fermionic wave
function acquire different amplification factors in such a way that
the total enhancement of the cross section is approximately
independent of the fermionic charge, $q$ (or, equivalently, of $\sin^2
\theta_w$). 

Elastic scattering is independent of the string radius for both
electroweak and semilocal strings
(for integral flux there is only a mild dependence on the radius
coming from the logarithmic suppresion factor in the Everett cross
section). 
Since the cross section is like that of $U(1)$ strings, we would
expect electroweak and semilocal strings to interact with the
surrounding plasma in a way that is analogous to topological
strings.

\section{Electroweak strings and baryon number}
\label{ewstringbaryonnumber}

As first shown by Adler \cite{Adl69}, and, Bell and Jackiw
\cite{BelJac69}, currents that are conserved in a classical field 
theory may not be conserved on quantization of the theory. 
In the GSW model, one such current is the baryon number
current and the anomalous current conservation equation is:
\begin{equation}
\partial_\mu j^\mu _{B} = {{N_F} \over {32\pi^2}}
 [ -g^2 W^a _{\mu \nu} {\tilde W}^{a \mu \nu} +
                         {g'}{}^2 Y_{\mu \nu} {\tilde Y}^{\mu \nu} ]. 
\label{anomalyeq} 
\end{equation} 
where $j^\mu _B$ is the expectation value of the baryon number
current operator $\sum_s b_s :{\bar \psi} \gamma^\mu \psi:$ 
where the sum is over all the species of fermions labeled by $s$, 
$\psi$ is the fermion spinor and $b_s$ is the baryon number for 
species $s$ and the operator product is normal ordered.
Also, $N_F$ denotes the number of families, and tilde the dual of
the field strengths. 

The anomaly equation can be integrated over all space leading to 
\begin{equation}
\Delta Q_{B} = {{N_F} \over {32\pi^2}}
     \int dt d^3 x
 [ -g^2 W^a _{\mu \nu} {\tilde W}^{a \mu \nu} +
                         {g'}{}^2 Y_{\mu \nu} {\tilde Y}^{\mu \nu} ]
= \Delta Q_{CS}.
\label{intanomeq}
\end{equation}
with,
\be
Q_{CS} =  
{{N_F} \over {32\pi^2}}  \int  d^3 x \epsilon_{ijk}
 \biggl [ g^2 \biggl ( W^{a ij} W^{a k} - {g \over 3}
                       \epsilon_{abc} W^{ai} W^{bj} W^{ck} \biggr ) -  
          {g'}{}^2 Y^{ij} Y^{k} \biggr ] \ .
\label{qcs}
\ee
Here, $\Delta ( \cdot )$ denotes the difference of the
quantities evaluated at two different times, $Q_B$ is the baryonic
charge and the surface currents and integrals at infinity are assumed
to vanish.  $Q_{CS} $ is called the Chern-Simons, 
or topological, charge and can be evaluated if
we know the gauge fields. The left-hand side of eq. (\ref{intanomeq})
evaluates the baryon number by counting the fermions directly. We
describe the evaluation of both the right- and left-hand side for
fermions on certain configurations of Z-strings in the following
subsections. Finally, in Sec.
\ref{cosmologicalapplications} 
we briefly comment on possible applications to cosmology.

\subsection{Chern-Simons or topological charge}
\label{chernsimons}

We will be interested in the Chern-Simons charge contained in
configurations of Z-strings. Then, we set all the gauge fields 
but for the Z-field to zero 
in the expression for the Chern-Simons charge, yielding
\begin{equation}
Q_{CS} = N_F {{\alpha ^2 } \over {32\pi^2}} \cos(2\theta_w)
                        \int d^3 x {\vec Z}\cdot {\vec B}_Z  
\label{qcsbrief}
\end{equation}
where, ${\vec B}_Z$ denotes the magnetic field in the Z gauge
field: $B^i_Z = \epsilon^{ijk} \partial_jZ_k $. 

The terms on the right-hand side
have a simple interpretation in terms of a concept called ``helicity''
in fluid dynamics \cite{BerFie84}.  Essentially, if a fluid flows with
velocity $\vec v$ and vorticity $\vec \omega = \vec \nabla \times \vec
v$, then the helicity is defined as:
\be
h = \int d^3 x {\vec v} \cdot {\vec \omega}
\label{fluidhelicity}
\ee
Since the helicity measures the velocity flow along the direction of
vorticity, it measures the corkscrew motion (or twisting) of the fluid 
flow. A direct analog is defined for magnetic fields:
\be
h_B = \int d^3 x {\vec A} \cdot {\vec B}
\label{magnetichelicity}
\ee
which is of the same form as the terms appearing in (\ref{qcsbrief}). Hence
the Chern-Simons charge measures the twisting of the magnetic lines of
force. The helicity associated with the Z field alone is given by:
\be
H_Z = \int d^3 x {\vec Z}\cdot {\vec B}_Z \ .
\label{zhelicity}
\ee
If we think in terms of flux tubes of $Z$ magnetic field, $H_Z$ measures
the sum of the link and twist number of these tubes:
\be
H_Z = L_Z + T_Z \ .
\label{linktwist}
\ee
For a pair of unit winding Z flux tubes that are linked
once as shown in Fig. \ref{linkedloops} the helicity is:
\be
H_Z =  2 F_Z ^2
\label{helicityoflink}
\ee
where, $F_Z$ is the magnetic flux in each of the two tubes 
Note that the helicity is positive for the
strings shown in Fig. \ref{linkedloops}.  If we reversed the direction
of the flux in one of the loops, the magnitude of $H_Z$ would be the
same but the sign would change. For the Z-string, we also know that
\be
F_Z = {{4\pi} \over g_z}
\label{zfluxforlink}
\ee
and so eq. (\ref{qcsbrief}) yields \cite{VacFie94}: 
\be
Q_{CS}  = N_F \cos (2\theta_w ) \ .
\label{qcsresult}
\ee

\begin{figure}[tbp]
\caption{\label{linkedloops} A pair of linked loops.}
\vskip 1 truecm
\epsfxsize = \hsize \epsfbox{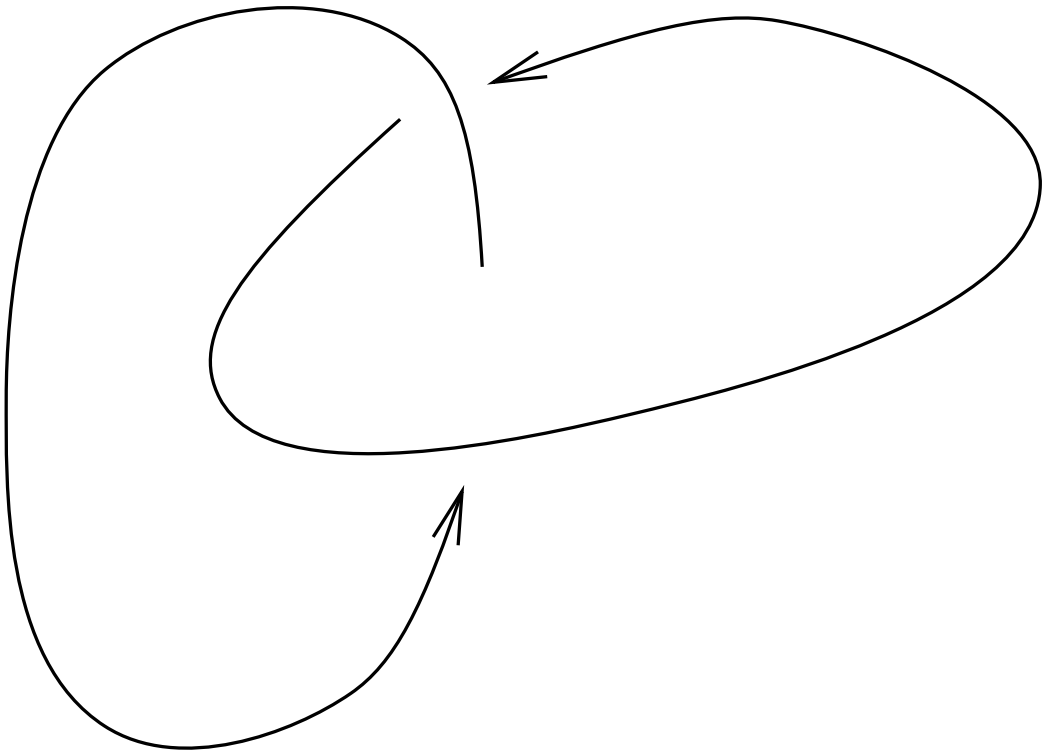}
\end{figure}

\subsection{Baryonic charge in fermions}
\label{baryoninfermion}

The baryon number associated with linked loops of Z-string has been
evaluated in Ref. \cite{GarVac95} by studying the
fermionic zero modes on such loops. 
This corresponds to
evaluating the left-hand side of eq. (\ref{intanomeq}) directly in terms
of the fermions that carry baryon number. The calculation involves adding
the baryonic charges of the infinite Dirac sea of fermions living on the
string together with zeta function regularization. 

To understand why the linking of loops leads to non-trivial effects,
note that the quarks and leptons have a non-trivial Aharanov-Bohm
interaction with the Z-string. So the Dirac sea of fermions on a loop 
in Fig. \ref{linkedloops} is affected by the Z-flux in the 
second loop. This shifts the level of the Dirac sea in the ground
state leading to non-trivial baryonic and other charges.

Instead of considering the linked loops as shown in Fig. \ref{linkedloops}
it is simpler to consider a large circular loop of radius 
$a \rightarrow \infty$ in the $xy$-plane threaded by $n$ straight infinite 
strings along the $z$-axis (Fig. \ref{threading}). Then the fermionic
wave-functions take the form:
\begin{equation}
\psi_L = e^{- i(E_p t - p \sigma )} \psi_L^{(0)}(r) \ , \ \ \
\psi_R = e^{- i(E_p t - (p-n/a) \sigma )} \psi_R^{(0)}(r) \
\label{propmodes}
\end{equation}
where the functions with superscript $(0)$ are the zero mode profile
functions described in Sec. \ref{zeromodes} and $\sigma$ is a
coordinate along the length of the circular loop.  From these
wavefunctions, the dispersion relation for a zero mode fermion on the
circular loop is 
\begin{equation}
\omega_k = \epsilon_i  ( k - qZ  ) \ .
\label{omega}
\end{equation}
where, $q$ is the Z-charge of the fermion, $\epsilon_i$ is defined in
eq. (\ref{epsi1}),
$\omega$ is related to the energy $E$ by $\omega \equiv  aE$, and
$k$ to the momentum $p$ by $k \equiv a p \in {\rm \angle \!\!\! Z}$. 
$Z$ is the component of the gauge field along the circular loop 
multiplied by $a$ and is given by,
\begin{equation}
Z \equiv {{2 n} \over {g_z}} \ .
\label{zna}
\end{equation}
The crucial property of
the dispersion relation is that, if there is an Aharanov-Bohm
interaction between the Z-string and the fermion, $\omega_k$ cannot
be zero for any value of $k$ since $k$ is an integer but $qZ$ is not.

The Z- , A- and baryon number (B) charges of the leptons and quarks
are shown in Table \ref{tab:effluents}. Note that we use $2q_Z/g_z$
to denote the Z-charge
and this is identical to the eigenvalue of the operator ${\bf q}$
defined in eq. \ref{zcharge} and also to $q$ used in the previous
section.

\begin{figure}[tbp]
\caption{\label{threading} A circular Z-string loop of radius $a$
threaded by $n$ Z-strings.}
\vskip 1 truecm
\epsfxsize = \hsize \epsfbox{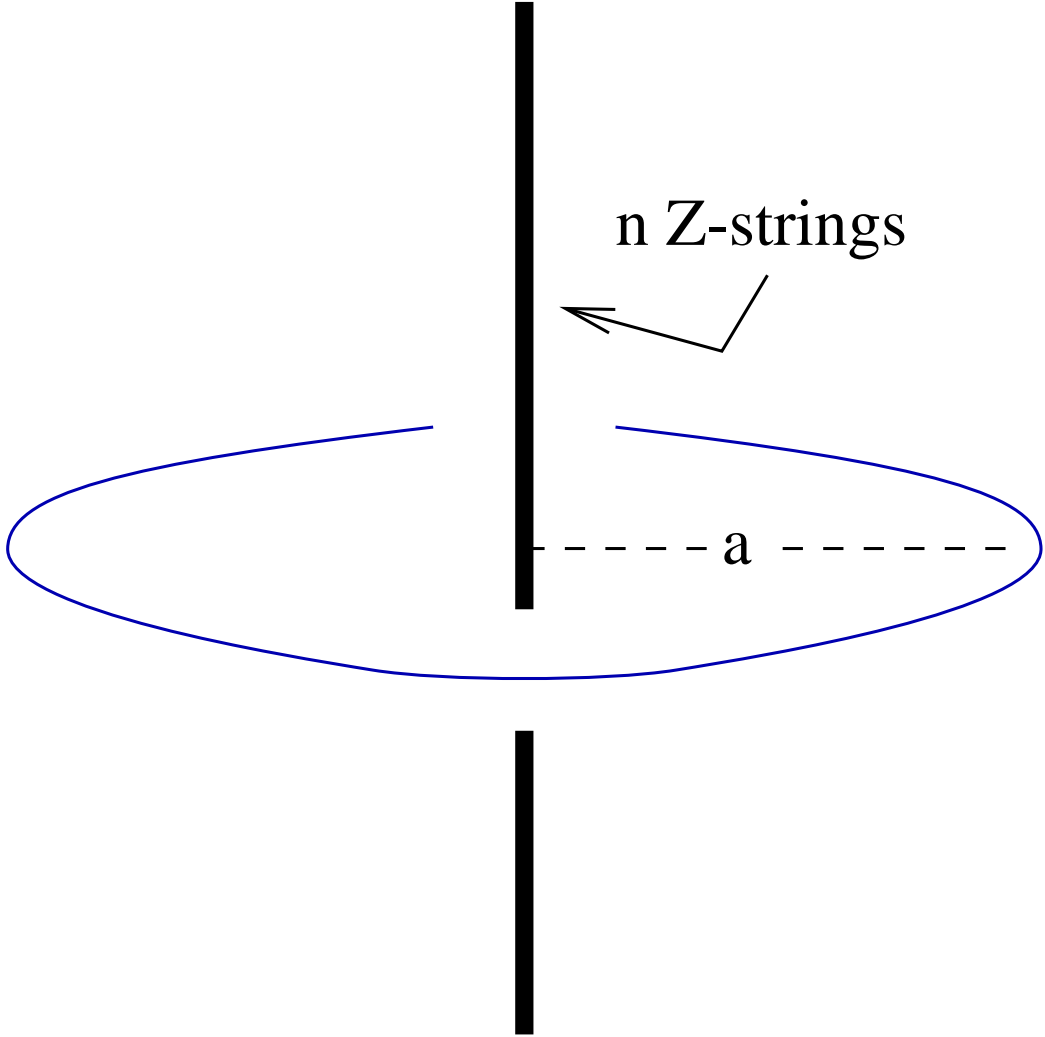}
\end{figure}

\begin{table*}[hbt]
\setlength{\tabcolsep}{1.5pc}
\newlength{\digitwidth} \settowidth{\digitwidth}{\rm 0}
\catcode`?=\active \def?{\kern\digitwidth}
\caption{Summary of $Z-$, electric and baryonic charges
for the leptons and quarks. The charges $q_Z$ are for the 
left-handed fermions and
$s^2 \equiv \sin^2 \theta_w$.}
\smallskip
\label{tab:effluents}
\begin{tabular*}{\textwidth}{@{}l@{\extracolsep{\fill}}cccc}
\hline
                 & \multicolumn{1}{c}{$\nu_L$}
                 & \multicolumn{1}{c}{$e$}
                 & \multicolumn{1}{c}{$d$}
                 & \multicolumn{1}{c}{$u$}         \\
\hline
\smallskip
$2q_Z/g_z$       &-1  &$    1-2s^2 $ &$1-{{2s^2}/ 3}$ 
                                          &$-1+{{4s^2}/ 3}$    \\
$q_A/e$             &0   &-1          & $-1 /3$ 
                                          & $2 / 3$     \\
$q_B$               &0   &0           &$1/ 3$
                                          & $1/ 3$      \\
\hline
\multicolumn{5}{@{}p{120mm}}{}
\end{tabular*}
\end{table*}

The energy of the fermions is found by
summing over the negative frequencies - that is, the Dirac sea - and so
the energy $E$ due to a single fermion species is:  
\begin{equation} E =
{1 \over a} \sum \omega_k
  = \epsilon_i {1 \over a} \sum_{k=k_F}^{-\epsilon_i  \infty} 
                       ( k - q Z )
\label{energy}
\end{equation}
where, $k_F$ denotes the Fermi level - the value of $k$ for 
the highest filled state. Therefore we need to sum a series 
of the type:
\begin{equation}
S = \sum_{k=k_F}^{\infty} ( k - q Z ) 
  = \sum_{k=0}^{\infty} ( k+k_F - q Z ) \ .
\label{series}
\end{equation}
The sum is found using zeta function regularization:
\begin{equation}
S = \zeta (-1, k_F -q Z) =
 - {1 \over {12}} - {1 \over 2} (k_F-qZ)(k_F-q Z-1)
\label{sum}
\end{equation}
With this result, the energy contribution from the $i^{th}$ species
of fermions takes the form:
\begin{equation}
E_i = -{1 \over {24a}} + 
{1 \over {2a}} \biggl [ 
      k_F^{(i)} - q_i Z + {\epsilon_i \over 2} 
               \biggr ] ^2
\equiv -{1 \over {24a}} + {1 \over {2a}} K_i ^2 \ .
\label{ei}
\end{equation}
Adding the contributions due to different
members of a single fermion family, we get 
\begin{equation} 
2 a E = K_\nu
          ^2 +K_e^2 + 3K_d^2 + 3 K_u^2 \ .  
\label{gene} 
\end{equation}

Next we can calculate the angular momentum of the fermions in the circular
loop background.  The system has rotational symmetry about the $z-$axis
and this enables us to define the generalized angular momentum operator as
the operator that annihilates the background
field configuration \cite{JamPerVac92}: 
\begin{equation} 
M_z = L_z + S_z + n I_z 
\label{gam} 
\end{equation} where,
\begin{equation} 
L_z = -i {\bf 1} {{\partial \ } \over {\partial \varphi}} \ , 
\label{lz} 
\end{equation} 
$S_z$ is the spin operator, and, the isospin operator is given in
terms of the $U(1)$ (hypercharge) and $SU(2)$ charges - $q_1$ and $q_2$
respectively - of the field in question:  
\begin{equation} I_z = {1 \over 2} \biggl [
      \biggl ( {{2q_2} \over g} \biggr ) T^3 -
      \biggl ( {{2q_1} \over {g'}} \biggr ) {\bf 1} 
                  \biggr ] \ .
\label{iz}
\end{equation}
The isopin operator acts via a commutator bracket on the gauge fields
and by ordinary matrix multiplication on the Higgs field and fermion
doublets. 

We are interested in the angular momentum of the chiral fermions on the
circular loop which lies entirely in the $xy$-plane. The fermions in
the zero modes therefore have $S_z =0$. (The spin of the fermions is
oriented along their momenta which lies in the $xy$-plane.) The action 
of $L_z$ is found by acting on the fermion wave-functions such as in 
eq. (\ref{propmodes}) (remembering to let $n \rightarrow -n$ for the 
neutrino and up quark).
The action of $I_z$ is found by using the charges of the fermions
given in the GSW model defined in Sec. \ref{fermionicsector}. 
We then find:
\begin{equation}
M_z \pmatrix{\nu_L\cr e_L\cr} = 
  \pmatrix{(k^{(\nu )} +n)\nu_L \cr k^{(e)} e_L\cr} \ , \ \ \ 
M_z \pmatrix{u_L\cr d_L\cr} =
  \pmatrix{(k^{(u)} +{n/ 3})u_L \cr 
           (k^{(d)} -{{2n}/ 3}) d_L\cr} \ , 
\label{mzpsil}
\end{equation}
\begin{equation}
M_z e_R = k^{(e)} e_R \ , \ \ 
M_z u_R = (k^{(u)} +{n/ 3}) u_R \ , \ \ 
M_z d_R = (k^{(d)} - {{2n}/ 3}) d_R \ 
\label{mzul}
\end{equation}
where the $k^{(i)}$ are defined above eqn. (\ref{omega}).
Now summing over states, as in the case of the energy, we find the
total generalized angular momentum of the fermions on the circular
loop:
\begin{equation}
{\cal M} = 
  {1\over 2} \biggl [ k_F ^{(\nu )} +n +{1\over 2} \biggr ] ^2 -
 {1\over 2} \biggl [ k_F ^{(e)}-{1\over 2} \biggr ]^2 -
{3\over 2} \biggl [ k_F ^{(d)}-{{2n}\over 3} -{1\over 2}\biggr ] ^2 +
{3\over 2} \biggl [ k_F ^{(u)}+{{n}\over 3} +{1\over 2}\biggr ] ^2 \ .
\label{totgam}
\end{equation}
Note that though the gauge fields do not enter explicitly in the 
generalized angular momentum, they do play a role in 
determining the angular
momentum of the ground state through the values of the Fermi levels. 

The calculation of the electromagnetic and baryonic charges and
currents on the linked loops is similar but has a subtlety. 
To find the total charge, a sum over the charges in all filled states 
must be done. This leads to a series of the kind:
\begin{equation}
S_q = \sum_{k=k_F}^{ \infty} 1 \ .
\label{sq1}
\end{equation}
To regularize the divergence of the series, it is written as
\begin{equation}
S_q =\lim_{\lambda\to 0} \sum_{k=k_F}^{ \infty} (k-q Z)^{\lambda} \ .
\label{sq2}
\end{equation}
The subtlety is that the gauge invariant combination
$k-qZ$ is used as a summand rather than $k$ or some other 
gauge non-invariant expression \cite{Man85}. 
Once again zeta function regularization is used to get:
\begin{equation}
S_q =  \sum_{k=0}^{\infty} (k+k_F -q Z)^0  =
 \zeta(0, k_F-q Z ) =
 - \biggl [ k_F -qZ - {1 \over 2} \biggr ] \ .
\label{sqdone}
\end{equation}
With this result,  the contribution to the charge due to fermion $i$ is:
\begin{equation}
Q_i = \epsilon_i {\bar q}_i 
      \biggl [ k_F^{(i)} -q_i Z + {\epsilon_i \over 2} \biggr ]
     = \epsilon_i {\bar q}_i K_i
\label{qi}
\end{equation}
where, ${\bar q}_i$ is the charge 
carried by the $i^{th}$ fermion of the kind that we wish
to calculate. 
(Note that ${\bar q}_i$ can represent any charge - electric,
baryonic {\it etc.} - and is, in general, different from the 
Z-charge $q_i$.) 

The currents along the string are given by 
${\bar \psi} \gamma^z \psi$ where $\gamma^z$ is given in eq. (\ref{gamma2}). 
This gives
\begin{equation}
J_i = \epsilon_i {{Q_i} \over {2\pi a}} \ .
\label{ji}
\end{equation}

By adding the contributions due to each variety of fermion,
expressions for the energy, angular momentum,
charges and currents for one loop threaded by $n$ have been found
in \cite{GarVac95}. These results are reproduced in
Table \ref{tab:results}.  It is reassuring to note that in the ground
state, the baryon number of the single loop is given by 
$nN_F \cos2\theta_W$ in agreement
with the calculation of the Chern-Simons number.

\begin{table*}[hbt]
\setlength{\tabcolsep}{.99pc}
\catcode`?=\active \def?{\kern\digitwidth}
\caption{Expressions for the energy, generalized angular momentum, 
charges and currents in terms of $x =  2n\sin^2\theta_w /3$. We have 
omitted the multiplicative factor $N_F$ in all the expressions for 
convenience.
}
\smallskip
\label{tab:results}
\begin{tabular*}{\textwidth}{@{}l@{\extracolsep{\fill}}cccc}
\hline
\smallskip
                 & \multicolumn{1}{c}{$x \in (0,1/3)$}
                 & \multicolumn{1}{c}{$(1/3,1/2)$}
                 & \multicolumn{1}{c}{$(1/2,2/3)$}
                 & \multicolumn{1}{c}{$(2/3,1)$}         \\
\hline
\smallskip
$aE$       &$12x^2-6x+1$  &$12x^2-9x+2$  &$12x^2-15x+5$
                                          &$12x^2-18x+7$    \\
$M$             &0   &$n-1$& $2-n$ & 0     \\
$Q_A /e$               &0   &-1           & +1 & 0      \\
$B$         &$-3x+1$  &$-3x+1$   &$-3x+2$ &$-3x+2$   \\
$2\pi a J_A /e$    &$-8x+2$  &$-8x+3$   &$-8x+5$  &$-8x+6$ \\
$2\pi a J_B$       &$-x$     &$-x$      &$1-x$   &$1-x$   \\
\hline
\multicolumn{5}{@{}p{120mm}}{}
\end{tabular*}
\end{table*}

\begin{figure}[tbp]
\caption{\label{evsx} The energy of the ground state of linked
loops versus $x = 2n\sin^2\theta_w /3$.}
\vskip 1 truecm
\epsfxsize = \hsize \epsfbox{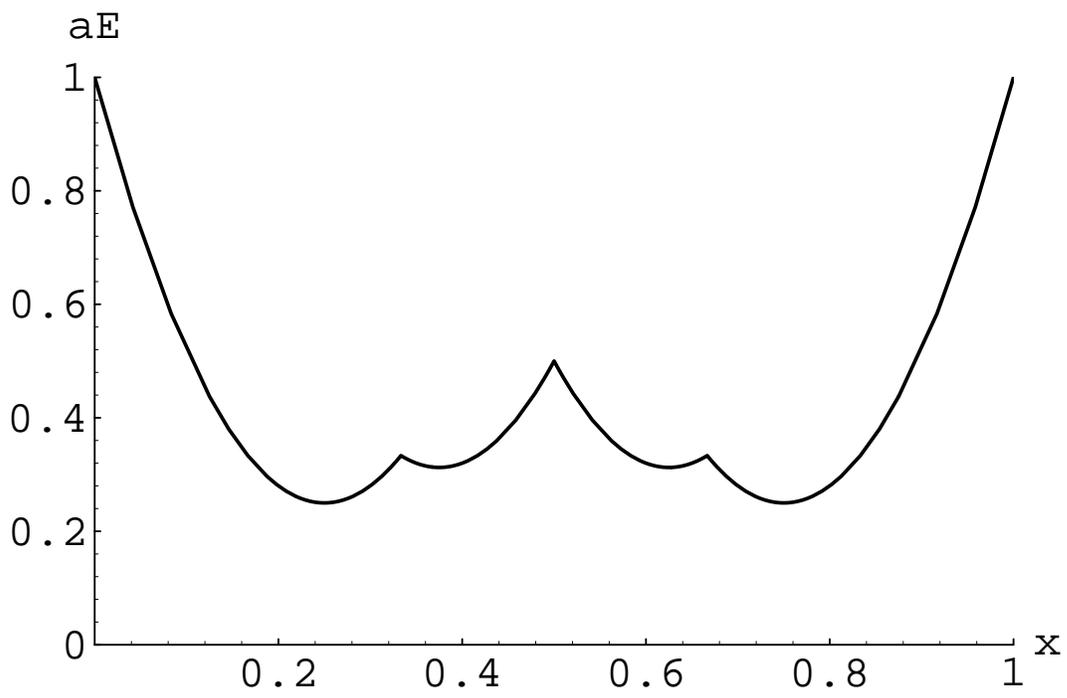}
\end{figure}

The energy of the fermionic ground state shows a complicated
dependence on $x$ as is demonstrated in Fig. \ref{evsx}.
Note that $E(x)$ 
does not have a monotonic dependence on $x$ and the energy of strings
that are linked $n$ times bears no simple relation to those
linked $m$ times. In particular, the energy does not continue
to decrease as we consider strings that have higher linkage.
The lowest energy possible, however, is when $x=1/4$ and
for $n=1$, this corresponds to $\sin^2 \theta_w = 3/8$, 
which is also the value set by Grand Unified models. It is not
clear if this is simply a coincidence or if there is some deeper
underlying reason \cite{Kep98}.

\subsection{Dumbells}
\label{dumbells}

In his 1977 paper, Nambu discussed the possible occurrence of
electroweak monopoles and strings in particle accelerators.
There are two issues in this discussion: the first is the
production crossection of solitonic states in particle collisions,
and the second is the signatures of such states if they are
indeed produced in an accelerator. The answer to the first 
question is not known though it is widely believed that
the process is suppressed not only by the large amount of energy
required but also due to the coherence of the solitonic
state. The second question was addressed by Nambu \cite{Nam77}
and he estimated the energy and lifetime of electroweak strings
that may be possible to detect in accelerators.

To find the energy of a Z-string segment, Nambu 
treated the monopoles at the ends 
as hollow spheres of radius $R$ inside which
all fields vanish. A straightforward variational calculation 
in units of $\eta \approx 246$ GeV then
gives the monopole mass
\be
M = {4 \pi \over 3 e} \sin ^{5/2}\theta_w \sqrt{m_H \over m_W} 
\ee
and radius
\be
R = \sqrt{ \sin\theta_w \over m_H m_W}
\ee
The string segment is approximated by a cylindrical tube with 
uniform $Z$ magnetic flux with all other fields vanishing. This 
gives 
\be \rho =
{2 \over \sqrt{m_H m_Z}}  \ \ , \qquad\qquad 
\tau = \pi  \left(
{m_H \over m_Z} \right ) 
\ee
for the core radius and string tension, respectively.

Now, if
the monopoles are a distance $l$ apart, the total energy of the 
system is
\be E = 2M -{Q^2 \over 4\pi l} + \tau l
\ee
which is clearly minimised by $l=0$ {\it i.e.} the string can
minimize its energy by collapsing. The tendency to collapse
can be countered by a centrifugal barrier if the string segment
(``dumbell'') is rotating fast enough about a perpendicular axis.
The energy and angular momentum of 
a relativistic dumbell has been estimated
by Nambu to be:
\be
E \sim  \half \pi l \tau \qquad 
L \sim {1\over 8} \pi l^2 \tau
\ee
where,
\be
{l \tau \over 2 M} = {v^2 \over 1-v^2} 
\ee
with $v \sim 1$ being the velocity of the poles.  The expressions for
$E$ and $L$ imply the existence of asymptotic Regge
trajectories, 
\be
L \sim \alpha_0 ' E^2
\ee
with slope 
\be
\alpha_0 ' = {1 \over {2\pi \tau}}  \sim  
\left( {m_Z \over m_H} \right)  {\rm TeV}^{-2}\ .
\ee
which, if found, would be a signature of dumbells.

The orbiting poles at the ends of the rotating dumbell will
radiate electromagnetically and this energy loss provides an
upper bound to the lifetime of the configuration.
An estimate of the radiated power from the analysis of
synchrotron radiation in classical electrodynamics 
({\it eg.} see \cite{Jac75}) gives 
\be
P \sim {{8\pi} \over 3} \times 137 \left ({\tau \over M}\right) ^2
\sin^4 \theta_w 
\ee
Therefore the decay width $\Gamma = P/E$ is given by
\be
\Gamma \simeq {E \over L}  
\ee
and for large angular momentum, can lead to significant lifetimes
(compared to $E^{-1}$). 

To obtain numerical estimates, note that the above estimates
are valid only if the dumbell length is much greater than the
width of the Z-string. This imposes a lower bound on the angular
momentum:
\be
L >> {\pi \over 2} \times 137 \sin^2 \theta_w \cos^2\theta_w
 \sim 36
\ee

Using the relation
between the energy and the angular momentum, such an object has 
$E >> 6 (m_H/m_Z)^{1/2}$ TeV.

The estimates above assume that the lifetime of the
dumbell is dictated by the energy emission in photons. In 
reality, there are other decay channels as well, though it is 
likely that these will be comparitively suppressed since 
the photon is the only massless boson present in the system.
The dumbell can also decay by fragmenting due to field-theoretic
instabilities of the kind discussed in 
Sec. \ref{stability}. These may be
suppressed due to the finite size of the dumbell, and as
Nambu points out, due to the angular momentum of the dumbell
\footnote{In the stability analysis for  a
finite piece of string of length $L$, the eigenvalues of the stability
equation are shifted by a contribution of order $\pi^2/ L^2$ with
respect to the infinitely long case, thus for sufficiently short
segments the radial decay mode could become stable.  Longitudinal
collapse might then be stabilized by rotation, as explained above}.
A careful analysis of these factors has not yet been performed
and is a vital open problem that may become experimentally
relevant with the next generation of accelerators.

\subsection{Possible cosmological applications}
\label{cosmologicalapplications}

The role of electroweak strings in cosmology depends on their abundance
during and after the electroweak phase transition. If this abundance is
negligible, electroweak strings may at best only be relevant in future 
accelerator experiments (see Sec. \ref{dumbells}). If, however, there 
is a cosmological
epoch during which segments and loops of electroweak strings were 
present, they could impact on two observational consequences: the first
is the presence of a primordial magnetic field, and the second is the
generation of a cosmological baryon number. What is perhaps most
remarkable is that the two consequences might be related - the
baryonic density of the universe would be related to the helicity of the 
primordial magnetic field \cite{Vac94,Rob89}.

(i) {\it Primordial magnetic fields:}
A gas of electroweak segments is necessarily accompanied by a gas
of electroweak monopoles. The eventual collapse and disappearance of
electroweak strings removes all the electroweak monopoles but the
long range magnetic field emanating from the monopoles is
expected to remain trapped
in the cosmological plasma since that is a very good electrical conductor.
This will then lead to a residual primordial magnetic field in the present
universe. 

A quantitative estimate of the resulting primordial magnetic field
cannot be made with confidence but a dimensional estimate is
possible. An estimate for the average flux through an area $L^2 =
N^2/T^2$, where $N$ is a dimensionless number that relates the
length scale of interest, $L$, to the 
cosmological thermal correlation length $T^{-1}$,
was obtained in \cite{Vac91,Vac94}, and
then translated into the average magnetic field through that area.
The result is:
\be
B\vert_{area} \sim T^2 /N \ .
\label{magneticfieldresult}
\ee
(Magneto-hydrodynamical considerations
provide a lower bound $\sim 10^{12}$ cms on $L$ at the present epoch.)
It is important to remember that the above is an areal ({\it i.e.} flux)
average, defined by \cite{EnqOle93}
\be
B\vert_{area} \equiv \biggl \langle \biggl 
( {1\over A} \int d{\vec S} \cdot {\vec B}
                                            \biggr ) ^2 \biggr \rangle ^{1/2}
\label{averagebdefn}
\ee
where the surface integral is over an area $A$ and 
$\langle \cdot \rangle$ denotes ensemble averaging.

(ii) {\it Baryon number:}
A gas of electroweak string segments and loops would, in general, 
contain some helicity density of the Z-field. When the electroweak strings
eventually annihilate, it is possible that the helicity gets converted
into baryon number \cite{VacFie94,Vac94}. 
However, in Ref. \cite{FarGolGutRajSin95,FarGolLueRaj96}
it is argued that fractional quantum numbers of a soliton 
are unrelated to the number of particles produced when the soliton
decays.  Instead, only the change in the winding of the Higgs field
in a process that
starts out in the vacuum and ends up in the vacuum can be related to the
particle number. This would imply that we would have to consider the formation
of electroweak strings together with their decay before we can find the 
resulting baryon number. Such a calculation has not yet been attempted.

An interesting question is to consider what happens to the helicity in
the Z-field after the strings disappear. One possibility is that the
helicity gets transferred to a frozen-in residual magnetic field after
the strings have decayed. To see this, consider a linked pair of loops
as in Fig. \ref{linkedloops}. The strings can break by nucleating
monopole-antimonopole pairs, and then the string segments can shrink,
finally leading to monopole annihilation.  If this process happens in
the early universe, the loops will be surrounded by the ambient plasma
which will freeze-in the magnetic field lines. Hence, after the
strings have disappeared, we will be left with a linked pair of
magnetic field lines.  In other words, the original helicity in the
Z-field has been transferred to helicity in the A-field. This argument
relies on the freezing-in of the magnetic field emanating from the
monopoles and in the real setting the physics can be much more
complicated.  However, a connection between the baryon abundance of
the universe and the properties of a primordial magnetic field seems
tantalizing.  

{\it Stable strings at the electroweak scale:}
If in more exotic models, strings at the electroweak scale were
stable and had the superconducting properties discussed above, they could
be responsible for baryogenesis \cite{Bar95} and the presence of 
primary antiprotons in cosmic rays \cite{StaVac96}. The
production of antiprotons follows on realizing that any strings tangled
in the galactic plasma would be moving across the galactic magnetic field.
In the rest frame of the string, the changing magnetic field 
causes an electric field along the string according to Faraday's law.
The electric field along the string raises the levels of the u- and 
d-quark Dirac seas (see Fig. \ref{zeromode}), as well as the electron 
Dirac sea (not shown in the figure).
This means that the electric field produces quarks and leptons on
the string. The electric charges of the particles are in the ratio
$e:u:d::-1:+2/3:-1/3$ and the rate of production of these particles
due to the applied electric field is proportional to the charges.
Furthermore, the quarks come in three colors and so for every electron
that is produced, $3\times 2/3 =2$ u-quarks and $3\times 1/3=1$ d-quark 
are also produced. As a result, the net electric charge produced is
$1\times (-1)+2\times (2/3)+1\times (-1/3) = 0$.
However, net baryon number $2\times (1/3)+1\times (1/3)=1$
is produced because the quarks carry baryonic
charge $1/3$ while the baryonic charges of the leptons vanish.
Depending on the orientation of the string, either baryons or antibaryons 
will be produced. Some of these would then be emitted from the string and 
would arrive on earth as cosmic rays.

{\it Formation of strings in the electroweak phase transition.} 
Early attempts to understand the formation rates of electroweak
strings were made in \cite{Vac94} based on the statistical
mechanics of strings. The estimates indicate that a density of
strings will be formed immediately after the phase transition.
However, the application of string statistical mechanics to
electroweak strings may not be justified and so other avenues
of investigation are needed. An alternative approach to
study electroweak string formation was taken by Nagasawa and Yokoyama
\cite{NagYok96}. They assumed a thermal distribution of 
scalar field values and gradients, and estimated the probability of
obtaining a string-like scalar field configuration.  The conclusion
was that electroweak vortex formation in a thermal system is totally
negligible.  One possible caveat is that the technique used in
\cite{NagYok96} ignores the effect of gauge fields, which we know are
significant in the formation of related objects such as semilocal
strings.  In \cite{SafCop97}, Saffin and Copeland have evolved the
classical equations of motion to study the formation of electroweak
strings, and they found the presence of the gauge fields led
to larger string densities than one would have inferred from the
scalar fields alone, at least when $\sin^2 \theta_w = 0$. 
However, this study does not directly address the question of string
formation in a phase transition because no measure has been placed on
the choice of initial conditions and their choice may be too
restrictive. Most recently, a promising development has taken place
\cite{Che98} - calculations in lattice gauge theory have been done to
study the electroweak phase transition and there is evidence that
electroweak strings will form.  Further studies along these lines will
provide important and quantitative insight into the formation of
electroweak strings.

Using the results on the formation of semilocal strings, we
can gain some intuition about the formation of
electroweak strings in the region of parameter space close to the
semilocal limit (the region of stability in Fig. \ref{stabilityregion}).
We have seen that semilocal strings with $\beta <1$ have a non-zero
formation rate, increasing as $\beta \to 0$. Initially short segments
of string are seen to grow and join nearby ones because this reduces
the gradient energy at the ends of the strings. The ends of
electroweak strings are proper magnetic monopoles, and therefore the
scalar gradients are cancelled much more efficiently by the gauge
fields, but as $\sin \theta_w \to 0$ the cores of the monopoles 
get larger and larger, and they 
could begin to overlap with nearby monopoles, so it is possible that
short segments of electroweak string will also grow into longer ones.

\begin{figure}[tbp]
\caption{\label{zeromode} 
The dispersion relations for the u and d
quark zero modes are shown. The filled states are denoted by solid
circles while dashes denote unfilled states. For convenience, periodic
boundary conditions are assumed along the string and so the
momentum takes on discrete values. 
}
\vskip 1 truecm
\epsfxsize = \hsize \epsfbox{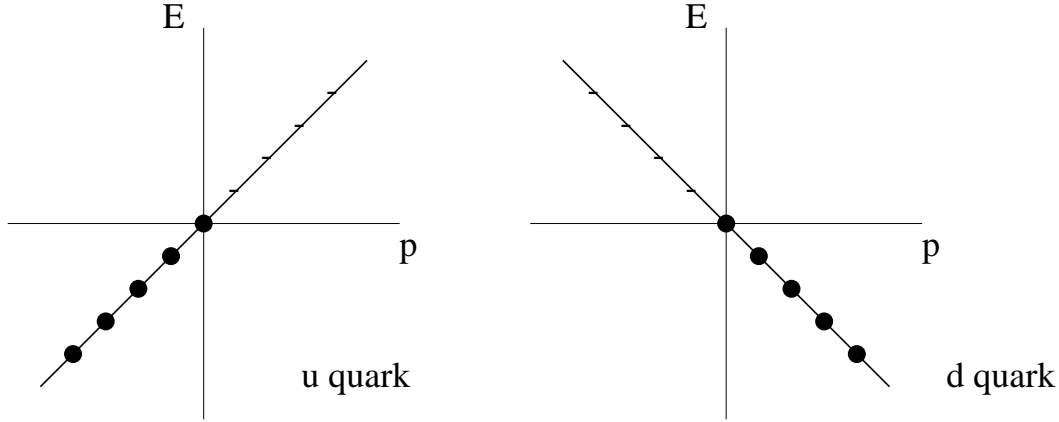}
\vskip 0.5 truecm
\end{figure}

\section{Electroweak strings and the sphaleron}
\label{ewstringandsphaleron}

The sphaleron is a classical solution in the GSW model
that carries baryon number $N_F/2$, where $N_F$ is the number 
of fermion families \cite{Man83,KliMan84}.
For $\theta_w =0$, the asymptotic form of the sphaleron Higgs field is:
\be
\Phi_{sph} = 
               \pmatrix{\cos\theta\cr \sin\theta ~ e^{i\varphi}\cr} \ .
\label{phisphaleron}
\ee
while the gauge fields continue to be given by 
eq. (\ref{gwbarmua})-(\ref{gpbarymu}) in which ${\bar \Phi}$ should
be replaced by $\Phi_{sph}$. (Note that the hypercharge gauge field
vanishes for $\theta_w =0$.)
Inside the sphaleron, the Higgs field vanishes at one point. The
sphaleron also has a magnetic dipole moment that has been 
evaluated for small values of $\theta_w$.
The reason that the sphaleron is important for particle physics
is that its energy defines the minimum energy required for the classical 
violation of baryon number in the GSW model. 

As has already been described in Sec. \ref{ewstringbaryonnumber}, 
non-trivial baryon number can be associated with linked and twisted
segments of electroweak string. Further, for specific values of the
link and twist, the baryon number of a configuration of Z-strings
can also be $N_F/2$. This raises the question: are sphalerons related
to Z-string segments? 

An early paper to draw a connection between the various solutions
in the GSW model is Ref. \cite{FujOtsToy89}. 
In \cite{VacFie94,HinJam94,Vac94,Hin94}, however, a direct correspondence 
between the field configuration of the Z-string and the sphaleron
was made. 

\subsection{Content of the sphaleron}
\label{sphaleroncontent}

In \cite{HinJam94} Hindmarsh and James evaluated the magnetic charge 
density and current density within the sphaleron. A subtlety in this 
calculation is that there is no unique definition of the electromagnetic
field when the Higgs field is not everywhere in the vacuum. The choice
adopted in \cite{HinJam94} (and also the choice in this review) is 
\be
F_{ij}^{em} = \sin\theta_w W_{ij}^{a} n^a + \cos\theta_w Y_{ij} \ .
\label{fijem}
\ee
The evaluation of the magnetic charge density (which is proportional
to the divergence of the magnetic field strength) clearly shows that
the sphaleron contains a region with positive magnetic charge density
and a region with negative magnetic charge density.  
Furthermore, the total charge in, say,
the positive charge region agrees with the magnetic charge of
a monopole. In addition, there is a flux of Z magnetic field connecting
the two hemispheres. This would seem to confirm that the sphaleron 
consists of a Z-string segment.
However, this is not the full picture. In addition to the string segment, 
Hindmarsh and James find that the electric current is non-zero in
the equatorial region and is in the azimuthal (${\hat e}_\varphi$)
direction.

\subsection{From Z-strings to the sphaleron}
\label{stringstosphaleron}

The scalar field configuration for a finite segment of Z-string was given
in Sec. \ref{dumbells}:
\begin{equation}
\Phi_{m\bar m} = \pmatrix{\cos(\Theta /2) \cr 
                          \sin(\Theta /2) ~ e^{i\varphi} \cr }
\label{phimbarm}
\end{equation}
where,
\be
\cos\Theta \equiv \cos\theta_m - \cos\theta_{\bar m} +1 
\label{angledefn}
\ee
and the angles $\theta_m$ and $\theta_{\bar m}$ are measured
from the monopole and antimonopole respectively, as shown 
in Fig. \ref{coords}.

\begin{figure}[tbp]
\caption{\label{coords} Definition of the coordinate angles $\theta_m$
and $\theta_{\bar m}$. The azimuthal angle, $\varphi$, is not shown.
}
\vskip 1 truecm
\epsfxsize = \hsize \epsfbox{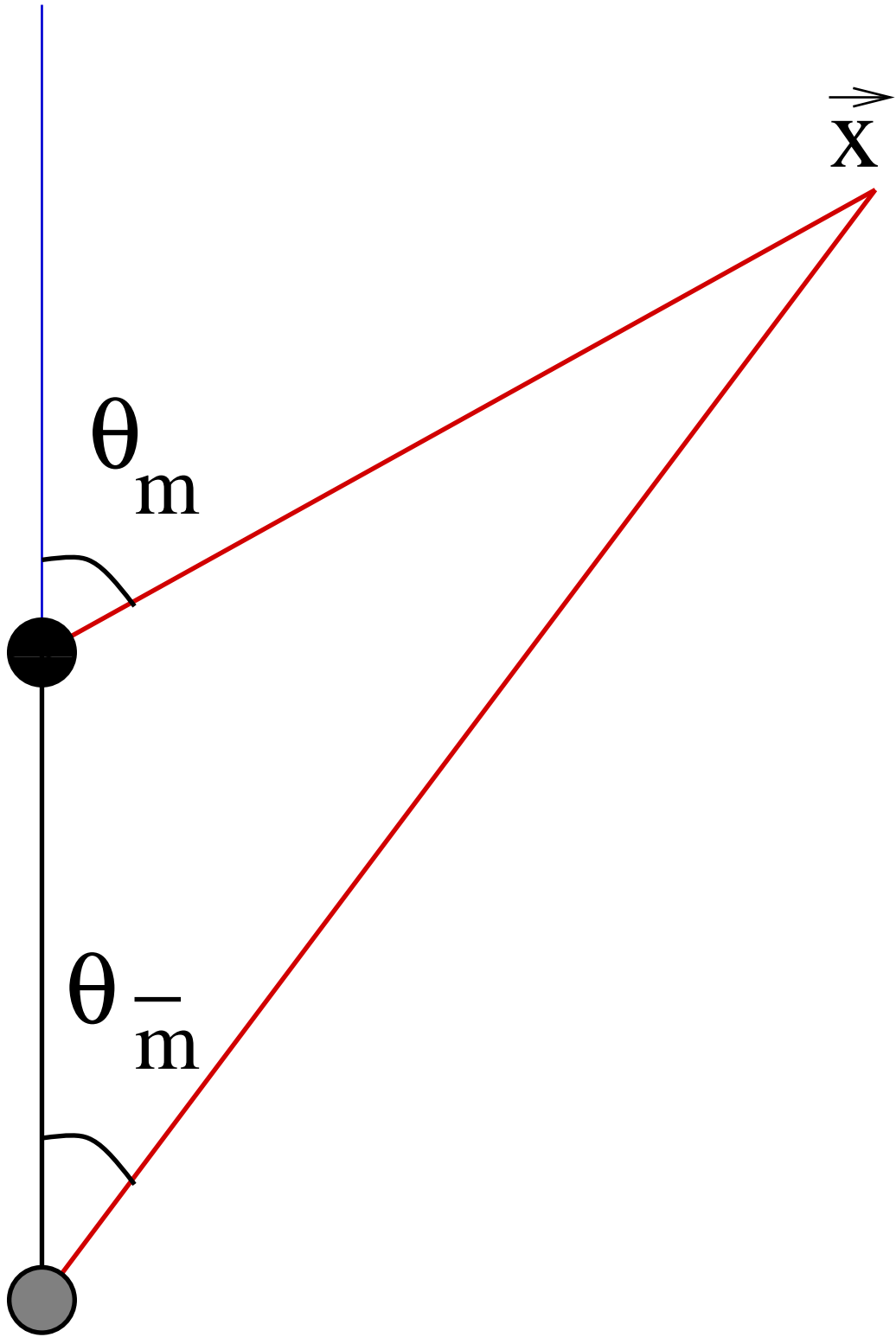}
\end{figure}

It is straightforward to check that (\ref{phimbarm}) yields the monopole 
field configuration close to the monopole ($\theta_{\bar m} \rightarrow 0$)
and the antimonopole configuration close to the antimonopole 
($\theta_{m} \rightarrow \pi$). It also yields a string singularity
along the straight line joining the monopole and antimonopole 
($\theta_m = \pi , ~ \theta_{\bar m} = 0$).
However, there are other Higgs field 
configurations that also describe monopoles and antimonopoles:
\begin{equation}
\Phi_m = e^{i\gamma} \pmatrix{ \cos(\theta_m /2) \cr
                     \sin(\theta_m /2) ~e^{i\varphi }\cr } \ , \ \ \ \
\Phi_{\bar m} = e^{i\gamma} \pmatrix{ \sin(\theta_{\bar m} /2) \cr 
                          \cos(\theta_{\bar m} /2) ~ e^{i\varphi}} \ .
\label{phimphibarm}
\end{equation}
 Next consider the Higgs field configuration:
\begin{equation}
\Phi_{m \bar m} (\gamma ) = 
\pmatrix{
            \sin(\theta_m /2) \sin(\theta_{\bar m} /2) e^{i\gamma}
          + \cos(\theta_m /2) \cos(\theta_{\bar m} /2) \cr 
   \sin(\theta_m /2) \cos(\theta_{\bar m} /2) e^{i\varphi }
 - \cos(\theta_m /2) \sin(\theta_{\bar m} /2) e^{i(\varphi - \gamma )} \cr
}
\label{twistedzstring}
\end{equation}
together with the gauge fields given by 
eq. (\ref{gwbarmua})-(\ref{gpbarymu}) with ${\bar \Phi}$ replaced
by $\Phi_{m\bar m} (\gamma )$. When we take the limit
$\theta_{\bar m} \rightarrow 0$ we find the monopole configuration
(with $\gamma =0$) and when we take $\theta_m \rightarrow \pi$ the 
configuration is that of an antimonopole (with arbitrary $\gamma$) 
provided we perform the spatial rotation
$\varphi \rightarrow \varphi + \gamma$. Note that the asymptotic gauge 
fields agree since these are determined by the Higgs field.
The monopole and antimonopole in (\ref{twistedzstring}) also have the 
usual string singularity joining them. This means that the configuration 
in eq. (\ref{twistedzstring}) describes a monopole and antimonopole 
pair that are joined by a Z-string segment that is twisted by an 
angle $\gamma$.  The Chern-Simons number of one such segment can be 
calculated \cite{VacFie94} and is
\be
Q_{CS} = N_F \cos2\theta_w ~ {{\gamma} \over {2\pi}} \ .
\label{qcsonesegment}
\ee
If $\gamma = \pi /\cos(2\theta_w )$ then the Chern-Simons number 
of the twisted segment of string is $N_F /2$ and is precisely that of the 
sphaleron. 

Given that the segment with twist $\pi /\cos(2\theta_w )$ has 
Chern-Simons number equal to that 
of the sphaleron, it is natural to ask if some deformation 
of it will yield the sphaleron. This deformation is not hard to guess for
the $\theta_w = 0$ case. In this case, if we let the segment size shrink
to zero, we have $\theta_m = \theta_{\bar m} = \theta$ and the Higgs field 
configuration of eq. (\ref{twistedzstring}) gives:
\be
\Phi_{m \bar m} (\gamma = \pi ) = 
               \pmatrix{\cos\theta\cr \sin\theta ~ e^{i\varphi}\cr} \ .
\label{phimbarmatpi}
\ee
This is exactly the scalar field configuration of the sphaleron
for $\theta_w =0$ (eq. (\ref{phisphaleron})). 
Note that the asymptotic gauge fields continue to be given by 
eq. (\ref{gwbarmua})-(\ref{gpbarymu}) and satisfy the requirement
that the covariant derivatives of the Higgs field vanish.

Encouraged by this successful connection in the $\theta_w =0$ case, it 
was conjectured in \cite{VacFie94,Vac94} that the sphaleron can also be 
obtained by collapsing a twisted segment of Z-string with 
Chern-Simons number $N_F /2$  {\it for any $\theta_w$}. If true, this
would mean that the asymptotic Higgs field configuration, $\Phi_S$, for the
sphaleron for arbitrary $\theta_w$ is given by
\be
\Phi_S = \pmatrix{
 \sin^2(\theta /2) ~ e^{i\gamma_S} + \cos^2(\theta /2) \cr 
 \sin(\theta /2) \cos(\theta /2) ~ e^{i\varphi} (1 - e^{-i\gamma_S})\cr }
\label{phisphaleronthetaw}
\ee
where $\gamma_S = \pi /\cos(2\theta_w )$. 

The twisting of the magnetic field lines in the sphaleron configuration
has been further clarified in \cite{Hin94}. The direction of magnetic 
field lines is shown for a dumbell in Fig. \ref{marksintra2} and for 
a ``stretched'' sphaleron in Fig. \ref{marksintra3}. (The asymptotic
fields for the stretched sphaleron are identical to those for the 
sphaleron and the twisted Z-string.) In the stretched sphaleron 
case, the magnetic field line twists around the vertical string segment 
by an angle $\pi$ (for $\theta_w \rightarrow 0$) as one goes from 
monopole to antimonopole. This twist provides non-trivial Chern-Simons
number to the configuration \cite{VacFie94}.

\begin{figure}[tbp]
\caption{\label{marksintra2} The thick solid line is the location
of the Z-string for a dumbell configuration and the dashed curves
lie in the equatorial plane and are drawn to guide the eye. The 
dotted lines depict lines of magnetic flux. The arrows show the
orientation of the vector 
${\hat n} \propto - \Phi^{\dag} {\vec \tau} \Phi$.
}
\vskip 1 truein
\epsfxsize = \hsize \epsfbox{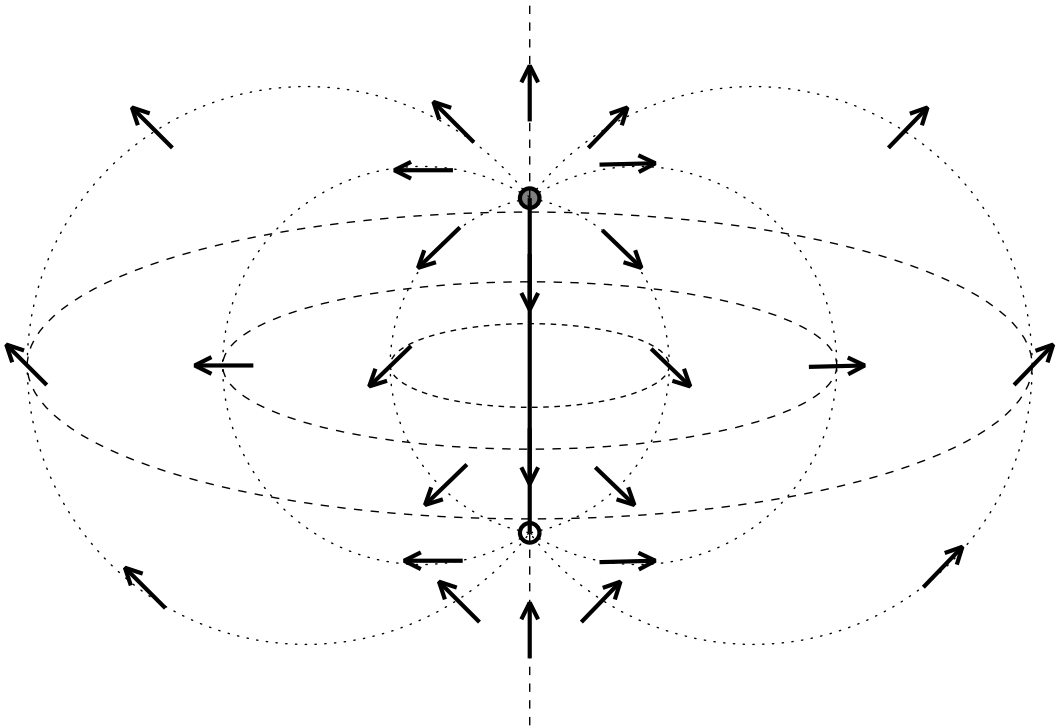}
\end{figure}

\begin{figure}[tbp]
\caption{\label{marksintra3} The field configuration for a stretched
sphaleron as in Fig. \ref{marksintra2}. Only one magnetic field line 
is shown. 
}
\vskip 1 truein
\epsfxsize = \hsize \epsfbox{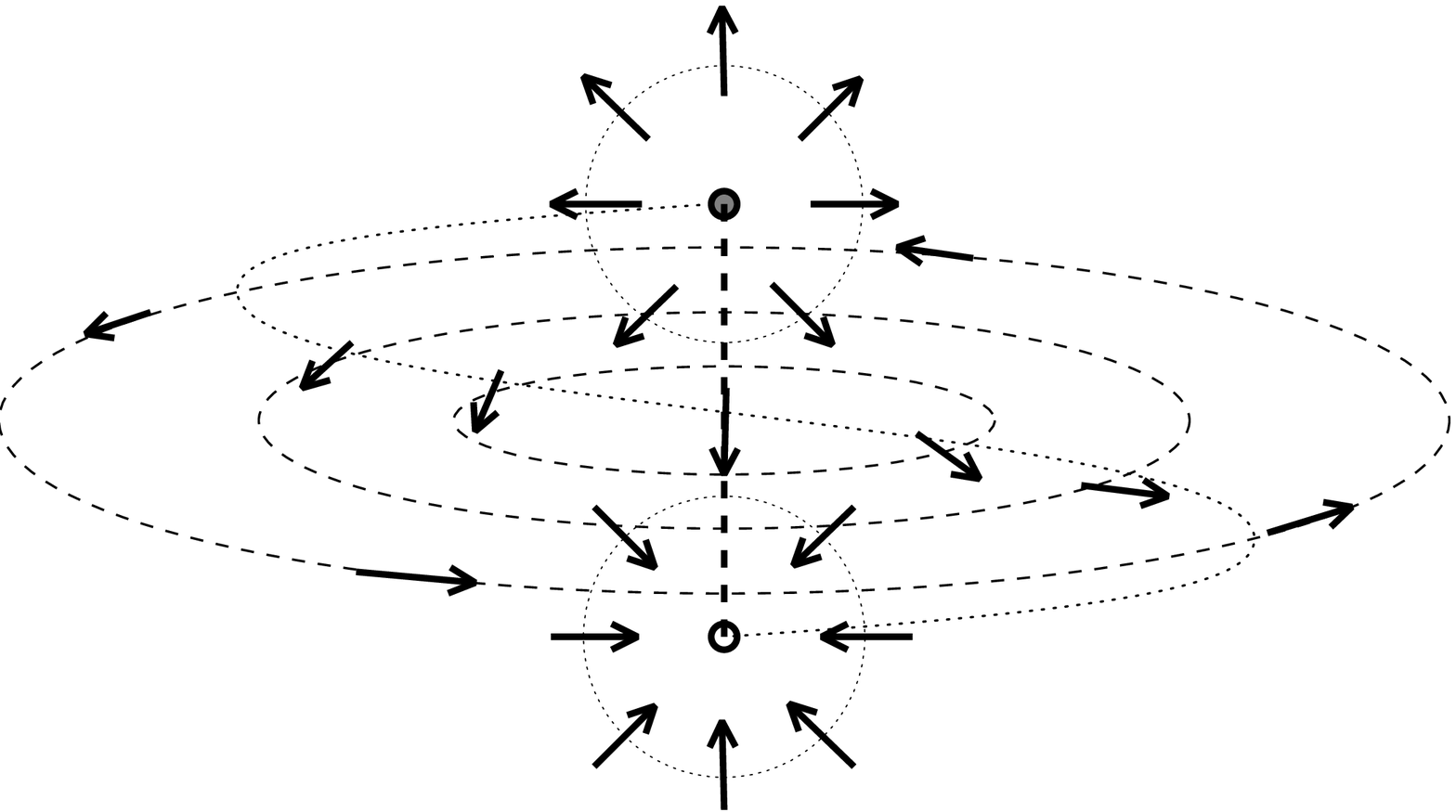}
\end{figure}

On physical grounds it seems reasonable that there should be a
critical value of twist at which one can get a static solution for a
Z-string segment. This is because the segment likes to shrink under
its own tension but the twist prevents the shrinkage and is equivalent
to a repulsive force between the monopole and antimonopole. (This idea
owes its origin to Taubes \cite{Tau82} who discovered a solution
containing a monopole and an antimonopole in an O(3) model in which
the Coulomb attraction is balanced by the relative misorientation of
the magnetic poles.) Then, if the string is sufficiently twisted, the
attractive force due to the tension and the repulsive force due to the
twist will balance and a static solution can exist. So far we have
been assuming that the only dynamics of the segment is towards
collapsing or expanding of the string segment.  However, since we are
dealing with twisted segments, we should also include the rotational
dynamics associated with twisting and untwisting.  So, while any twist
greater than a certain critical twist might successfully prevent the
segment from collapsing, only a special value of the twist can give a
static solution to the rotational dynamics. Furthermore, we expect
that this solution will be unstable towards rotations that twist and
untwist the string segment. This would be the unstable mode of the
sphaleron. 

Similar connections between the W--string and the sphaleron 
have also been constructed in \cite{AxeJohNieTor96}.

\section{The $^3$He analogy}
\label{he3section}

The symmetry structure of $^3$He closely resembles the electroweak
symmetry group and hence we expect the analog of electroweak strings
to exist in $^3$He \cite{VolWol90,Vol92,VolVac96}. 
Indeed, this analog is called the $n=2$ vortex.
We now explain this correspondence in greater detail.

\subsection{Lightning review of $^3$He}
\label{lightening}

$^3$He nuclei have spin $1/2$ and two such nuclei form a Cooper
pair which is the order parameter for the system. Unlike $^4$He,
the pairing is a spin triplet ($S=1$) as well as an orbital
angular momentum triplet ($L=1$). As a result there are $3\times 3$
components of the wavefunction of the Cooper pair - that is the
order parameter has 9 complex components. Hence, the order parameter
is written as a 3 by 3 complex valued matrix: $A_{\alpha i}$
with $\alpha$ (spin index) and $i$ (spatial index) ranging from 1 to 3. 

At temperatures higher than a few milli Kelvin the system is 
invariant under spatial rotations ($SO(3)_L$) as well as rotations 
of the spin degree of freedom of the Cooper pair ($SO(3)_S$). Another 
symmetry is under overall phase rotations of the wavefunction ($U(1)_N$)
and the corresponding conserved charge is particle number ($N$). 
Hence the symmetry group is:
\be
G = SO(3)_L \times SO(3)_S \times U(1)_N \ .
\label{he3symmetry}
\ee

There are several possible phases of $^3$He corresponding to different
expectation values of the order parameter. In the A-phase, the orbital
angular momenta of the Cooper pairs are all aligned and so are the
spin directions. This corresponds to
\be
A_{\alpha i} = \Delta_0 {\hat d}_\alpha \psi_i
\label{he3aorderparameter}
\ee
where $\Delta_0 \sim 10^{-7}$ eV is the temperature dependent gap
amplitude, the real unit vector ${\hat d}_\alpha$ is the spin part 
of the order parameter, and 
\be
\psi_i = {{{\hat m}_i + i{\hat n}_i} \over {\sqrt{2}}}
\label{psiimini}
\ee
with ${\hat m}$ and ${\hat n}$ being orthogonal unit vectors,
is the orbital part of the order parameter. 
This expectation value of the order parameter
leads to the symmetry breaking:
\be
G \rightarrow U(1)_{S_3}\times U(1)_{L_3-N/2} \times Z_2\ .
\label{he3asymbreaking}
\ee
The reason why a $U(1)$ subgroup of $SO(3)_L \times U(1)_N$ survives
the symmetry breaking can be derived from the expectation value
in eq. (\ref{he3aorderparameter}). A spatial rotation of the order 
parameter is equivalent to a phase rotation of $\psi_i$ and this phase 
can be absorbed by a corresponding $U(1)_N$ rotation of the order parameter.
Hence, just as in the electroweak case, a diagonal $U(1)$ subgroup
remains unbroken. The $U(1)_{S_3}$ survives since rotations about
the ${\hat d}$ axis leave the order parameter invariant. The non-trivial
element of the residual discrete $Z_2$ symmetry corresponds to a sign 
inversion of both $\psi_i$ and ${\hat d}$. A depiction of the A-
and B- phases is shown in Fig. \ref{he3phases} (after \cite{Liu82}).

\begin{figure}[tbp]
\caption{\label{he3phases} Depiction of the A- and B- phases of
$^3$He. In the A-phase, the spin orientations of 
all the Cooper pairs are parallel and so are the orbital orientations.
In the B-phase, the relative orientation of the spin and
orbital orientations are fixed in all the Cooper pairs but
neither the spin nor the orbital orientations of the various 
Cooper pairs are aligned.
}
\vskip 1truecm
\epsfxsize = \hsize \epsfbox{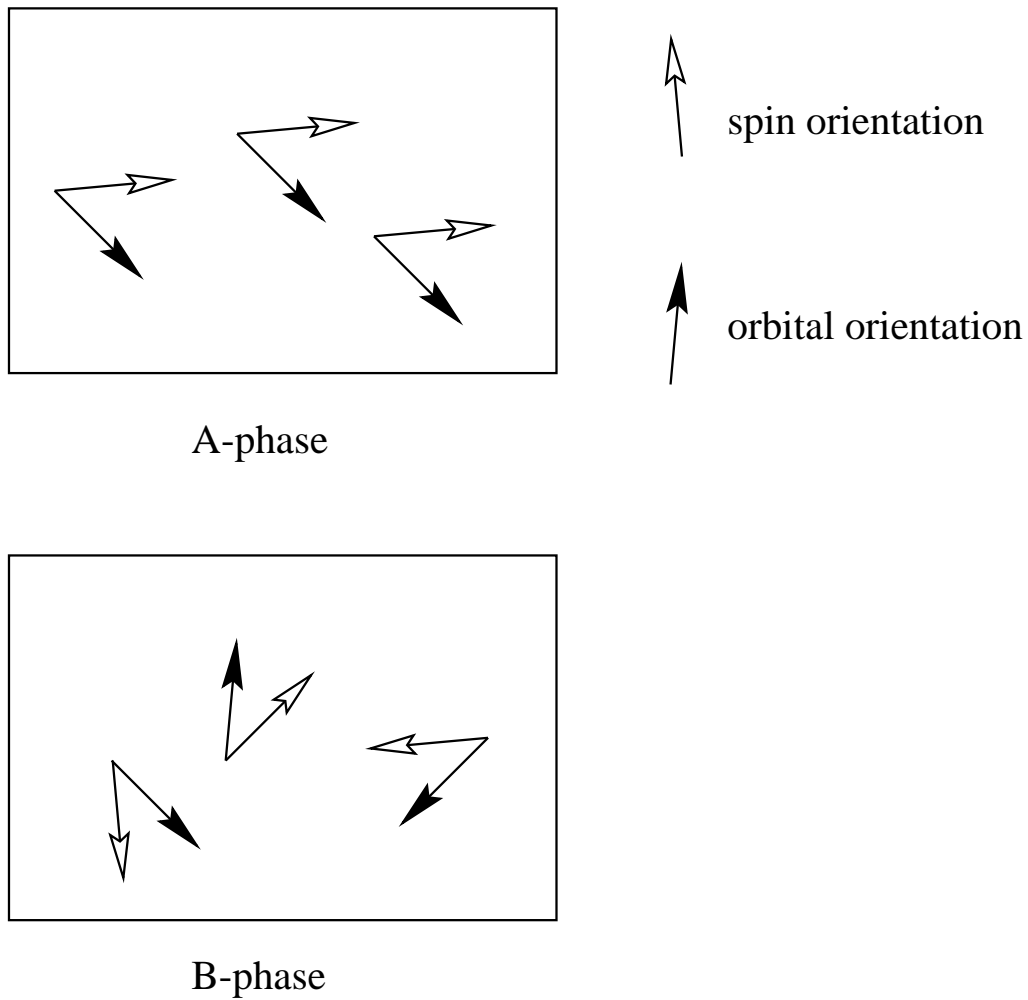}
\end{figure}

In the B-phase, neither the orbital angular momenta nor the spin 
directions of the different Cooper pairs are aligned. But the 
angle between the direction of the angular momenta and the spin
direction is fixed throughout the sample. Hence in the 
B-phase, independent rotations of the orbital angular momenta 
and of spin are no longer symmetries. However, a simultaneous 
rotation of both orbital angular momenta and spin remains an
unbroken symmetry. In other words, a diagonal subgroup of
$SO(3)_S \times SO(3)_L$ remains unbroken. Therefore, in the
B-phase the order parameter is written as:
\be
A_{\alpha i} = 3^{-1/2} e^{i\phi} 
                 R_{\alpha i} ({\hat n}, \theta )
\label{he3borderparameter}
\ee
where, $\phi$ is a phase and the $3\times 3$ matrix $R_{\alpha i}$
describes relative rotations of the spin and orbital degrees of
freedom about an axis ${\hat n}$ and by angle $\theta$. The
symmetry breaking pattern is:
\be
G \rightarrow SO(3)_{L_3+S_3} \ .
\label{he3bsymbreaking}
\ee
This symmetry breaking resembles the chiral symmetry breaking
transition studied in QCD (with two flavors of quarks) and may be 
useful for experimentally investigating phenomenon such as the 
formation of ``disoriented chiral condensates'' \cite{Bjo97}.
The B-phase does not resemble the electroweak model and hence
we will not discuss it any further. We shall also not
discuss the various other phases of $^3$He 
(for example, the ${\rm A}_1$ phase) which are known to occur. 
(For a useful chart of the phases, see Sec. 6.2 of 
Ref. \cite{VolWol90}.)

In addition to the continuous symmetries, there are a number of
discrete symmetries that arise in the phases of $^3$He. These are 
important for the classification of topological defects in $^3$He. 
A description may be found in \cite{SalVol87}.

\subsection{Z-string analog in $^3$He}
\label{zstringinhe3}

Clearly the A-phase closely resembles the electroweak symmetry
breaking because of the mixing of the generator of the non-Abelian
group ($SO(3)_L$) and the Abelian group ($U(1)_N$). The orbital 
part of the order parameter is responsible for this pattern of
symmetry breaking and hence $\psi_i$ plays the role of the electroweak
Higgs field $\Phi$. The connection, however, is indirect since
$\psi_i$ is a complex 3 vector while $\Phi$ is a complex doublet.
The idea is that the $^3$He-A real vector
\be
{\hat l}_{HeA} = i{ {{\vec \psi} \times {{\vec \psi}^{\dag}}} \over
              {{\vec \psi}^{\dag} {\vec \psi}} } =
               {\hat m} \times {\hat n}
\label{lhathea}
\ee
is analogous to the electroweak real vector
\be
{\hat l}_{ew} = - {{\Phi^{\dag} {\vec \tau} \Phi} \over
                   {\Phi^{\dag} {\vec \tau} \Phi} } \ .
\label{hlatew}
\ee
The electroweak Z-string is a non-topological solution for which
the Higgs field configuration is:
\be
\Phi = {\eta \over {\sqrt{2}}} f(r) e^{i\varphi}
           \pmatrix{0\cr 1\cr} \ .
\ee
For this configuration ${\hat l}_{ew} = {\hat z}$. 

The vacuum manifold $M_A$ of $^3$He-A has
\be
\pi_1 ( M_A ) = Z_4
\label{pi1he3a}
\ee
and hence there are topological $Z_4$ vortices in $^3$He-A. The
vortices occur in classes labeled by $n=\pm 1/2,1$. The vortices
with $n$ equal to an even integer are topologically equivalent to
the vacuum. The non-trivial topological vortices 
(labeled by $n=\pm 1/2,1$) cannot be the 
equivalent of the non-topological Z-string. However, the topologically
trivial $n=-2$ vortex is also seen in $^3$He-A. The order parameter
for this vortex is:
\be
A_{\alpha j}(\rho ,\varphi)=
\Delta_0 \hat z_{\alpha }[ \ 
e^{in\varphi}  f_1(\rho )(\hat x_j+i  \hat y_j)  +\ 
 e^{i(n+2)\varphi}  f_2(\rho )(\hat x_j-i  \hat y_j)]~~.
\label{opforneq2}
\ee
where $f_1(\rho )$ and $f_2(\rho )$ are two profile functions
with $f_1 (\infty )=1$, $f_2(\infty ) =0$, $f_1 (0) =0$ and 
$f_2 (0)$ depending on $n$. In correspondence
with the electroweak Z-string, the $n=2$ vortex 
has ${\hat l}_{HeA} = {\hat z}$. However, the order parameter
need not vanish at the center of the vortex for certain members
of the $n=2$ class of vortices. For example, with $n=-2$, we
may have $f_2(0)\ne 0$.

The $n=2$ vortex is not topological and can be
continuously deformed into the vacuum manifold. The configuration
at the terminus of the $n=2$ vortex is called the 
hedgehog or monopole ${\hat l}_{HeA} = {\hat r}$
(the radial unit vector). This texture is the
direct analog of the electroweak magnetic monopole 
(${\hat l}_{ew} = {\hat r}$) at the terminus of a Z-string.

The $n=2$ discontinuous vortex is unstable but even so 
has been observed in $^3$He. In the laboratory, the rotation of
the sample stabilizes the $n=2$ vortex. This seems to be closely
analogous to the result of Garriga and Montes \cite{GarMon95}
who find that electroweak strings can be stabilized by external 
magnetic fields (Sec. \ref{stabilitycontd}).

Before proceeding further, it is prudent to remind ourselves
of some important differences between the (bosonic sector of the) 
GSW model and $^3$He. The symmetries in $^3$He are all
global whereas the symmetries in the GSW model are all
local. So the $n=2$ discontinuous vortex is like a global analog
of the Z-string. Another important difference is in the discrete 
symmetries in the two systems. The symmetry structure of the GSW  
model is really $[SU(2)\times U(1)]/Z_2$
since the $Z_2$ elements ${\bf 1}$ and $- {\bf 1}$ which form
the center of $SU(2)$ also occur in $U(1)$. On the contrary,
the symmetry group of $^3$He-A has a multiplicative $Z_2$
factor which gives rise to the non-trivial topology of the
vacuum manifold.

It is important to note that we cannot expect $^3$He to provide an 
exact replica of the GSW model. However, the similar
structures of the two systems means that certain issues can be
experimentally addressed in the $^3$He context while they are
far beyond the reach of current particle physics experiments.
An issue of this kind is the baryon number anomaly in the
GSW model and the anomalous generation of momentum
in $^3$He.

As described in Sec. \ref{superconductivity}, there are fermionic 
zero modes on the Z-string and an electric field applied along the 
Z-string leads to the anomalous production of baryon number. What is the
corresponding analog in $^3$He? At first sight, $^3$He does not
have the non-Abelian gauge fields that the electroweak string has
and so it seems that the analogy is doomed. But this is not true.
The point is that the physics of fermionic zero modes has to do
with the dynamics of fermions on the {\em fixed} background of the 
Z-string. Likewise, in $^3$He we can be interested in the dynamics 
of quasiparticles in the fixed background of the $n=2$ vortex.
As far as the interaction of quasiparticles with the order
parameter background is concerned, one can think of the $^3$He-A
vortex as being due to a to a (fictitious) gauge field ${{Z^\prime}^\mu}$. 
Then the interaction of quasiparticles with the order parameter is of the 
form $j_\mu {{Z^\prime}^\mu}$ which is exactly analogous to the interaction
of quarks and leptons with the Z-boson. Just as in the electroweak
case, the $^3$He quasiparticles have zero modes on the vortex.
In close analogy with the scenario where the motion of a 
superconducting string through an external magnetic field
leads to currents along the string (Sec. \ref{cosmologicalapplications}), 
the velocity of the $^3$He vortex through the superfluid leads
to an anomalous flow of quasiparticles but this time in the
direction perpendicular to the vortex. This flow 
causes an extra force on the vortex as it moves through the
superfluid that can be monitored experimentally. Such a force was
measured in the Manchester experiment \cite{Bevetal97,Vol98}
and is in excellent agreement with theoretical predictions. 
Hence the Manchester experiment verifies the anomalous production of 
quasiparticle momentum on moving vortices and the corresponding
production of baryon number on electroweak strings moving through
a magnetic field.

\section{Concluding remarks and open problems}
\label{conclude}

Quantum field theory has been very successful in describing
particle physics. Yet the successes have mostly been relegated
to perturbative phenomena. A more spectacular level of success
will be achieved when our field theoretic description of
particle physics is confirmed at the non-perturbative level.
The first non-perturbative objects that are likely to be encountered 
in this quest are topological defects and their close cousins
that we have described in this review.

The search for topological defects can be conducted in 
accelerator experiments or in the cosmological realm via 
astronomical surveys. These searches are complementary - only
supermassive topological defects can be evident in astronomical
surveys, while only the lightest defects can potentially be produced
in accelerators. Foreseeable accelerator experiments give us
access only to topological defects at the electroweak symmetry breaking 
scale. So it is very important to understand the 
defects present in the standard electroweak model and all its
viable extensions. One may hope that the structure of defects
will yield important clues about the underlying symmetry of
the standard model.

With this hope, we have described wide classes of defects
present in field theories. These defects are not all topological
and this is relevant to the standard electroweak model which also
lacks the non-trivial topology needed to contain topological defects.
The absence of topology in the model means that the defect solutions
cannot be enumerated in topological terms and neither can their
stability be guaranteed. We have described, however, how the
existence of defect solutions may still be derived by examining
the topological defects occurring in subspaces of the model.
The electroweak defects can be thought of as being topological
defects that are embedded in the electroweak model.
 
The issue of stability of the defect solution is yet more involved 
and has not yet been fully resolved in the presence of fermions.
That the electroweak Z-string is stable for large $\theta_w$ 
(within the bosonic sector) was inspired by the discovery of
semilocal strings and their stability properties. The explicit
stability analysis of the electroweak string marks out the
region of parameter space in which the Z-string is stable.
Then it is clear that the Z-string is unstable for the parameters
of the standard model. In certain viable extensions of the standard
model and under some external conditions (such as an external
magnetic field), the standard electroweak Z-string can still be 
stable.

Even if the Z-string is unstable, it is possible that the lifetime
of segments of string is long enough so that they can be observed
in accelerators. This possibility was discussed in the first
paper on the subject by Nambu \cite{Nam77}. The discovery of
Z-string segments would truly be historic since it would confirm 
the existence of magnetic monopoles in particle physics.
However, the rate of formation of Z-string segments and their
lifetime has not yet been studied in detail. Some of the 
difficulties in this problem lie quite deep since they involve the 
connection of perturbative particle physics to the non-perturbative
solitonic features. Additionally, the influence of fermions
on electroweak strings needs further investigation.

Electroweak strings may play a cosmological role in the genesis
of matter over antimatter as is evident since configurations
of electroweak string have properties that are similar to the
electroweak sphaleron. The challenge here is to determine the
number density of electroweak strings formed during the electroweak
phase transition and their decay rate. Note that the formation
of topological strings has been under constant examination over
the last two decades and only now, with some experimental input,
are we beginning to understand their formation. The cosmological
formation of electroweak defects has not been addressed with as 
much vigour. Recently though, there have been spurts of activity
in this area, with lattice calculations beginning to
shed interesting insight \cite{Che98}. 
It is very likely that further lattice results will be
able to give quantitative information about the formation of
electroweak strings at the electroweak phase transition.

While particle physics experiments to detect electroweak
strings are quite distant, experiments in condensed matter
systems to study topological defects
are becoming more feasible and can be used to test
theoretical ideas that are relevant to both particle 
physics and condensed matter physics. Already there are 
experiments that test theories of the formation of 
topological vortices. We can also expect that condensed matter
experiments might some day test the formation of defects that
are not topological. The experiments on He$^3$ are most relevant
in this regard since it contains close analogs of electroweak 
strings. Furthermore, ideas relating to the behaviour of fermions
in the background of electroweak strings can also be tested in
the realm of He$^3$. This makes for exciting physics in the
years to come which will stimulate the growth of particle physics,
cosmology and both, theoretical and experimental, condensed matter
physics.

\section{Acknowledgements}

We are grateful to G. Volovik for very useful comments on section
\ref{he3section}, to M. Groves and W. Perkins for an early draft of
their paper, and to M. Hindmarsh for the sphaleron figures in section
\ref{ewstringandsphaleron}. AA thanks K. Kuijken and
L. Perivolaropoulos for help with some of the figures in sections
\ref{NO} and \ref{semilocal}, and J. Urrestilla for pointing out
several typos in an earlier draft.

This work was supported by a NATO Collaborative Research Grant CRG
951301, and our travel was also partially supported by NSF grants
PHY-9309364 (AA), as well as by UPV grant UPV 063.310-EB187/98 and
CICYT grant AEN-93-0435. TV was partially supported by the
Department of Energy, USA.
TV thanks the University of the Basque
Country, and AA thanks Case Western Reserve University, Pierre Van
Baal and Leiden University for their hospitality.

\end{document}